\documentclass[11pt,a4paper]{article}

\usepackage[utf8]{inputenc}
\usepackage[english]{babel}
\usepackage{amsmath,amsfonts,dsfont,amssymb,graphicx,rotating, setspace, booktabs, tikz, natbib, xcolor, enumerate, multirow, array, algorithm, algorithmic}
\usepackage[toc,page]{appendix}
\usepackage[left=2cm,right=2cm,top=3cm,bottom=3cm]{geometry}
\usepackage[skip=10pt, font=small, labelfont=bf]{caption}
\usepackage{hyperref}
\hypersetup{
    colorlinks=true,
    linkcolor=blue,
    citecolor=blue,
    filecolor=blue,      
    urlcolor=cyan,
}

\graphicspath{{figures/}}
\newtheorem{defn}{Definition}
\newtheorem{assumpt}{Assumption}
\newtheorem{theorem}{Theorem}
\newtheorem{example}{Example}

\DeclareMathOperator{\tr}{tr}
\DeclareMathOperator{\Y}{Y}
\DeclareMathOperator{\X}{X}
\DeclareMathOperator{\W}{W}
\DeclareMathOperator{\w}{w}
\DeclareMathOperator{\E}{E}

\newcommand{\bl}[1]{\mathbf{#1}}
\newcommand{\bs}[1]{\boldsymbol{#1}}

\newcommand{\info}[2]{\mathcal{#1}_{#2}}
\newcommand{\R}{R }
\newcommand{\keywords}[1]{\small \textbf{\textit{Keywords---}} #1}

\begin{document}
\author{Fiammetta Menchetti \\ Università di Firenze \\ \texttt{fiammetta.menchetti@unifi.it} \\
\and
Iavor Bojinov \\ Harvard Business School \\ \texttt{ibojinov@hbs.edu} \vspace{8pt} \\}
\title {Estimating the effectiveness of permanent price reductions for competing products using multivariate Bayesian structural time series models.}
\date{\today}

\maketitle

% ----------------------------------------------------------------------------------------
% NOTE 
%-----------------------------------------------------------------------------------------
% The manuscript is the latest file available, updated on Oct 2, 2020 
% (i.e., the same version that was sent to Arxiv)
% ----------------------------------------------------------------------------------------

\begin{abstract}
The Florence branch of an Italian supermarket chain recently implemented a strategy that permanently lowered the price of numerous store brands in several product categories. To quantify the impact of such a policy change, researchers often use synthetic control methods for estimating causal effects when a subset of units receive a single persistent treatment, and the rest are unaffected by the change. In our applications, however, competitor brands not assigned to treatment are likely impacted by the intervention because of substitution effects; more broadly, this type of interference occurs whenever the treatment assignment of one unit affects the outcome of another. This paper extends the synthetic control methods to accommodate partial interference, allowing interference within predefined groups but not between them. Focusing on a class of causal estimands that capture the effect both on the treated and control units, we develop a multivariate Bayesian structural time series model for generating synthetic controls that would have occurred in the absence of an intervention enabling us to estimate our novel effects. In a simulation study, we explore our Bayesian procedure’s empirical properties and show that it achieves good frequentists coverage even when the model is misspecified. We use our new methodology to make causal statements about the impact on sales of the affected store brands and their direct competitors. Our proposed approach is implemented in the CausalMBSTS \R package.

\end{abstract}


%focusing on the introduction of a new price policy by a supermarket chain. 

\newpage

\section{Introduction}
\label{sect:intro}
On October $4^\text{th}$, 2018, the Florence branch of a large Italian supermarket chain permanently lowered the price of $707$ store brands in several product categories. In the past, the firm had regularly used temporary promotions (discounting products for a brief period); however, the new permanent price reduction represented a significant strategic shift in its business model. The firm hypothesized that the lower price would expand its customer base, increasing sales and, ultimately, revenue. To evaluate the success of the new strategy, we model the permanent price reduction as a single persistent intervention. In this paper, we focus on the cookies product category and estimate the causal effect of permanently reducing the price of 10 store branded cookies on daily sales. 

A popular approach for obtaining estimates of causal effects from panel data with a single intervention is to use synthetic control methods (e.g., \cite{Abadie:Gardeazabal:2003,Abadie:Diamond:Hainmueller:2010,Abadie:Diamond:Hainmueller:2015,Brodersen:Gallusser:Koehler:Remy:Scott:2015}). Unlike traditional difference-in-difference methods, synthetic controls provide a more flexible framework as they directly impute the unobserved outcome for treated time series by combining data from multiple control series that were not directly impacted by the treatment but are, nevertheless, correlated with the counterfactual outcome \citep{ONeill:Kreif:Grieve:Sutton:Sekhon:2016}. In our supermarket study, daily wine sales could be a suitable control series because changes in the price of cookies are unlikely to affect wine sales; instead, the control series captures temporal trends that are useful in modeling how the sales of cookies would have evolved in the absence of a price reduction. More broadly, synthetic control methods have been successfully applied to evaluate the effectiveness of policy changes in healthcare \citep{Kreif:Grieve:Hangartner:Turner:Nikolova:Sutton:2016,Choirat:Dominici:Mealli:Papadogeorgou:Wasfy:Zigler:2018,Viviano:Bradic:2019}, economics \citep{Billmeier:Nannicini:2013,Abadie:Diamond:Hainmueller:2015,Dube:Zipperer:2015,Gobillon:Magnac:2016,Benmichael:Feller:Rothstein:2018}, marketing and online advertising \citep{Brodersen:Gallusser:Koehler:Remy:Scott:2015,Li:2019}, amongst others. 

Typically, synthetic control methods assume that there is no interference between experimental units; that is, the assignment any unit receives has no bearing on the outcome of any other unit \citep{Cox:1958}. However, there are many applications where this assumption is violated (e.g., \cite{Hudgens:Halloran:2008,Tchetgen:VanderWeele:2012}, and \cite{Basse:Feller:Toulis:2019}). In our study, for each store brand cookie, the firm identified a competitor brand that is a direct substitute differing primarily in the brand name. As traditional economic theory suggests, if two goods are substitutes, lowering the price of one will impact the sales of the other \citep{Nicholson:Snyder:2012}; therefore, any price changes to the store brand will impact the sales of the direct competitor, and vice-versa—violating the no interference assumption. Beyond the direct competitor, it is reasonable to assume that the price reduction will have negligible effects on the sales of other products. 

More broadly, the setting where units interfere within predefined groups without interfering across these groups is known as partial interference \citep{Sobel:2006} and has been extensively studied for cross-sectional data (e.g., \cite{Rosenbaum:2007}, \cite{Hudgens:Halloran:2008}, and \cite{Forastiere:Airoldi:Mealli:2020}). In panel settings, like the supermarket study, partial interference has received relatively less attention, partly because of the added complications induced by the temporal component. In practice, authors often sidestep the issue by aggregating units that are likely to interfere with each other, generating a single treated time series that now satisfies the no-interference assumption \citep{bojinov2020importance}. One obvious downside of this approach is the inherent loss of information and a decreased ability to detect heterogeneous treatment effects. 

To tackle this issue directly, we extend the synthetic control framework to partial interference setting by leveraging the extended potential outcomes that allow both spillovers across units and time \citep{Robins:1986,Robins:Greenland:Hu:1999,VanderWeele:2010,Bojinov:Shephard:2019,Bojinov:Rambachan:Shephard:2020}. We then define new classes of causal effects that capture the impact of an intervention on both the unit that received it and the units within the same group. To perform inference, we derive the multivariate version of the popular Bayesian structural time-series model for causal inference introduced in \cite{Brodersen:Gallusser:Koehler:Remy:Scott:2015}. Like its univariate counterpart, our model allows for a great deal of flexibility due to its ability to incorporate trends and seasonality effects. To fit the model, we provide a Markov chain Monte Carlo algorithm and describe how to use the resulting draws to estimate our causal effects; all algorithms are implemented in the CausalMBSTS \citep{causalMBSTSRpackage2020} \R package. We then use a small simulation study to investigate the frequentist properties of our proposed approach and our ability to use posterior predictive checks \citep{Rubin:1984} to assess the model fit. 

In the supermarket study, our framework treats every store-competitor pair jointly, allowing us to model the group-specific interference directly. To determine the intervention's impact, we use our Bayesian structural time-series model to estimate the causal effect at various time horizons. The results show that the new strategy had a minor, short-term impact on store brands' sales; interestingly, we do not detect significant effects on the competitor brands. In contrast, performing the equivalent aggregated analysis that ignores the interference incorrectly concludes that the price reduction positively affected sales. 
% Our results also suggest that an analysis that aggregates the units leads to misleading results because of the inherent loss of information.

Two papers consider our setup of partial-interference on panel or time series data; \citet{Cao:Dowd:2019} and \citet{Grossi:Lattarulo:Mariani:Mattei:2020}. \citet{Cao:Dowd:2019}, develops a model that requires that the impact of an intervention on one unit to the other is linear with an unknown parameter. Our paper imposes no such restriction, making it much more generally applicable. \citet{Grossi:Lattarulo:Mariani:Mattei:2020} formulation focuses on a context where only a single unit is intervened on, while the others are assigned to control. Since the treatment received by that unit may affect the outcomes of the other units, they rely on the partial interference assumption and identify different clusters such that the units belonging to different clusters do not interfere with each other. The inference is restricted to the group containing the treated unit, while the others form the ``donor pool'' used to construct synthetic controls by combining the donor outcomes. Our work presents a generalization of their specific context. Again, we study partial interference, but we allow for the existence of multiple treated units. By extending the univariate Bayesian structural time series model to the multivariate setting, we can also model the interference between units in the same cluster by explicitly modeling their dependence structure while transparently dealing with the surrounding uncertainty. 

The paper is structured as follows. In Section \ref{sect:causal_framework}, we present our causal framework, defining the treatment assignments, potential outcomes, causal effects, and our underlying assumptions. In Section \ref{sect:MBSTS}, we introduce the multivariate Bayesian structural time series model and explain how to apply it to our setting. In Section \ref{sect:simulation_study}, we detail a simulation study that tracks our approach's performance. In Section \ref{sect:empirical_analysis}, we provide the details of our supermarket study analysis. The final section presents our concluding remarks.

\section{Causal framework}
\label{sect:causal_framework}

In this section, we outline our framework for estimating the causal effect of an intervention in a panel setting with partial interference among statistical units. Throughout, we illustrate key concepts and definitions by leveraging our analyses of the Italian supermarket chain's new price policy. In our empirical example, the statistical units are grouped into pairs, and so we begin by introducing the notation for a bivariate outcome variable; we then provide extensions to general group sizes. We conclude the section by defining our causal effects. 

\subsection{Notation}
\label{subsect:notation}

Throughout, we use a superscript $s$ to denote the store brand and $c$  the competitor brand.  At time $t \in \{1, \dots, T \}$, for each pair $j \in \{1,\dots,J \}$, let $\W_{j,t}^{(s)}\in\mathcal{W}$ be the treatment assignment for the store brand, $\W_{j,t}^{(c)}\in \mathcal{W}$  be treatment assignment for the competitor brand, and $\bl{W}_{j,t} = (\W_{j,t}^{(s)}, \W_{j,t}^{(c)})\in\mathcal{W}^2$ the pair assignment. We mostly focus on the binary treatment case, where $\mathcal{W}=\{0,1\}$; following convention, we refer to ``1'' as treatment and ``0'' as control. In our supermarket study, each pair is assigned to one of four possible treatments: no permanent price reduction $\bl{W}_{j,t}=(0,0)$, both receive a permanent price reduction $\bl{W}_{j,t}=(1,1)$, store brand receives a permanent price reduction only $\bl{W}_{j,t}=(1,0)$, or competitor brand receives a permanent reduction only $\bl{W}_{j,t}=(0,1)$. We then define the assignment path for each pair as the matrix $\bl{W}_{j,1:T} = (\bl{W}_{j,1}, \dots, \bl{W}_{j,T})'\in \mathcal{W}^{2\times T}$, and the assignment panel that captures the assignments of all units throughout the study as $\bl{W}_{1:J,1:T} = (\bl{W}_{1,1:T}', \dots, \bl{W}_{J,1:T}')\in \mathcal{W}^{2J\times T}$. We will use this vector and matrix notation for other variables, but will sometimes drop the subscript if the dimensions are obvious from the context. Realizations of random variables will be denoted by their lower case; for example, $\bl{w}_{j,t}$ will denote a sample from $\bl{W}_{j,t}$. 

In the panel set up, the pairs can change their assignment at any point in time, but to keep our notation less cumbersome, we only focus on the case when there is a single persistent policy change, as was the case in our supermarket study. 

\begin{assumpt}[Single intervention]\label{assumption:single-intervention}
    We say pair $j$ received a single intervention, if there exists a $t_j^\ast \in \{1, \dots, T \}$ such that for all $t \le t_j^\ast$ we have $\bl{W}_{j,t}=(0,0)$ and for all $t,t' > t_j^\ast$ we have $\bl{W}_{j,t}=\bl{W}_{j,t'}$. If all pairs receive a single intervention, then we say the study is a single intervention panel study. For simplicity, we also assume that the intervention happen simultaneously, that is, $t_j^\ast=t_{j'}^\ast = t^\ast$. 
\end{assumpt}

We maintain Assumption \ref{assumption:single-intervention}, which allows us to drop the $t$ subscript from the treatment assignment so that $\bl{W}_j =( \W_j^{(s)},\W_j^{(c)}) \in \{0,1\}^{2}$ for all $t > t^\ast$ and $\bl{W}_j=(0,0)$ for $t\le t^\ast$. 

\subsubsection{Potential outcomes}
We now define the potential outcomes that describe what would be observed for a particular pair at a fixed point in time for a given assignment panel. Generally, the potential outcomes are a function of the full treatment panel (e.g., \cite{Bojinov:Rambachan:Shephard:2020}); however, restricting our attention to non-anticipating potential outcomes\footnote{Following \cite{Bojinov:Shephard:2019}, we say the potential outcomes are non-anticipating if the outcomes at time $t$ are not impacted by future treatment assignments. That is, the potential outcomes only depend on past or current treatment assignments. In our empirical setting, for $t <t^*$, this assumption would be violated if the knowledge of the upcoming price reduction changed present sales. For instance, consumers could have postponed their purchases leading to a decrease in sales before the intervention. We can, however, safely exclude this, as the supermarket chain did not advertise the upcoming permanent discount in advance.} and Assumption \ref{assumption:single-intervention} somewhat simplify the setup. 

Assuming the intervention occurred at time $t^\ast+1$, for each pair $j \in \{1,\dots,J \}$ at time $t \in \{1, \dots, t^\ast \}$, we observe an outcome $\bl{Y}_{j,t} = ( \Y_{j,t}^{(s)}, \Y_{j,t}^{(c)} )$, where $\Y_{j,t}^{(s)}$ is the outcome of the store brand and $\Y_{j,t}^{(c)}$ is the outcome of the competitor brand. In our application, the outcome of interest is the average hourly sales for each product.

For $t>t^\ast$, generally the outcomes depend on the treatment assignment matrix,  $\bl{Y}_{j,t}(\bl{w}_{1:J}) = ( \Y_{j,t}^{(s)}(\bl{w}_{1:J}), \Y_{j,t}^{(c)}(\bl{w}_{1:J}) )$. In our empirical application, the products within each pair are alike and only differ on their brand name and packaging; whereas, brands in different pairs differ on many characteristics (e.g., ingredients, flavor, or weight). Therefore, we assume that a price reduction of one brand will impact its sales and the sales of its direct competitor. This assumption represents a model of consumer behavior in which customers' selection of the cookie type is not driven by price but rather by individual preferences; the choice within cookie type is then impacted by the price.\footnote{Within our supermarket study, every store brand has its specific direct competitor; there are no cookies that belong to multiple pairs.} To connect the general setting to our empirical application, we assume that there is no interference across pairs.
 
\begin{assumpt}[Partial temporal no-interference]\label{assumption:partial-interference}
    For all $j\in \{1,\dots, J\}$, and $t\in \{t^\ast+1, \dots, T \}$ we assume that for any $\bl{w}_{1:J}, \bl{w}'_{1:J}\in \mathcal{W}^{2\times J}$ such that $\bl{w}_j = \bl{w}_j'$,
    \[
        \bl{Y}_{j,t}(\bl{w}_{1:J}) = \bl{Y}_{j,t}(\bl{w}'_{1:J}).  
    \]  
    
    This allows us to simplify out notation and write $\bl{Y}_{j,t}(\bl{w}_{1:J}) = \bl{Y}_{j,t}(\bl{w}_j) $.
\end{assumpt}

In our application, there are four potential outcome paths that can occur, corresponding to the four different assignments. For each store-competitor pair, we can combine the post-treatment outcomes to define four potential outcome paths or \textbf{potential outcome time series}, 
$$\bl{Y}_{j,t^*+1:T}(\bl{w}_{j}) = (\Y_{j,t^*+1:T}^{(s)}(\bl{w}_{j}),\Y_{j,t^*+1:T}^{(c)}(\bl{w}_{j})).$$
Note that, even though we dropped the $t$ script from the assignment, our setup implicitly assumes that the outcomes at time $t>t^\ast$ are a function of the assignment path. This ensures that the potential outcomes at two different points in time correspond to two different treatment paths and are not directly comparable. 

To connect the potential outcomes to the observed outcome, we assume that there is full compliance; that is, every pair receives the assigned treatment. In a causal inference setting for panel data, for each unit, there is only one observed potential outcome time series, whereas the others are all unobserved. Generally, we will denote the observed treatment as $\bl{w}^\text{obs}_j$ which then leads to the observed outcome $\bl{Y}_{j,t^\ast +1:T} = \bl{Y}_{j,t^\ast +1:T}(\bl{w}^\text{obs}_j)$. In our application, only the store brand receives the permanent price reduction making the observed outcome $\bl{Y}_{j,t^*+1:T} = \bl{Y}_{j,t^*+1:T}(1,0)$.

\subsubsection{Covariates}
For each pair and time point, we observe a vector of covariates $\bl{X}_{j,t} \in \mathcal{X}$ that are not impacted by the intervention. If the covariates were impacted by the treatment, then we would consider them as secondary outcomes. 

\begin{assumpt}[Covariates-treatment independence]\label{assumption:covariates-independence} 
Let $\bl{X}_{j,t}$ be a vector of covariates; for all $t > t^*$ and for all assignments $\bl{w}_{j},\bl{w}_{j}'\in \mathcal{W}^2$ we assume that
$$\bl{X}_{j,t}(\bl{w}_{j}) = \bl{X}_{j,t}(\bl{w}_{j}') \hspace{30pt} \forall j \in \{1,\dots,J \}. $$
\end{assumpt}

\vspace{5pt}
For example, in the supermarket study, we use the following covariates: weekend and holiday dummies; daily sales of products in categories unaffected by the price reduction; and the prices of both goods before the intervention. For all of these covariates, Assumption \ref{assumption:covariates-independence} is likely to be satisfied.  We include the prior price as it is a good predictor of sales had there not been an intervention. Note that the inclusion of the actual daily price after the reduction would have violated Assumption \ref{assumption:covariates-independence}. To check if the control series $\bl{X}_{j,1:T}$ are genuinely unaffected by the intervention, we can test if the time series exhibits a change at the intervention time.

\subsubsection{Assignment mechanism}
We now define the class of assignment mechanism (i.e., conditional distributions of the assignment given the set of potential outcomes, covariates, and past assignments) that will allow us to estimate the causal effects defined in the subsequent section. Our assumption has two parts. The first requires the assignment is individualistic; that is, the treatment of one pair has no bearing on another. The second requires the assignment is non-anticipating; that is, the assignment in a given period does not depend on future outcomes or covariates. \\
% \FM{A major change done in the displayed equation; in the above discussion, the word 'identify' was deleted.}

\begin{assumpt}[Non-anticipating individualistic treatment]\label{assumption:individualistic-treatment}
%The assignment mechanism is independent across pairs, at time $t^*+1$ for the $j$-th pair depends solely on its past outcomes and past covariates,
The assignment mechanism at time $t^*+1$ is independent across pairs and for the $j$-th pair depends solely on its past outcomes and past covariates,

\begin{small}
\begin{align*}
\Pr(\bl{W}_{1:J,t^*+1} = \bl{w}_{1:J,t^*+1}| \bl{W}_{1:J,1:t^*} & , \bl{W}_{1:J, t^*+2:T}, \bl{Y}_{1:J,1:T}(\bl{w}_{1:J,1:T}), \bl{X}_{1:J,1:T}) = \\
& = \prod_{j=1}^J \Pr(\bl{W}_{j,t^*+1} = \bl{w}_{j,t^*+1}| \bl{Y}_{j,1:t^*}(\bl{w}_{j,1:t^*}), \bl{X}_{j,1:t^*}).% \hspace{10pt} \forall j \in \{1,\dots,J \}    
\end{align*}

\end{small}
\end{assumpt}

The non-anticipating treatment assumption is the extension of the unconfounded assignment mechanism in a cross-sectional setting \citep{Imbens:Rubin:2015, Bojinov:Rambachan:Shephard:2020}. Assumption \ref{assumption:individualistic-treatment} is essential in ensuring that, conditional on past outcomes and covariates, any differences in the outcomes are attributable to the intervention.  

\subsubsection{Multivariate case}
\label{subsubsect:multivariate}
   
Our framework easily generalizes to groups of size $d_j>2$. For $j \in \{1,\dots,J \}$, let $\W_j^{i}\in \mathcal{W}$ be the treatment status of the $i^\text{th}$ unit inside the $j^\text{th}$ group, and let $\bl{W}_{j} = (\W_{j}^{(1)},\dots,\W_{j}^{(d_j)}) \in \mathcal{W}^{d_j}$ be the treatment status of the $j$-th group. Again, Assumption \ref{assumption:single-intervention}, allowed us to drop the subscript for time.   
We then define the outcome to be a $d_j$-variate vector, $\bl{Y}_t = (\Y_t^{(1)}, \dots, \Y_t^{(d_j)})$, for $t\le t^\ast$. 
Assuming that there is only partial interference, Assumption \ref{assumption:partial-interference}, the potential outcomes for $t>t^\ast$ for any $\bl{w}_j \in \{0,1\}^{d_j}$ are

$$\bl{Y}_{j,t}(\bl{w}_{j}) = (\Y_t^{(1)}(\bl{w}_{j}), \dots, \Y_t^{(d_j)}(\bl{w}_{j}) ).$$
Again, we can use the more compact notation to denote the potential outcome time series, $\bl{Y}_{j,t^*+1:T}(\bl{w}_{j})$. 
All other assumptions and definitions easily extend to the multivariate case.

\subsection{Causal estimands} 
\label{subsect:estimands}
% \FM{Major changes done in this section: the marginal and conditional effects have been shifted in Appendix \ref{appB_other_effects} and the example has been extended a little.}
% In a panel setting, the number of causal estimands increases substantially, as any contrast of potential outcomes has a causal interpretation. 
We now develop a new class of causal estimands, for which we define a contemporaneous effect (i.e., an instantaneous effect at each time point after the intervention), a cumulative effect (i.e., a partial sum of the contemporaneous effect), and an average temporal effect (i.e., a normalization of the cumulative effect). To simplify our notation, we will drop the subscript $j$ that tracks the group and focus on analyzing each multivariate time series separately; $d$ will then indicate the group size. Even though our goal is to estimate the heterogeneous effect on each pair, the definitions below are given for a general multivariate case where units define groups of size $d > 2$. For simplicity, for the remainder of the paper, we focus on the binary treatment case where $\mathcal{W} = \{0,1\}$. Generalizing to multiple treatments is straightforward but makes the notation more cumbersome.

Since we are following the potential outcome approach to causal inference, we restrict $t>t^\ast$ so that the causal effects are defined as comparisons between two potential outcomes. 

\begin{defn} 
% Let $\bl{Y}_t(\bl{w}) = (Y_t^{(1)}(\bl{w}), \dots, Y_t^{(p)}(\bl{w}))$ be a $p$-variate potential outcome series and let $\mathcal{W} \in \{0,1\}^p$ be the set of all possible treatment paths, $\mathcal{W} \{ \bl{w}_1, \dots, \bl{w}_i \dots, \bl{w}_n \}$ where $n=2^p$ and $\bl{w}_i = (w^{(1)},\dots,w^{(p)})$. 
For $\bl{w},\tilde{\bl{w}} \in \mathcal{W}^{d}$, the \textbf{general causal effect} of an assignment $\bl{w}$ compared to an alternative assignment $\tilde{\bl{w}}$ is

\begin{align}
\label{eqn:general}
\bs{\tau}_t(\bl{w},\tilde{\bl{w}}) & =  (\tau_t^{(1)}(\bl{w},\tilde{\bl{w}}), \dots, \tau_t^{(d)}(\bl{w},\tilde{\bl{w}}))  \\ \nonumber
& = ( \bl{Y}_t^{(1)}(\bl{w}) - \bl{Y}_t^{(1)}(\tilde{\bl{w}}),\dots, \bl{Y}_t^{(d)}(\bl{w}) - \bl{Y}_t^{(d)}(\tilde{\bl{w}})) = ( \bl{Y}_t(\bl{w}) - \bl{Y}_t(\tilde{\bl{w}}))
\end{align}

The cumulative general causal effect at time point $ t' > t^*$ is 

\begin{equation}
\label{eqn:cum_general}
\Delta_{t'}(\bl{w},\tilde{\bl{w}}) =  \sum\limits_{t=t^*}^{t'} \bs{\tau}_t (\bl{w},\tilde{\bl{w}})
\end{equation}

The temporal average general causal effect at time point $t'$ is 

\begin{equation}
\label{eqn:avg_general}
\bar{\bs{\tau}}_{t'}(\bl{w},\tilde{\bl{w}}) = \frac{1}{t'-t^*} \sum\limits_{t=t^*+1}^{t'} \bs{\tau}_t (\bl{w},\tilde{\bl{w}})= \frac{1}{t'-t^*}  \Delta_{t'}(\bl{w},\tilde{\bl{w}})
\end{equation}

\end{defn}

In a general $d$-variate case, the total number of general causal effects that we can estimate is $C_{2^d,2}$. 

\begin{example}
For the supermarket study, $d=2$ and $\mathcal{W}^2 = \{ (0,0),(0,1), (1,1), (1,0) \}$. 
The general causal effect, $\bs{\tau}_t((1,0),(0,0)) = \bl{Y}_t(1,0) - \bl{Y}_t(0,0)$, measures the units sold on day $t$ when only the store brand receives a permanently discount compared to the alternative scenario where neither receive a discount. 
The cumulative general causal effect $\Delta_{t'}((1,0),(0,0))$, obtained from summing the general causal effects, captures the total additional units sold due to the price reduction up to time $t'$. Finally, the temporal average general effect $\bar{\tau}_t((1,0),(0,0))$ measures the average daily change in units sold due to the new policy up to $t'$.  
\end{example}

% \textcolor{red}{}
% In our application, we focus on the general causal effect. Other two classes of effects (the marginal and the conditional causal effects) are introduced in Appendix \ref{appB_other_effects}. Their estimation, though, requires a strong set of assumptions (whose formalization is beyond the scope of this paper); as a result, the estimates could be less reliable but we still find it useful to provide some insights on them so to give a complete overview of the effects that it would be possible to estimate in a multivariate setting.
There are two natural extensions to the general causal effect: the marginal causal effect, which captures the impact of changing one unit's treatment averaged over other units' possible assignments, and the conditional causal effect, which captures the effect of changing one unit's treatment fixing the other units' assignments. We provide the details in  Appendix \ref{appB_other_effects}, as they are not of primary interest in our supermarket study. 

\section{Multivariate Bayesian Structural Time Series}
\label{sect:MBSTS}
We now outline our approach for estimation and inference of the causal effects defined in Section \ref{subsect:estimands}. We begin by deriving the multivariate Bayesian structural time series models (MBSTS), which are the multivariate extensions of the models used by \citet{Brodersen:Gallusser:Koehler:Remy:Scott:2015} and \citet{Choirat:Dominici:Mealli:Papadogeorgou:Wasfy:Zigler:2018}. Like their univariate versions, MBSTS models are flexible and allow for a transparent uncertainty incorporation. Flexibility comes from our ability to add sub-components (e.g., trend, seasonality, and cycle) that encapsulate the characteristics of the data.  Uncertainty is quantified through the posterior distribution, which we derive and provide a sampling algorithm. 

Estimation has two steps. First, we fit an MBSTS model for each pair in the period up to the intervention, $t \in \{1, \dots, t^*\}$. Second, we estimate the target causal effects by forecasting the unobserved potential outcomes in the period following the intervention, $t \in \{t^*+1, \dots, T\}$. This section mirrors the two steps by first describing the model priors and posterior inference followed by detailing the forecast and inference step. To improve the readability of the model equations, in Section \ref{subsect:model} we drop the explicit dependence of the outcome on the treatment status (writing $\bl{Y}_t$ to indicate $\bl{Y}_t(\bl{w})$)  because the model is fit using the data prior to the intervention when $\bl{W}_{t} = (0, 0)$ for all store-competitor pairs and all $t \leq t^*$. We resume the usual notation in Section \ref{subsect:estimation}, where we derive the posterior distributions of the causal estimands defined in Section \ref{subsect:estimands}. 

Throughout this section, we employ random matrices to simplify the notation and subsequent posterior inference by allowing us to avoid matrix vectorization. Recalling the notation introduced by \citet{Dawid:1981}, let $\bl{Z}$ be an $(n \times d)$ matrix with standard normal entries, then $\bl{Z}$ follows a \textit{standard matrix Normal distribution}, written $\bl{Z} \sim \mathcal{N}(I_n, I_d)$, where $I_n$ and $I_d$ are $(n \times n)$ and ($d \times d$) identity matrices (the entries of $\bl{Z}$ are, therefore, independent). More generally, throughout the rest of paper, $\bl{Y} \sim \mathcal{N}(\bl{M},\bs{\Lambda}, \bs{\Sigma})$ indicates that $\bl{Y}$ follows a matrix normal distribution with mean $\bl{M}$, row variance-covariance matrix $\bs{\Lambda}$ and column variance-covariance matrix $\bs{\Sigma}$. Finally, a $d$-dimensional vector ($n=1$) following a multivariate standard Normal distribution will be indicated as $\bl{Z} \sim N_d (\bs{0},I_d)$ and $\mathcal{IW}(\nu, \bs{S})$ will denote an Inverse-Wishart distribution with $\nu$ degrees of freedom and scale matrix $\bs{S}$.

\subsection{The model}
\label{subsect:model}
Two equations define the MBSTS model. The first one is the ``observation equation'' that links the observed data $\bl{Y}_t$ to the state vector $\bs{\alpha}_t$, that models the different components in the data (such as trend, seasonality, or cycle), and covariates, which increase the counterfactual series' prediction accuracy. The second one is the ``state equation'' that determines the state vector's evolution across time. 

\begin{align}
\label{eqn:model}
\nonumber
\underbrace{\bl{Y}_t}_{1 \times d} & =  \underbrace{\bl{Z}_t}_{1\times m} \underbrace{\bs{\alpha}_t}_{m\times d} + \underbrace{\bl{X}_t}_{1\times P} \underbrace{\bs{\beta}}_{P\times d}+ \underbrace{\bs{\varepsilon}_t}_{1\times d} , & \bs{\varepsilon}_t & \sim N_d(\bs{0}, H_t\bs{\Sigma}) & \\
\underbrace{\bs{\alpha}_{t+1}}_{m\times d} & = \underbrace{\bl{T}_t}_{m\times m} \underbrace{\bs{\alpha}_t}_{m\times d} + \underbrace{\bl{R}_t}_{m\times r} \underbrace{\bs{\eta}_t}_{r\times d} ,  & \bs{\eta}_t & \sim \mathcal{N}(\bs{0}, \bl{C}_t, \bs{\Sigma}), \hspace{10pt} & \bs{\alpha}_1 \sim \mathcal{N}(\bl{a}_1, \bs{P}_1, \bs{\Sigma}).
\end{align}
For all $t \leq t^*$, $\bs{\alpha}_t$ is matrix of the $m$ states of the $d$  different time series and $\bs{\alpha}_1$ is the starting value;  $\bl{Z}_t$ is a vector selecting the states entering the observation equation; $\bl{X}_t$ is a vector of regressors;\footnote{Notice that this parametrization assumes the same set of regressors for each time series but still allows the coefficients to be different across the $d$ time series.} $\bs{\beta}$ is matrix of regression coefficients; and $\bs{\varepsilon}_t$ is a vector of observation errors. For the state equation, $\bs{\eta}_t$ is a matrix of the $r$ state errors (if all states have an error term, then $r = m$); $\bl{T}_t$ is a matrix defining the equation of the states components (e.g., in a simple local level model $\bl{T}_t = 1$); and $\bl{R}_t$ is a matrix selecting the rows of the state equation with non-zero error terms. Under our specification, we assume that $\bs{\varepsilon}_t$ and $\bs{\eta}_t$ are  mutually independent and independent of $\bs{\alpha}_1$. We denote the variance-covariance matrix of the dependencies between the time series by

$$\bs{\Sigma} = 
\begin{bmatrix}
\sigma^2_{1} & \sigma_{12} & \cdots & \sigma_{1d} \\
\sigma_{21} & \sigma^2_{2} & \cdots & \sigma_{2d} \\
\vdots & \vdots & \ddots & \vdots \\
\sigma_{d1} & \sigma_{d2} & \cdots &  \sigma^2_{d}
\end{bmatrix}.
\hspace{20pt}
$$
then $H_t$ is the variance of the observation error at time $t$; to simplify the notation we can also define $\bs{\Sigma}_{\varepsilon} = H_t \bs{\Sigma}$. Finally, $\bl{C}_t$ is an $(r \times r)$ matrix of dependencies between the states disturbances and, since we are assuming that different states are independent, $\bl{C}_t$ is a diagonal matrix. Indeed, we can also write $\bs{\eta}_t \sim N_d (\bs{0}, \bl{Q}_t)$ where $\bl{Q}_t$ is the Kronecker product of $\bl{C}_t$ and $\bs{\Sigma}$, denoted by $\bl{Q}_t = \bl{C}_t \otimes \bs{\Sigma}$. Furthermore, different values in the diagonal elements of $\bl{C}_t$ allow each state disturbance to have its own $(d \times d)$ variance-covariance matrix $\bs{\Sigma}_r$.\footnote{The notation $H_t \bs{\Sigma}$ and $c_r \bs{\Sigma}$ means that the dependence structure between the $d$ series is the same for both $\bs{\varepsilon}_t$ and $\bs{\eta}_t$; furthermore, when $H_t$ and $\bl{C}_t$ are known, the posterior distribution of $\bs{\alpha}_t$ is available in closed form \citep{West:Harrison:2006}. Instead, we employ a simulation smoothing algorithm to sample from the posterior of the states and in Section \ref{subsect:posteriors} we derive posterior distributions for $\bs{\Sigma}_{\varepsilon}$ and $\bs{\Sigma}_r$ in the general case of unknown $H_t$ and $\bl{C}_t$.} In short, 

$$
\bl{Q} = \bl{C}_t \otimes \bs{\Sigma}_{\varepsilon} =
\begin{bmatrix}
c_1 \bs{\Sigma} & 0 & \cdots 0 \\
0 & c_2 \bs{\Sigma} & \cdots & 0 \\
\vdots & \vdots & \ddots & \vdots \\
0 & 0 & \cdots &  c_r \bs{\Sigma} 
\end{bmatrix}
=
\begin{bmatrix}
\bs{\Sigma}_1 & 0 & \cdots 0 \\
0 & \bs{\Sigma}_{2} & \cdots & 0 \\
\vdots & \vdots & \ddots & \vdots \\
0 & 0 & \cdots &  \bs{\Sigma}_{r} 
\end{bmatrix}.
$$

To build intuition for the different components of the MBSTS model, we find it is useful to consider an example of a simple local level model.

\begin{example}\label{ExampleMMnoST}
The multivariate local level model is characterized by a trend component evolving according to a simple random walk without a seasonality component and Normally distributed disturbance terms. 

\begin{align}
%\label{eqn:example_model}
\bl{Y}_t & =  \bs{\mu}_t + \bs{\varepsilon}_t & \bs{\varepsilon}_t \sim N_d(\bs{0},H_t \bs{\Sigma})\\ \nonumber
\bs{\mu}_{t+1} & =  \bs{\mu}_t + \bs{\eta}_{t, \mu} & \bs{\eta}_{t,\mu} \sim N_d(\bs{0},c_1 \bs{\Sigma})\\ \nonumber
\end{align}

We can recover the general formulation outlined in (\ref{eqn:model}) by setting $\bs{\alpha}_t = \bs{\mu}_t$ and $\bl{Z}_t = \bl{T}_t = \bl{R}_t = 1$. Figure \ref{fig:example_plot}, provides a graphical representation of what a sample from this model would look like when $d=2$.

\begin{figure}[h!]
    \centering
    \caption{The figure shows $200$ observations sampled from a multivariate local level model with $d=2$. In our empirical application, $\Y_1$ and $\Y_2$ would denote the number of units sold of the store and competitor brands.}
    \label{fig:example_plot}
    \includegraphics[scale = 0.7]{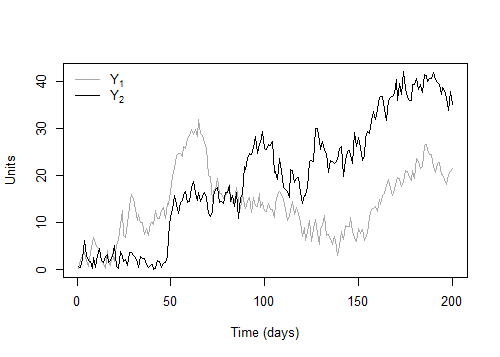}
\end{figure}

\end{example}

% \FM{Major change here, I added the following extra paragraph}
Unlike the previous example, the data in our supermarket study exhibit a weekly pattern (see Section \ref{sect:empirical_analysis}). The following MBSTS models is an extension of Example \ref{ExampleMMnoST} that includes a seasonal component.

% making the model in Example \ref{ExampleMMnoST}, provide a poor fit. Because of the flexibility of the MBSTS model, it is relatively easy to incorporate both a trend and a seasonal component:

% ; thus, for the empirical analysis we use an MBSTS model with both a trend and a seasonal component, which can be written as follows, 

\begin{align}
\label{eqn:appl_model}
\bl{Y}_t & = \bs{\mu}_t + \bs{\gamma}_t + \bl{X}_t \bs{\beta} + \bs{\varepsilon}_t & \bs{\varepsilon}_t \sim N_d(\bs{0},H_t \bs{\Sigma})\\ \nonumber
\bs{\mu}_{t+1} & = \bs{\mu}_t + \bs{\eta}_{t, \mu} & \bs{\eta}_{t,\mu} \sim N_d(\bs{0},c_1 \bs{\Sigma})\\ \nonumber
\bs{\gamma}_{t+1} & = - \sum\limits_{s=0}^{S-2} \bs{\gamma}_{t-s} + \bs{\eta}_{t,\gamma} & \bs{\eta}_{t,\gamma} \sim N_d(\bs{0},c_2 \bs{\Sigma}),
\end{align}
where $\bl{Y}_t = (\Y_t^{(s)}, \Y_t^{(c)})$ is a bivariate vector of the units sold by the store brand, $\Y_t^{(s)}$, and the units sold by the corresponding competitor brand, $\Y_t^{(c)}$; $\bs{\mu}_t$ and $\bs{\gamma}_t$ denote, respectively, the trend and seasonal components; and $\bl{X}_t$ is the vector of covariates satisfying Assumption \ref{assumption:covariates-independence}. Finally, $\bs{\eta}_{t, \mu}$, $\bs{\eta}_{t, \gamma}$ are the state errors having variance-covariance matrices $\bs{\Sigma}_1 = c_1 \bs{\Sigma}$,  $\bs{\Sigma}_2 = c_2 \bs{\Sigma}$, and $S=7$ is the weekly seasonal period. We selected the trend plus seasonal model based on the results of our posterior predictive checks; as detailed in Appendix \ref{appA_postpred}, posterior predictive checks are a viable tool to assess model performance.
% In the next section, we describe how to derive the prior distributions for this model. 

\subsubsection{Prior elicitation} 
\label{subsect:priors}

The unknown parameters of Model (\ref{eqn:model}) are the variance-covariance matrices of the error terms and the matrix of regression coefficients $\bs{\beta}$. Since both the observation and state errors are normally distributed, we use a conjugate Inverse-Wishart prior for their variance-covariance matrices. Generally, the MBSTS model can handle dynamic covariate coefficients. However, in our supermarket study, the relationship between covariates and the outcome is likely stable over time, and so we use a matrix normal prior, $\bs{\beta}  \sim \mathcal{N}(\bl{b}_0,\bl{H}, \bs{\Sigma}_{\varepsilon})$. 

In our application, we have a large pool of possible controls but believe that only a small subset is useful. We can incorporate such a sparsity assumption by setting $\bl{b}_0 = 0$ and introducing a selection vector $ \bs{\varrho} = (\varrho_1, \dots, \varrho_P)'$, with $\varrho_p \in \{0,1 \}$, $p \in [1,\dots,P]$. Then, $\bs{\beta}_p=0$ when $\varrho_p = 0$, meaning that the corresponding row of $\bs{\beta}$ is set to zero and the regressor $\X_p$ is excluded from our model; when $\varrho_p = 1$ then we include $\X_p$ in our model. This is known as Spike-and-Slab prior and it can be written as
$$\Pr(\bs{\beta}, \bs{\Sigma}_{\varepsilon}, \bs{\varrho}) = \Pr(\bs{\beta}_{\varrho} | \bs{\Sigma}_{\varepsilon}, \bs{\varrho})\Pr(\bs{\Sigma}_{\varepsilon}| \bs{\varrho})\Pr(\bs{\varrho}).$$
We model each element of $\bs{\varrho}$ as an independent Bernoulli random variable with parameter $\pi$. 

Let $\bs{\theta} = ( \nu_{\varepsilon}, \nu_{r}, \bs{S}_{\varepsilon}, \bs{S}_{r} , \bl{X}_{1:t^*})$ be the vector of known parameters and matrices and  $\bl{X}_{\varrho}$ and $\bl{H}_{\varrho}$ the selected regressors and the variance-covariance matrix of the corresponding rows of $\bs{\beta}$,  the full set of prior distributions at time $t \leq t^*$ is:

\begin{align*}
\bs{\varrho} | \bs{\theta} & \sim \prod\limits_{p = 1}^P \varrho_p (1-\pi)^{1-\varrho_p}, \\
\bs{\Sigma}_{\varepsilon}| \bs{\varrho}, \bs{\theta} & \sim \mathcal{IW} (\nu_{\varepsilon}, \bl{S}_{\varepsilon}), \\
\bs{\beta}_{\varrho} | \bs{\Sigma}_{\varepsilon}, \bs{\varrho} , \bs{\theta} & \sim \mathcal{N}(\bs{0}, \bl{H}_{\varrho}, \bs{\Sigma}_{\varepsilon}), \\
\bs{\alpha}_{t}| \bl{Y}_{1:t-1}, \bs{\Sigma}_{\varepsilon}, \bs{\Sigma}_r , \bs{\theta} & \sim \mathcal{N}(\bl{a}_{t}, \bs{P}_{t},\bs{\Sigma}),\\
\bs{\Sigma}_r | \bs{\theta} & \sim \mathcal{IW} (\nu_{r}, \bl{S}_{r}) .
\end{align*}

For setting the prior hyperparameters, \citet{Brown:Vannucci:Fearn:1998} suggest using $\nu_{\varepsilon} = d+2$, the smallest integer value such that the expectation of $\bs{\Sigma}_{\varepsilon}$ exists. We use a similar strategy for $\nu_{r}$. As for the scale matrices of the Inverse-Wishart distributions, in our empirical analysis we set

$$\bl{S}_{\varepsilon} = \bl{S}_r  = \begin{bmatrix}
s^2_1 & s_1 s_2 \rho \\
s_1 s_2 \rho & s^2_2 
\end{bmatrix},$$
where, $s^2_1$,$s^2_2$ are the sample variances of the store and the competitor brand respectively and $\rho$ is a correlation coefficient that can be elicited by incorporating our prior belief on the dependence structure of the two series. Finally, we set $\bl{H}_{\varrho} = (\bl{X}_{\varrho}' \bl{X}_{\varrho})$, which is the Zellner's g-prior \citep{Zellner:Siow:1980}.

\vspace{15pt}

\subsubsection{Posterior Inference}
\label{subsect:posteriors}
Let $\tilde{\bl{Y}}_{1:t^*} = \bl{Y}_{1:t^*} - \bl{Z}_{1:t^*} \bs{\alpha}_{1:t^*}$ indicate the observations up to time $t^*$ with the time series component subtracted out. The full conditional distributions are given by,

\begin{align}
\bs{\beta}_{\varrho} | \tilde{\bl{Y}}_{1:t^*}, \bs{\Sigma}_{\varepsilon}, \bs{\varrho}, \bs{\theta} & \sim  \mathcal{N} (\bl{M}, \bl{W}, \bs{\Sigma}_{\varepsilon}) \label{eqn:full_beta},\\
\bs{\Sigma}_{\varepsilon}|\tilde{\bl{Y}}_{1:t^*}, \bs{\varrho}, \bs{\theta} & \sim  \mathcal{IW}(\nu_{\varepsilon}+t^*, \bl{SS}_{\varepsilon})  \label{eqn:full_sigma},\\
\bs{\Sigma}_r | \bs{\eta}_{1:t^*}^{(r)}, \bs{\theta} & \sim  \mathcal{IW} (\nu_{r} + t^*, \bl{SS}_{r}), \label{eqn:full_sigma_eta} 
\end{align}
where $\bl{M} = (\bl{X}_{\varrho}' \bl{X}_{\varrho}+\bl{H}_{\varrho}^{-1})^{-1}\bl{X}_{\varrho}' \tilde{\bl{Y}}_{1:t^*}$, $\bl{W}=(\bl{X}_{\varrho}' \bl{X}_{\varrho}+\bl{H}_{\varrho}^{-1})^{-1}$, $\bl{SS}_{\varepsilon} = \bl{S}_{\varepsilon} + \tilde{\bl{Y}}_{1:t^*}' \tilde{\bl{Y}}_{1:t^*} - \bl{M}'\bl{W}^{-1}\bl{M}$, $\bl{SS}_{r} = \bl{S}_{r} + \bs{\eta}_{1:t^*}^{'(r)}\bs{\eta}_{1:t^*}^{(r)}$, and $\bs{\eta}_{1:t^*}^{(r)}$ indicates the disturbances up to time $t^*$ of the $r$-th state. Full proof of relations (\ref{eqn:full_beta}),(\ref{eqn:full_sigma}), and (\ref{eqn:full_sigma_eta}) is given in Appendix \ref{appB_inference}.

To sample from the joint posterior distribution of the states and model parameters, we employ a Gibbs sampler in which we alternate sampling from the distribution of the states given the parameters and sampling from the distribution of the parameters given the states (see Algorithm \ref{algorithm} in Appendix \ref{appB_inference}).

\subsubsection{Prediction and estimation of causal effects}
\label{subsect:prediction}

Let $\bs{\vartheta} = (\bs{\alpha}_{1:t^*}, \bs{\beta}_{\varrho}, \bs{\Sigma}_{\varepsilon}, \bs{\Sigma}_r, \bs{\varrho})$ be the vector of states and model parameters. We can use the joint posterior distribution $\Pr(\bs{\vartheta}| \bl{Y}_{1:t^*})$ to make in-sample and out-of-sample forecasts by drawing from the posterior predictive distribution. This process is particularly straightforward for in-sample forecasts.

To sample a new vector of observations $\bl{Y}_{1:t^*}^{new}$ given the observed pre-intervention data $\bl{Y}_{1:t^*}$, we note that,
%We can use the joint posterior distribution to make in-sample and out-of-sample forecasts by drawing from the posterior predictive distribution.  This process is particularly straightforward for in-sample forecasts.  

%Let $\bs{\vartheta} = (\bs{\alpha}_{1:t^*}, \bs{\beta}_{\varrho}, \bs{\Sigma}_{\varepsilon}, \bs{\Sigma}_r, \bs{\varrho})$ be the vector of states and model parameters. To sample a new vector of observations $\bl{Y}_{1:t^*}^{new}$ given the observed data $\bl{Y}_{1:t^*}$, we note that,
\begin{small}
\begin{align}
\label{eqn:post_pred_distrib}
\Pr(\bl{Y}_{1:t^*}^{new} | \bl{Y}_{1:t^*}) & =  \int \Pr(\bl{Y}_{1:t^*}^{new}, \bs{\vartheta}|\bl{Y}_{1:t^*})d\vartheta = \int \Pr(\bl{Y}_{1:t^*}^{new} | \bl{Y}_{1:t^*}, \bs{\vartheta})\Pr(\theta | \bl{Y}_{1:t^*}) d \bs{\vartheta} \\ \nonumber
& =  \int \Pr(\bl{Y}_{1:t^*}^{new} | \bs{\vartheta}) \Pr(\bs{\vartheta}| \bl{Y}_{1:t^*}) d \bs{\vartheta},
\end{align}
\end{small}
where the last equality follows because  $\bl{Y}_{1:t^*}^{new}$ is independent of $\bl{Y}_{1:t^*}$ conditional on $\bs{\vartheta}$. We then obtain in-sample forecasts from the posterior predictive distribution by substituting the Gibbs draws from $\Pr(\bs{\vartheta}|\bl{Y}_{1:t^*})$ into the model equations (\ref{eqn:model}). We typically use in-sample forecasting for model checking.

To predict the counterfactual time series in the absence of an intervention, we need out-of-sample forecasts. Forecasting future observations given the model estimated on the pre-intervention data is still relative straightforward; except, the new values are no longer independent of $\bl{Y}_{1:t^*}$ given $\bs{\vartheta}$. To see this, consider the vector $\bs{\vartheta}' = (\bs{\alpha}_{t^*+k},\dots,\bs{\alpha}_{t^*+1}, \bs{\vartheta})$ and let $\bl{Y}_{t^*+k}^{new}$ denote the $k$-step ahead forecast after the intervention. Then, 
%Drawing from the predictive posterior distribution is still relative straightforward, except, the new samples are no longer independent of $\bl{Y}_{1:t^*}$ given $\bs{\vartheta}$. 
%To see this, consider the vector $\bs{\vartheta}' = (\bs{\alpha}_{t^*+k},\dots,\bs{\alpha}_{t^*+1}, \bs{\vartheta})$. Then,

\begin{small}
\begin{align*}
\Pr(\bl{Y}_{t^* +k}^{new} | \bl{Y}_{1:t^*}) & = \int \Pr(\bl{Y}_{t^* +k}^{new}, \bs{\vartheta}' | \bl{Y}_{1:t^*})d \bs{\vartheta}' = \int \Pr(\bl{Y}_{t^* +k}^{new}, \bs{\alpha}_{t^*+k},\dots,\bs{\alpha}_{t^*+1}, \bs{\vartheta}|\bl{Y}_{1:t^*}) d \bs{\vartheta}' = \\
& = \int \Pr(\bl{Y}_{t^* +k}^{new}| \bs{\alpha}_{t^*+k}, \dots, \bs{\alpha}_{t^*+1}, \bs{\vartheta},\bl{Y}_{1:t^*})\Pr(\bs{\alpha}_{t^*+k} | \bs{\alpha}_{t^*+k-1}, \dots, \bs{\alpha}_{t^*+1}, \bs{\vartheta},\bl{Y}_{1:t^*}) \cdots \\
& \hspace{27pt} \cdots \Pr(\bs{\alpha}_{t^*+1}|\bl{Y}_{1:t^*}, \bs{\vartheta})\Pr(\bs{\vartheta}|\bl{Y}_{1:t^*})d \bs{\vartheta}'
\end{align*}
\end{small}
To make out-of-samples forecasts, respecting the dependence structure highlighted above, we substitute the existing draws from $\Pr(\bs{\vartheta}| \bl{Y}_{1:t^*})$, obtained by the Gibbs sampler, into the model equations (\ref{eqn:model}), thereby updating the states and sampling the new sequence $\bl{Y}_{t^*+1}^{new}, \dots, \bl{Y}_{t^*+k}^{new}$. 

\subsection{Causal effect estimation}
\label{subsect:estimation}
We can now estimate the causal effects defined in Section \ref{subsect:estimands} by using the MBSTS model to predict the counterfactual outcomes. Below, we focus on the general causal effect given in equation (\ref{eqn:general}) ; the details for the marginal and conditional effects are in Appendix \ref{appB_other_effects}. 

% Let $\Pr (\bl{Y}_t(\bl{w}) | \bl{Y}_{1:t^*}(\bl{w}))$ and $\Pr (\bl{Y}_t(\tilde{\bl{w}}) | \bl{Y}_{1:t^*}(\tilde{\bl{w}}))$ with $t > t^*$ be the out-of-samples draws from the posterior predictive distribution of the outcome under the treatment assignments $\bl{w}, \tilde{\bl{w}} \in \mathcal{W}^d$. Then, 

% \begin{equation}
% \label{eqn:post_general}
%     \Pr(\bs{\tau}_t(\bl{w},\tilde{\bl{w}})|\bl{Y}_{1:t^*}(\bl{w}), \bl{Y}_{1:t^*}(\tilde{\bl{w}})) = \Pr (\bl{Y}_t(\bl{w}) | \bl{Y}_{1:t^*}(\bl{w})) - \Pr (\bl{Y}_t(\tilde{\bl{w}}) | \bl{Y}_{1:t^*}(\tilde{\bl{w}}))
% \end{equation}

% Considering two treatment paths $\tilde{\bl{w}}, \tilde{\bl{w}}' \in \{0,1\}^d$, let $\Pr (\bl{Y}_t(\tilde{\bl{w}}) | \bl{Y}_{1:t^*}(\bl{w}))$ and $\Pr (\bl{Y}_t(\tilde{\bl{w}}') | \bl{Y}_{1:t^*}(\bl{w}))$ with $t > t^*$ be the posterior predictive distribution of the outcome under the two treatment assignments. Then, 
% \begin{equation}
% \label{eqn:post_general}
%     \Pr(\bs{\tau}_t(\tilde{\bl{w}},\tilde{\bl{w}}')|\bl{Y}_{1:t^*}(\bl{w})) = \Pr (\bl{Y}_t(\tilde{\bl{w}}) | \bl{Y}_{1:t^*}(\bl{w})) - \Pr (\bl{Y}_t(\tilde{\bl{w}}') | \bl{Y}_{1:t^*}(\bl{w}))
% \end{equation}
Recall that $\bl{Y}_{1:t^*}(0,0)$ is the observed pre-intervention data. For two treatments $\bl{w}, \tilde{\bl{w}} \in \{0,1\}^d$, let $\Pr (\bl{Y}_{t^*+k}(\bl{w}) | \bl{Y}_{1:t^*}(0,0))$ and $\Pr (\bl{Y}_{t^*+k}(\tilde{\bl{w}}) | \bl{Y}_{1:t^*}(0,0))$
%, with $t > t^*$, 
be the posterior predictive distributions of the outcome at time $t^*+k$ under the two treatment assignments.

%For two treatments $\tilde{\bl{w}}, \tilde{\bl{w}}' \in \{0,1\}^d$, let $\bl{Y}_t^{new}(\tilde{\bl{w}})$ and $\bl{Y}_t^{new}(\tilde{\bl{w}}')$ be the out-of-samples forecasts from the posterior predictive distribution at time $t > t^*$.

Then, for each draw from the posterior predictive distributions, we set
\begin{equation}
\label{eqn:post_general}
    \bs{\tau}_{t^*+k}^{new}(\bl{w},\tilde{\bl{w}}) = \bl{Y}_{t^*+k}^{new}(\bl{w}) -
     \bl{Y}_{t^*+k}^{new}(\tilde{\bl{w}})
\end{equation}
\noindent yielding samples from the posterior distribution of the general causal effect. Samples from the posterior distributions of the cumulative general effect and the temporal average general effect at $t' > t^*$ can be derived from (\ref{eqn:post_general}) as follows: 
\begin{equation}
    \label{eqn:post_cum_general}
    \Delta_{t'}^{new}(\bl{w},\tilde{\bl{w}}) = \sum\limits_{t = t^*+1}^{t'} \bs{\tau}_t^{new}(\bl{w},\tilde{\bl{w}})
\end{equation}

\begin{equation}
    \label{eqn:post_avg_general}
   \bar{\bs{\tau}}_t^{new}(\bl{w}, \tilde{\bl{w}})= \frac{1}{t'-t^*} \Delta_{t'}^{new}(\bl{w},\tilde{\bl{w}})
\end{equation}

% \begin{equation}
%     \label{eqn:post_cum_general}
%     \Pr(\Delta_{t'}(\bl{w},\tilde{\bl{w}}) | \bl{Y}_{1:t^*}(\bl{w}), \bl{Y}_{1:t^*}(\tilde{\bl{w}})) = \sum\limits_{t = t^*+1}^{t'} \Pr(\bs{\tau}_t(\bl{w},\tilde{\bl{w}})|\bl{Y}_{1:t^*}(\bl{w}), \bl{Y}_{1:t^*}(\tilde{\bl{w}}))
% \end{equation}

% \begin{equation}
%     \label{eqn:post_avg_general}
%     \Pr(\bar{\bs{\tau}}_t(\bl{w}, \tilde{\bl{w}}) | \bl{Y}_{t^*}(\bl{w}), \bl{Y}_{t^*}(\tilde{\bl{w}})) = \frac{1}{t'-t^*}\Pr(\Delta_{t'}(\bl{w},\tilde{\bl{w}}) | \bl{Y}_{t^*}(\bl{w}), \bl{Y}_{t^*}(\tilde{\bl{w}}))
% \end{equation}

Having samples from posterior distributions of the causal effects, we can easily compute posterior means and $95\%$ credible intervals.

\begin{example}
In our supermarket study, we are interested in estimating the general causal effect of the permanent price reduction on the store-competitor pair, $\bs{\tau}_t((1,0),(0,0)) = \bl{Y}_t(1,0)-\bl{Y}_t(0,0)$, with $t > t^*$. For a positive integer $k$, $\bl{Y}_{t^*+k}(1,0)$ is the observed outcome post-intervention and the predictive posterior distribution of the counterfactual outcome in the absence of intervention is $\Pr (\bl{Y}_{t^*+k}(0,0) | \bl{Y}_{1:t^*}(0,0))$. To get samples from the posterior distribution of the general causal effect at time $t^*+k$ we draw multiple times from $\Pr (\bl{Y}_{t^*+k}(0,0) | \bl{Y}_{1:t^*}(0,0))$, i.e., $\bs{\tau}_{t^*+k}^{new}((1,0),(0,0)) = \bl{Y}_{t^*+k}(1,0) - \bl{Y}_{t^*+k}^{new}(0,0)$.

%The posterior distribution of the general causal effect at time $t^*+k$ is then $\Pr(\bs{\tau}_{t^*+k}((1,0),(0,0))|\bl{Y}_{1:t^*}(0,0)) = \bl{Y}_{t^*+k}(1,0) - \Pr (\bl{Y}_{t^*+k}(0,0) | \bl{Y}_{1:t^*}(0,0))$.
\end{example}

Notice that \eqref{eqn:post_general}, \eqref{eqn:post_cum_general}, and \eqref{eqn:post_avg_general} do not require $\bl{Y}_t(\bl{w})$ or $\bl{Y}_t(\tilde{\bl{w}})$ to be observed in the post-intervention period. However, estimation of unobserved potential outcomes other than $\bl{Y}_t(0,0)$ requires a stronger set of modelling assumptions, making the inference less reliable. The marginal and the conditional effects are of secondary importance and are included in Appendix \ref{appB_other_effects}.
% In our application, we are mostly interested in estimating the general effect $\bs{\tau}_t((1,0),(0,0)) = \bl{Y}_t(1,0)-\bl{Y}_t(0,0)$, where $\bl{Y}_t(1,0)$ is the observed outcome post-intervention. 

In practice, to obtain reliable estimates of the causal effects, the assumed model has to describe the data adequately. Therefore, we recommend checking model adequacy through the use of posterior predictive checks \citep{Rubin:1981, Rubin:1984, Gelman:Carlin:Stern:Dunson:Vehtari:Rubin:2013}. Under our setup, we can also show that the above procedure yields unbiased estimates of the general causal effect and, in turn, the marginal and conditional effects. A detailed description of posterior predictive checks and the discussion of our estimators' frequentist properties are given, respectively, in Appendix \ref{appB_postpred} and \ref{appB_unbiased}.

\subsubsection{Combining results}
\label{subsubsect:meta_analysis}

To estimate an average across the various store-competitor pairs, we can combine the separate estimates through a meta-analysis.\footnote{We avoid using a full joint model across different pairs for computational feasibility. Moreover, as there is no interference across the store-competitor pairs, we can factorize the joint distribution into a product of marginals that can be analyzed separately.} 
% In the previous section, we obtained a posterior distribution of the general causal effect for each analyzed cookie pair. As a result, combining the estimates of the individual pairs is a lot more intuitive. 
For example, denote the temporal average causal effect of the permanent price reduction on the $j$-th cookie pair is $\bar{\bs{\tau}}_{j,t}((1,0), (0,0))$ with posterior distribution $\Pr(\bar{\bs{\tau}}_{j,t}((1,0), (0,0))|\bl{Y}_{1:t^*}(0,0))$ given in (\ref{eqn:post_avg_general}). We can define the summary temporal average effect across all $j$ pairs,

\begin{equation}
    \label{eqn:tau_summary}
    \bar{\bar{\bs{\tau}}}_t ((1,0), (0,0)) = \frac{1}{J} \sum\limits_{j = 1}^J \bar{\bs{\tau}}_{j,t}((1,0), (0,0)).
\end{equation}
To obtain samples from the posterior distribution of $\bar{\bar{\bs{\tau}}}_t ((1,0), (0,0))$, we aggregate the posterior samples from each of the $j$ temporal average causal effect. 
% \begin{equation}
%     \label{eqn:post_tau_summary}
%     \Pr(\bar{\bar{\bs{\tau}}}_t ((1,0), (0,0)) |\bl{Y}_{1:t^*}(0,0)) = \frac{1}{J} \sum\limits_{j = 1}^J \Pr(\bar{\bs{\tau}}_{j,t}((1,0), (0,0))|\bl{Y}_{1:t^*}(0,0)).
% \end{equation}

%For example, let $\bar{\bs{\tau}}_{j,t}(\bl{w}, \tilde{\bl{w}})$ be the temporal average causal effect on the $j$-th cookie pair and assume we estimated a posterior distribution for each $j$ as in (\ref{eqn:post_avg_general}). Then, we can define the summary temporal average effect across all $j$ pairs and its posterior distribution as,

% \begin{equation}
%     \label{eqn:tau_summary}
%     \bar{\bar{\bs{\tau}}}_t (\bl{w}, \tilde{\bl{w}}) = \frac{1}{J} \sum\limits_{j = 1}^J \bar{\bs{\tau}}_{j,t}(\bl{w}, \tilde{\bl{w}})
% \end{equation}

% \begin{equation}
%     \label{eqn:post_tau_summary}
%     \Pr(\bar{\bar{\bs{\tau}}}_t (\bl{w}, \tilde{\bl{w}}) |\bl{Y}_{1:t^*}(\bl{w}), \bl{Y}_{1:t^*}(\tilde{\bl{w}})) = \frac{1}{J} \sum\limits_{j = 1}^J \Pr(\bar{\bs{\tau}}_{j,t}(\bl{w}, \tilde{\bl{w}})|\bl{Y}_{1:t^*}(\bl{w}), \bl{Y}_{1:t^*}(\tilde{\bl{w}})).
% \end{equation}

% In words, to combine the estimated temporal average effect of the individual cookie pairs we can directly average across their posterior distributions. 

\section{Simulation study}
\label{sect:simulation_study}
We now describe a simulation study exploring the frequentist properties of our proposed approach for correctly specified models and a misspecified model. The results suggest that the misspecification only leads to a minor drop in performance and that posterior predictive checks are viable approaches to assess model adequacy.
% As expected, our model performs well under the correct model and minor misspecification; more importantly, we show that posterior predictive checks are viable approaches to assess model adequacy.

\subsection{Design}
The simulation study is specifically designed to resemble our supermarket example. As described in Section \ref{subsect:model}, we use an MBSTS model with both a trend and a seasonal component. The simulated data is then generated according to model (\ref{eqn:appl_model}),
\begin{align*}
%\label{eqn:sim_model}
\bl{Y}_t & = \bs{\mu}_t + \bs{\gamma}_t + \bl{X}_t \bs{\beta} + \bs{\varepsilon}_t & \bs{\varepsilon}_t \sim N_d(\bs{0},H_t \bs{\Sigma})\\ \nonumber
\bs{\mu}_{t+1} & = \bs{\mu}_t + \bs{\eta}_{t, \mu} & \bs{\eta}_{t,\mu} \sim N_d(\bs{0},c_1 \bs{\Sigma})\\ \nonumber
\bs{\gamma}_{t+1} & = - \sum\limits_{s=0}^{S-2} \bs{\gamma}_{t-s} + \bs{\eta}_{t,\gamma} & \bs{\eta}_{t,\gamma} \sim N_d(\bs{0},c_2 \bs{\Sigma}),
\end{align*}
where $\bl{Y}_t = (\Y_1,\Y_2)$ is a bivariate time series, $\bs{\mu}_t$ is a trend component evolving according a random walk and $\bs{\gamma}_t$ is a seasonal component with period $S=7$. We further set $H_t = 1$, $c_1 = 3$, $c_2 = 2$ and 
$\bs{\Sigma} = \begin{bmatrix}
1 & -0.3 \\ -0.3 & 1 
\end{bmatrix}$. 
We then assume a regression component formed by two covariates, $X_1 \sim f(x)$, with $f(x) = 1-x + N(0,0.5)$ and $X_2 \sim N(2,0.3)$, with coefficient $\bs{\beta}$ sampled from a matrix-normal distribution with mean $\bl{b}_0 = \bs{0}$ and $\bl{H} = I_P$. 

To estimate the causal effect, we use two different models for inference: a correctly specified model with both trend and seasonal components (M1) and a misspecified model with only the seasonal part (M2). For both models, we choose the following set of hyperparameters: $\nu_{\varepsilon} = \nu_{r} = 4$; $\bl{S}_{\varepsilon} = \bl{S}_{r} = 0.2
 \begin{bmatrix}
  s^2_1     &  s_1 s_2 \rho \\
  s_1 s_2 \rho  &  s^2_2
 \end{bmatrix}$, where $s^2_1$ and $s^2_2$ are the sample variances of $\Y_1$ and $\Y_2$, respectively, and $\rho = -0.8$ is a correlation coefficient reflecting our prior belief of their dependence structure; and Zellner's g-prior for the variance-covariance matrix of $\bs{\beta}$. 
 
To make our simulation close to our empirical application, we generated $1,000$ data sets in a fictional time period starting January 1, 2018 and ending June 30, 2019. We model the intervention as taking place on January 2, 2019, and assume a fixed persistent contemporaneous effect; for example, the series goes up by $+10\%$ and stays at this level throughout. To study the empirical power and coverage, we tried $5$ different impact sizes ranging from $+1\%$ to $+100\%$ on $\Y_1$ and from $-1\%$ to $-90\%$ on $\Y_2$. After generating the data, we estimated the effects using both M1 and M2, for a total of $10,000$ estimated models (one for each data set, impact size and model type), each having $1,000$ draws from the resulting posterior distribution. Finally, we predicted the counterfactual series in the absence of intervention for three-time horizons, namely, after $1$ month, $3$ months, and $6$ months from the intervention.  

We evaluate the performance of the models in terms of: 
\begin{enumerate}
\item length of the credible intervals around the temporal average general effect $\bar{\tau}_t((1,0),(0,0))$;
\item absolute percentage estimation error, computed as $\frac{|\hat{\bar{\tau}}_t((1,0),(0,0)) - \bar{\tau}_t((1,0),(0,0))|}{\bar{\tau}_t((1,0),(0,0))};$
\item interval coverage, namely, the proportion of the true pointwise effects covered by the estimated $95\%$ credible intervals.
\end{enumerate}
% We focus on the percentage estimation error because without normalizing the bias different effect sizes are not immediately comparable. To see this, consider that a small bias for estimating a substantial effect is better than that same bias when trying to estimate a small effect.
We focus on the percentage estimation error because different effect sizes are not immediately comparable without normalizing. For example, a small bias when estimating a large effect is better than the same bias when estimating a much smaller effect.   
\subsection{Results} 

\begin{figure}[h!]
    \centering
    \caption{Average absolute percentage error ($\pm$ 2 s.e.m) at the first time horizon under M1 and M2 for the impact sizes $\geq 10\%$ ($Y_1$) and $\leq -10\%$ ($Y_2$).}
    \label{fig:sim_error}
    \includegraphics[scale=0.6]{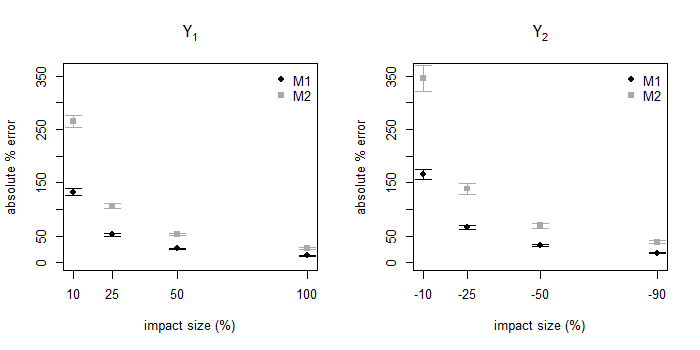}
\end{figure}

\begin{figure}[h!]
    \centering
    \caption{Average proportion of credible intervals excluding zero ($\pm$ 2 s.e.m) at the first time horizon under M1 and M2 for all impact sizes.}
    \label{fig:sim_interval}
    \includegraphics[scale=0.6]{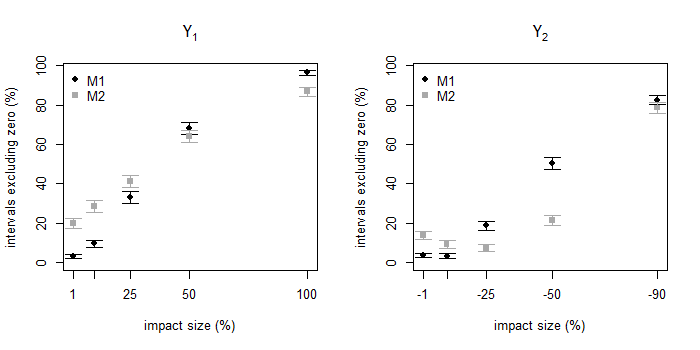}
\end{figure}

Table \ref{tab:sim_ci_tseas} reports the average interval length for M1 and M2 across the different effect sizes and time horizons. As expected, the length of credible intervals estimated under M1 increases with the time horizon. In contrast, for M2, the interval length is stable across time as the model lacks a trend component. Figure \ref{fig:sim_error} shows the absolute percentage error decreases as the effect size increases because small effects are more difficult to detect. To confirm this claim, in Figure \ref{fig:sim_interval}, we report the percentage of times we detect a causal effect over the $1,000$ simulated data sets. Under M1 for the two smallest effect sizes—which exhibit the highest estimation errors—we rarely correctly conclude that a causal effect is present. However, when the effect size increases, we can detect the presence of a causal effect at a much higher rate. The results under M2  are somewhat counterintuitive as, even though the model is misspecified, smaller effects are more readily detected. This phenomenon occurs because of the smaller credible intervals; that is, for small effect sizes, our results are biased with low variance, which means we often conclude there is an effect.

Table \ref{tab:sim_cover_tseas} reports the average interval coverage under M1 and M2. The coverage under M2 ranges from $82.0\%$ to $88.6\%$, which is lower than the desired $95 \%$. In contrast, the frequentists coverage under M1 is at the nominal  $95\%$ for both $\Y_1$ and $\Y_2$.

\begin{table}[h!]
\centering
\caption{Length of credible intervals around the temporal average general effect, $\bar{\bs{\tau}}_t((1,0),(0,0))$ estimated under M1 and M2 for each effect size and time horizon.}
\label{tab:sim_ci_tseas}
\begin{tabular}{llrrrrrrrr} \toprule
  &  & \multicolumn{2}{c}{1 month} & & \multicolumn{2}{c}{3 months} & & \multicolumn{2}{c}{6 months} \\ \cmidrule{3-4} \cmidrule{6-7} \cmidrule{9-10}
&  $\bar{\bs{\tau}}_t((1,0),(0,0))$  & $Y_1$ & $Y_2$ & & $Y_1$ & $Y_2$ & & $Y_1$ & $Y_2$ \\ \midrule 
\multirow{5}{*}{M1} & $(1.01,0.99)$ & 20.93 & 21.10 & & 27.62 & 27.80 & & 46.58 & 46.28 \\ 
 & $(1.10,0.90)$ & 21.34 & 21.37 & & 28.09 & 28.15 & & 46.98 & 46.89 \\ 
 & $(1.25,0.75)$ & 21.33 & 21.30 & & 28.18 & 28.09 & & 47.11 & 46.97 \\ 
 & $(1.50,0.50)$ & 21.30 & 21.31 & & 28.11 & 28.11 & & 47.02 & 46.91 \\ 
 & $(2.00,0.10)$ & 21.38 & 21.25 & & 28.24 & 28.06 & & 47.12 & 46.90 \\ \midrule
\multirow{5}{*}{M2} & $(1.01,0.99)$ & 30.39 & 30.39 & & 30.40 & 30.41 & & 30.48 & 30.47  \\ 
 & $(1.10,0.90)$ & 30.48 & 30.48 & & 30.50 & 30.50 & & 30.57 & 30.58 \\ 
 & $(1.25,0.75)$ & 30.48 & 30.46 & & 30.51 & 30.49 & & 30.60 & 30.58 \\ 
 & $(1.50,0.50)$ & 30.45 & 30.43 & & 30.47 & 30.46 & & 30.55 & 30.54 \\ 
 & $(2.00,0.10)$ & 30.49 & 30.49 & & 30.52 & 30.51 & & 30.60 & 30.57 \\\bottomrule
\end{tabular}
\end{table}

\begin{table}[h!]
\centering
\caption{Interval coverage under M1 and M2 for each effect size and time horizon.}
\label{tab:sim_cover_tseas}
\begin{tabular}{llrrrrrrrr} \toprule
 &   & \multicolumn{2}{c}{1 month} & & \multicolumn{2}{c}{3 months} & & \multicolumn{2}{c}{6 months} \\ \cmidrule{3-4} \cmidrule{6-7} \cmidrule{9-10}
& $\bar{\bs{\tau}}_t((1,0),(0,0))$    & $Y_1$ & $Y_2$ & & $Y_1$ & $Y_2$ & & $Y_1$ & $Y_2$ \\ \midrule
 \multirow{5}{*}{M1} & $(1.01,0.99)$ & 96.0 & 95.0 & & 96.1 & 95.3 & & 96.0 & 96.3 \\ 
&  $(1.10,0.90)$ & 95.9 & 94.9 & & 96.0 & 95.2 & & 95.9 & 96.3 \\ 
&  $(1.25,0.75)$ & 96.0 & 95.0 & & 96.0 & 95.3 & & 96.0 & 96.2 \\ 
&  $(1.50,0.50)$ & 96.1 & 94.9 & & 96.1 & 95.2 & & 96.1 & 96.2 \\ 
&  $(2.00,0.10)$ & 95.9 & 95.0 & & 96.1 & 95.3 & & 96.0 & 96.3 \\  \midrule
 \multirow{5}{*}{M2} &  $(1.01,0.99)$ & 86.8 & 88.4 & & 85.5 & 87.2 & & 82.0 & 84.6  \\
&  $(1.10,0.90)$ & 87.0 & 88.5 & & 85.7 & 87.3 & & 82.1 & 84.7 \\ 
&  $(1.25,0.75)$ & 87.0 & 88.6 & & 85.7 & 87.3 & & 82.1 & 84.7 \\ 
&  $(1.50,0.50)$ & 86.9 & 88.6 & & 85.6 & 87.3 & & 82.0 & 84.7 \\ 
&  $(2.00,0.10)$ & 86.9 & 88.6 & & 85.7 & 87.3 & & 82.1 & 84.6  \\\bottomrule
\end{tabular}
\end{table}

Overall, the simulation results suggest that when the model is correctly specified, the proposed approach performs well in estimating the causal effect of an intervention. Conversely, when the model is misspecified, the estimation error increases and the credible intervals do not achieve the required coverage; however, the results are likely to provide practitioners with useful insights.

In practice, we recommend assessing our model's adequacy before performing substantive analysis by using posterior predictive checks. In Appendix \ref{appA_plots} we provide examples results obtained under M1 (Figures \ref{fig:sim_M1_y1} and  \ref{fig:sim_M1_y2}) and posterior predictive checks under both M1 and M2. From the observation of Figures \ref{fig:sim_M1_ppc} and \ref{fig:sim_M2_ppc}, we can immediately see that M1 yields a better approximation of the empirical density of the simulated data and lower residual autocorrelation than M2. 

%From the observation of Figures \ref{fig:sim_M1_ppc} and \ref{fig:sim_M2_ppc}, we can immediately see that M1 yields a better approximation of the empirical density of the simulated data and lower residual autocorrelation than M2.   

\section{Empirical analysis}
\label{sect:empirical_analysis}

We now describe the results of our empirical application, where we analyze the efficacy of a strategic shift by an Italian supermarket chain to permanently reduce the price of a selected subset of store brands in its Florence stores. The firm's primary objective was to increase the customer base and sales. The policy change affected $707$ products in several categories; below, we provide the details for the ``cookies'' category. 

\subsection{Data \& methodology}

Among the $284$ items in the ``cookies'' category, there are $28$ store brands, of which $10$ were selected for a permanent price reduction ranging from $-3.5\%$ to $-23.2\%$ (the median was $-11.8\%$). For each store brand, the supermarket chain identified a direct competitor brand, thereby defining $10$ pairs of cookies.
%\footnote{Because of missing data on one of the competitor brands, we perform the analysis on $10$ pairs.}
Those in the same pair are almost identical except for their brand name. In contrast, cookies belonging to different pairs differ on many characteristics (e.g., ingredients, target market, and weight). As discussed earlier, in this setup, the permanent discount on a store brand is likely to impact its direct competitor but is unlikely to affect the sales of the cookies in different pairs, allowing us to justify the partial temporal no-interference assumption. 

\begin{figure}
\caption{\textbf{Store brands}. Starting from the left: time series of the average unit sold per hour; evolution of price per unit; autocorrelation function. The price plot shows the permanent price reduction after the intervention date (indicated by the vertical dashed line)} %, i.e., the price of the store brands is lowered and stays at this level throughout.}
\label{fig:tp_des}
\centering
\includegraphics[scale=0.6]{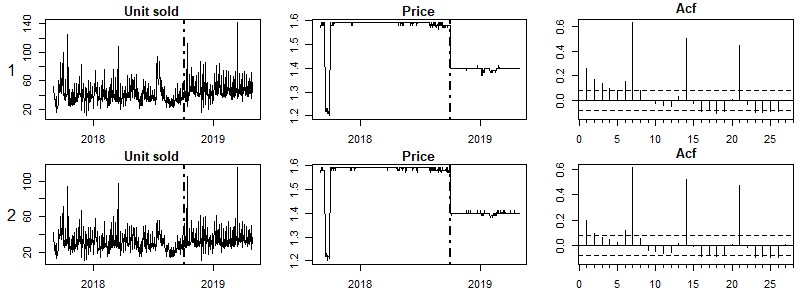}

\end{figure}

\begin{figure}
\caption{\textbf{Competitor brands}. Starting from the left: time series of the average unit sold per hour; evolution of price per unit; evolution of price relative to the store brand (the dashed horizontal line indicates a relative price equal to $1$); autocorrelation function. The price plot shows the temporary promotions these brands are subject to, i.e., both before and after the intervention date (indicated by the vertical dashed line) the price of competitor brands is reduced for a while and then bounces back to the original level.}
\label{fig:sp_des}
\centering
\includegraphics[scale=0.6]{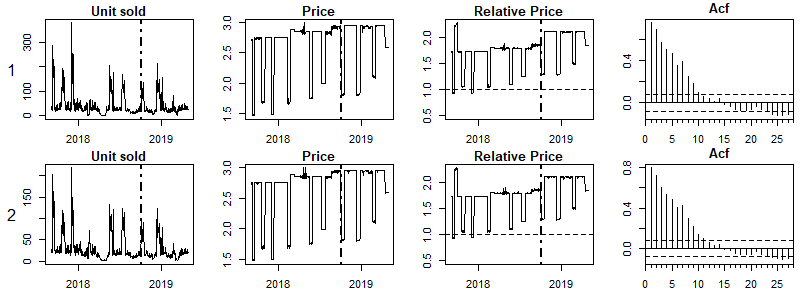}
\end{figure}

Our data consists of daily sales for all cookies from September 1, 2017, until April 30, 2019. Our outcome variable is the average units sold per hour—computed as the number of units sold daily divided by the number of hours that the stores stay open. We focus on hourly average sales because Italian regulations dictate that the supermarket chain only operates for a limited number of hours on Sundays; this discrepancy leads to a considerable difference in daily sales. As an example, Figure \ref{fig:tp_des} shows the time series of daily units sold by two store brands, their price, and the autocorrelation function. The plots show a strong weekly seasonal pattern. Figure \ref{fig:sp_des} exhibits the same plots for two competitor brands.\footnote{The equivalent plots for all the remaining store and competitor brands are provided in Appendix \ref{appA_plots}.} The occasional drops in the price series are from temporary promotions run regularly by the supermarket chain. In our data, the competitor brands are subject to several promotions during the analysis period. However, those differ from the permanent price reduction on their temporary nature and the regular frequency. As our goal is to evaluate the effectiveness of the store's policy change--—a permanent price reduction--—we will not consider temporary promotions as interventions. There is also considerable visual evidence that the store brands' intervention influenced the competitor cookies' pricing strategy. Indeed, all competitor brands (except for brand $10$) received a temporary promotion matching the time of the intervention, suggesting that competitors may have reacted to the new policy.\footnote{See Figure \ref{fig:sp_des_all} in Appendix \ref{appA_plots}.} 

Under partial temporal no-interference, we fit an MBSTS model for each pair; we also use covariates to improve the prediction of the counterfactual series. In particular, the set of regressors include: two dummies taking value $1$ on Saturday and Sunday, the former being the most profitable day of the week, whereas on the latter stores operate reduced hours; a holiday dummy taking value $1$ on the day before and after a national holiday, accounting for consumers' tendency to shop more before and after a closure day; a set of synthetic controls selected among one category (e.g., wine sales) that did not receive active treatment. Including covariates should increase prediction accuracy in the absence of intervention, but suitable covariates must respect two conditions: they should be good predictors of the outcome before the intervention, and they must satisfy Assumption \ref{assumption:covariates-independence}. As a result, the unit prices can not be part of our models; nevertheless, they are important drivers of sales, especially during promotions \citep{Neslin:Henderson:Quelch:1985,Blattberg:Briesch:1995,Pauwels:Hanssens:Siddarth:2002}. We solved this issue by using the ``prior price,'' which is equal to the actual price up to the intervention, and then it is set equal to the last price before intervention (which is the most reliable estimate of the price without an intervention). 

Finally, to speed up computations, the set of synthetic controls is selected in two steps: first, we select the best ten matches among the $260$ possible control series in the ``wines'' category by dynamic time warping;\footnote{Dynamic time warping (DTW) is a technique for finding the optimal alignment between two time series. Instead of minimizing the Euclidean distance between the two sequences, it finds the minimum-distance warping path, i.e., given a matrix of distances between each point of the first series with each point of the second series, contiguous set of matrix elements satisfying some conditions. For further details see \citet{Keogh:Ratanamahatana:2005,Salvador:Chan:2007}. Implementation of DTW has been done with the R package \texttt{MarketMatching} \citep{MarketMatching}.} then, we group them with the other predictors and perform multivariate Bayesian variable selection. 

Each model is estimated in the period before the intervention; then, as described in Section \ref{subsect:prediction}, we predict the counterfactual series in the absence of intervention by performing out-of-sample forecasts. Next, we estimate the intervention's causal effect at three different time horizons: one month, three months, and six months from the treatment day. This allows us to determine whether the effect persists over time or quickly disappears.

\subsection{Results}
\label{subsect:results}
We now present the results for the best MBSTS model with both a trend and seasonality component. The model was selected amongst an array of possible alternatives using posterior predictive checks; see Appendix \ref{appA_plots} for the details, and Appendix \ref{appA_postpred} for a description of the other models tried. Convergence diagnostics are provided in Appendix \ref{appB_convergence}.

\begin{figure}[h!]
\centering
\caption{Pointwise causal effect of the permanent price reduction on the fourth store-competitor pair at $1$ month, $3$ months and $6$ months after the intervention.}
\label{fig:causal_effect}
\includegraphics[scale=0.6]{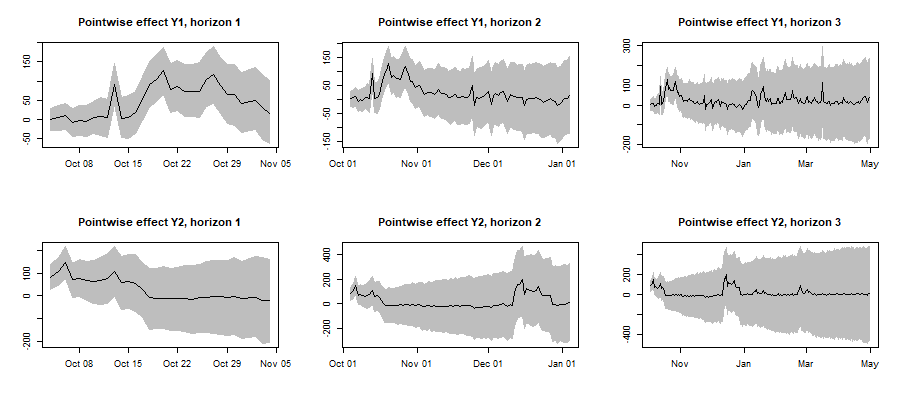} \\
\end{figure}

The estimates of the temporal average general effect, reported in Table \ref{tab:general_effect}, reveal the presence of three significant causal effects --- where the $95 \%$ credible intervals do not include $0$ --- on the store brands belonging to pairs $4$,$7$ and $10$ at the first time horizon. Interestingly, we do not find a significant effect on the competitor brands in the same pairs, most likely because, during the intervention period, competitor brands were subject to multiple temporary promotions that might have reduced the negative impact of the permanent discount on store brands. Furthermore, Italian supermarket chains have introduced store brands products only in recent years; so, despite the price reduction on store brand cookies, some consumers may still prefer the competitor cookie because of subjective factors, such as brand loyalty. Another important result is that after the initial surge in sales, we cannot detect a significant effect for longer time horizons. Figure \ref{fig:causal_effect} plots the pointwise general effect $\hat{\bs{\tau}}_t((1,0),(0,0))$ for the fourth pair at each time horizon, that is, the difference between the observed series and the predicted counterfactual computed at every time point. See Appendix \ref{appA_plots} for additional plots.

Overall, these results suggest that the firm's strategic change had a minor impact on the store brands' sales. Furthermore, since we do not detect an effect after the first month, it seems that this intervention failed to significantly and permanently impact sales. Of course, as we showed in the simulation study, there could have been a small effect that our model was unable to detect. However, since the firm needed a significant boost in sales to make up for the loss in profits due to the price reduction, we can conclude that this policy was ineffective. 
%This result is robust to different prior assumptions, see Appendix \ref{appB_sensitivity} for a detailed sensitivity analysis. 
This result is robust to different prior assumptions (see Appendix \ref{appB_sensitivity} for detailed sensitivity analysis) and to modifications in the set of covariates. In particular, we obtain similar results when, instead of using the individual prices, we include among the predictors the difference in price or the price ratio between the store and competitor brand (see Tables \ref{tab:general_effect_diff} and \ref{tab:general_effect_ratio} in Appendix \ref{appB_addres}). 

An alternative analysis strategy is to aggregate the sales of store and competitor brands and treating each aggregate as a univariate time series. However, this procedure leads to a loss of information, providing misleading results that could drive the analyst to make the wrong decision. To show that, we estimated the causal effect using the univariate BSTS models on a range of different aggregated sales. We report the results for three: the average sales of the brands in the same pair, the average sales of all store brands, and the average sales of all store and competitor brands. The average is computed as the total number of units sold daily by all products in the aggregate divided by the opening hours. Notice that we did not consider the aggregate of the competitor brands alone. This is because it would have required the prediction of the counterfactual series under treatment. 

Like the multivariate analysis, for each aggregate, we used a model that contained a trend and seasonality component as well as a set of covariates. The covariates included the three dummies (described earlier), aggregate sales of all wines, and the prior price—computed by averaging the prior prices of all cookies in each aggregate. Table \ref{tab:univ_effect} shows the results of the univariate analysis. We find evidence of a positive effect on the tenth pair at the first and second-time horizons and a positive effect on the eighth pair at the first horizon. In addition, the estimated effects on the store brands aggregate and the store-competitor aggregate are both positive and significant for the first time horizon. To provide a comparison with these last two aggregates, Table \ref{tab:univ_effect} reports the summary temporal average effect on all cookie pairs obtained by combining the individual estimates with a meta-analysis, as described in Section \ref{subsubsect:meta_analysis}. The summary effect on the store brands is positive and significant at the first time horizon, and, interestingly, it is in line with the estimated effect on the store brands aggregate from the univariate analysis. However, with the univariate analysis, we cannot isolate the effect on the competitor brands, and we would have erroneously concluded that the new policy had a positive impact on the store-competitor aggregate. In contrast, the meta-analysis shows that the effect on competitor brands is not significant. Overall, despite a similar result for the tenth pair, we would have reached the wrong conclusions for pairs $4$,$7$, and $8$, and would have reported the misleading finding of an overall positive impact on the sales. 

To further illustrate the range of possible causal estimands in a multivariate setting, we also estimated the marginal and the conditional effects. The results, given in Table  \ref{tab:mean_marginal_effect} and Table \ref{tab:cond_effect} in Appendix \ref{appB_other_effects}, show three significant marginal effects on the sales of store brands and little evidence of a conditional effect.

\begin{table}[h!]
\centering
\caption{Temporal average general causal effects of the new price policy on the ten store (s) - competitor (c) pairs computed at three time horizons. In this table, $\hat{\bar{\bs{\tau}}}_t$ stands for the general effect $\hat{\bar{\bs{\tau}}}_t((1,0),(0,0))$.}
\label{tab:general_effect}
\begin{tabular}{rrrrrrrrrrrrr} \toprule
 & & \multicolumn{3}{c}{$1$ month} & & \multicolumn{3}{c}{$3$ months} & & \multicolumn{3}{c}{$6$ months} \\ 
 & & $\hat{\bar{\bs{\tau}}}_t$ & $2.5\%$ & $97.5\%$ & & $\hat{\bar{\bs{\tau}}}_t$ & $2.5\%$ & $97.5\%$ & & $\hat{\bar{\bs{\tau}}}_t$ & $2.5\%$ & $97.5\%$ \\ \cmidrule{3-5} \cmidrule{7-9} \cmidrule{11-13}
\multirow{2}{*}{(1)}   
& s & 6.97  & -24.25  & 38.47  & & 4.68  & -44.00 & 53.61 & & 6.99 & -65.91 & 79.55 \\ 
& c & 24.89 & -101.30 & 153.64 & & 17.49 & -193.06 & 219.08 & & 5.09 & -307.48 & 309.00 \\ 
  \multirow{2}{*}{(2)}
& s & 7.02  & -14.79 & 28.90 & & 4.92 & -30.20  & 38.56 & & 6.56 & -44.17 & 58.01 \\ 
& c & 14.71 & -62.26 & 99.44 & & 8.92 & -119.33 & 144.72 & & 0.92 & -205.51 & 201.82 \\ 
  \multirow{2}{*}{(3)}
& s & 7.94 & -14.08 & 32.26 & & 5.30 & -31.95 & 41.38 & & 7.82 & -48.46 & 62.50 \\ 
& c & 15.42 & -62.17 & 90.81 & & 11.06 & -113.64 & 132.60 & & 4.84 & -189.44 & 197.55 \\
  \multirow{2}{*}{(4)}
& s & \bf{47.84} & \bf{4.71} & \bf{96.82} & & 22.65 & -52.13 & 96.38 & & 23.73 & -88.10 & 131.67 \\ 
& c & 28.86 & -77.93 & 135.93 & & 20.91 & -151.05 & 190.01 & & 11.20 & -256.88 & 279.74 \\ 
  \multirow{2}{*}{(5)}
& s & 4.11 & -46.65 & 54.64 & & 7.57 & -76.37 & 91.02 & & 11.75 & -111.67 & 136.65 \\ 
& c & 45.47 & -63.13 & 154.24 & & 16.68 & -156.03 & 188.67 & & 9.42 & -263.47 & 280.16 \\ 
  \multirow{2}{*}{(6)}
& s & 9.53 & -14.45 & 33.68 & & 11.76 & -28.33 & 51.70 & & 13.58 & -45.97 & 74.20 \\ 
& c & 25.64 & -37.88 & 93.36 & & 6.71 & -104.80 & 113.12 & & 4.13 & -163.82 & 164.96 \\  
  \multirow{2}{*}{(7)}
& s & \bf{78.19} & \bf{0.15} & \bf{154.08} & & 34.45 & -82.11 & 151.65 & & 29.48 & -149.12 & 206.10 \\ 
& c & 182.70 & -221.16 & 600.08 & & 102.61 & -581.90 & 769.52 & & 80.62 & -951.26 & 1069.94 \\
  \multirow{2}{*}{(8)}
& s & 25.23 & -28.60 & 78.16 & & 23.34 & -67.87 & 109.37 & & 17.07 & -115.20 & 145.12 \\ 
& c & 15.91 & -15.15 & 47.53 & & 6.03 & -44.60 & 60.30 & & 3.82 & -73.60 & 82.80 \\  
  \multirow{2}{*}{(9)}
& s & 40.29 & -9.84 & 90.38 & & 15.37 & -64.38 & 97.76 & & 12.07 & -108.11 & 136.44 \\ 
& c & 17.17 & -30.76 & 68.56 & & 1.20 & -79.88 & 84.48 & & 2.81 & -118.55 & 127.05 \\  
  \multirow{2}{*}{(10)}
& s & \bf{12.43} & \bf{1.35} & \bf{23.64} & & 9.64 & -8.07 & 27.98 & & 5.30 & -22.02 & 32.67 \\ 
& c & 0.04 & -9.36 & 9.79 & & 1.92 & -13.22 & 17.72 & & 4.00 & -18.33 & 27.03 \\  
\bottomrule
\end{tabular}
\end{table}

%\vspace{5cm}

\begin{table}[t!]
\centering
\caption{Univariate temporal average causal effect ($\hat{\bar{\tau}}_t$) at three time horizons of the new price policy on: i) aggregated sales (pairs 1-10); ii) the store brands aggregate (SA); iii) the store - competitor aggregate (SCA). The last two lines show, separately for the store brands (META-S) and the competitor brands (META-C), the summary temporal average effect combined with a meta-analysis.}
\label{tab:univ_effect}
\begin{tabular}{lrrrrrrrrrrrr} \toprule
  & \multicolumn{3}{c}{$1$ month} & & \multicolumn{3}{c}{$3$ months} & & \multicolumn{3}{c}{$6$ months} \\ 
  & $\hat{\bar{\tau}}_t$ & $2.5\%$ & $97.5\%$ & & $\hat{\bar{\tau}}_t$ & $2.5\%$ & $97.5\%$ & & $\hat{\bar{\tau}}_t$ & $2.5\%$ & $97.5\%$ \\ \cmidrule{2-4} \cmidrule{6-8} \cmidrule{10-12}
 Pair 1   & 16.65 & -36.89 & 64.97 & & 12.46 & -73.66 & 93.47 & & 6.97  & -115.80 & 130.39 \\
 Pair 2   & 9.85  & -25.50 & 42.76 & & 4.56  & -54.77 & 62.29 & & -0.24 & -85.55  & 85.37 \\
 Pair 3   & 11.20 & -29.89 & 48.21 & & 8.66  & -58.13 & 73.73 & & 6.25  & -90.95  & 107.34 \\
 Pair 4   & 36.86 & -4.18  & 75.70 & & 22.78 & -46.31 & 87.32 & & 18.50 & -76.66  & 119.12 \\
 Pair 5   & 29.05 & -40.13 & 88.51 & & 11.51 & -102.42& 121.54& & 10.70 & -158.37 & 186.19 \\ 
 Pair 6   & 16.86 & -14.59 & 44.80 & & 4.09  & -50.47 & 57.12 & & 5.40  & -74.01  & 88.53 \\ 
 Pair 7   & 120.86& -129.59& 352.65& & 75.54 & -272.11& 393.52& & 57.87 & -568.82 & 687.77 \\ 
 Pair 8   & \bf{20.06} & \bf{4.95}   & \bf{34.39} & & 12.59 & -11.39 & 36.03 & & 8.91  & -25.75  & 42.42 \\ 
 Pair 9   & 28.58 & -0.03  & 55.95 & & 8.51  & -38.36 & 54.54 & & 9.53  & -56.66  & 78.61 \\
 Pair 10  & \bf{7.29}  & \bf{4.19}   & \bf{10.00} & & \bf{6.63}  & \bf{1.64}   & \bf{10.94} & & 5.75  & -1.49   & 12.17 \\ \midrule
 SA       & \bf{25.01} & \bf{10.08}  & \bf{39.04} & & 15.04 & -8.80  & 37.56 & & 15.52 & -19.30  & 49.19 \\
SCA      & \bf{34.56} & \bf{8.55}   & \bf{58.78} & & 19.98 & -20.53 & 58.62 & & 16.16 & -44.40  & 78.19 \\ \midrule
 META-S & \bf{23.95}  & \bf{3.62} & \bf{45.32} & & 13.97 & -18.39 & 47.89 & & 13.43 & -34.05 & 67.37 \\
 META-C & 37.08 & -34.98 & 106.39 & & 19.35 & -100.10 & 133.78 & & 12.68 & -163.61 & 184.61 \\
 \bottomrule
\end{tabular}
\end{table}

\vspace{50pt}
\section{Conclusion}

This paper presents a causal analysis of the effectiveness of a new pricing strategy implemented by an Italian supermarket chain. The results suggest that the policy change had a minor impact on the store brands' sales and little evidence of a detrimental effect on competitor brands. Our findings relayed on a new methodology for analyzing the effectiveness of a single persistent intervention in the presence of partial interference. Interestingly, we showed that methods that fail to account for the interference lead to incorrect results that overestimate the price reduction's effectiveness. 

We believe that our approach brings several contributions to the nascent stream of literature on synthetic control methods in panel settings with interference. First, we derived a wide class of new causal estimands. Second, MBSTS allows us to model the interference between units in the same group by explicitly modeling their dependence structure and, simultaneously, ensuring a transparent way to deal with the surrounding uncertainty. Finally, the approach is flexible, and the underlying distributional assumptions can be tested in a very natural way by posterior inference.

\clearpage

\bibliography{references}

\newpage

\begin{appendices}

% \newgeometry{bottom=2cm, top=2cm, left=2cm, right=2cm}
\section{}
\label{appA}

\newgeometry{bottom=2cm, top=2cm, left=2cm, right=2cm}

\subsection{Additional plots}
\label{appA_plots}

\begin{figure}[h]
 \centering
 \caption{\textbf{Store brands}. Starting from the left: time series of the average unit sold per hour; evolution of price per unit; autocorrelation function. The price plot shows the permanent price reduction after the intervention date (indicated by the vertical dashed line)}
 \includegraphics[scale=0.5]{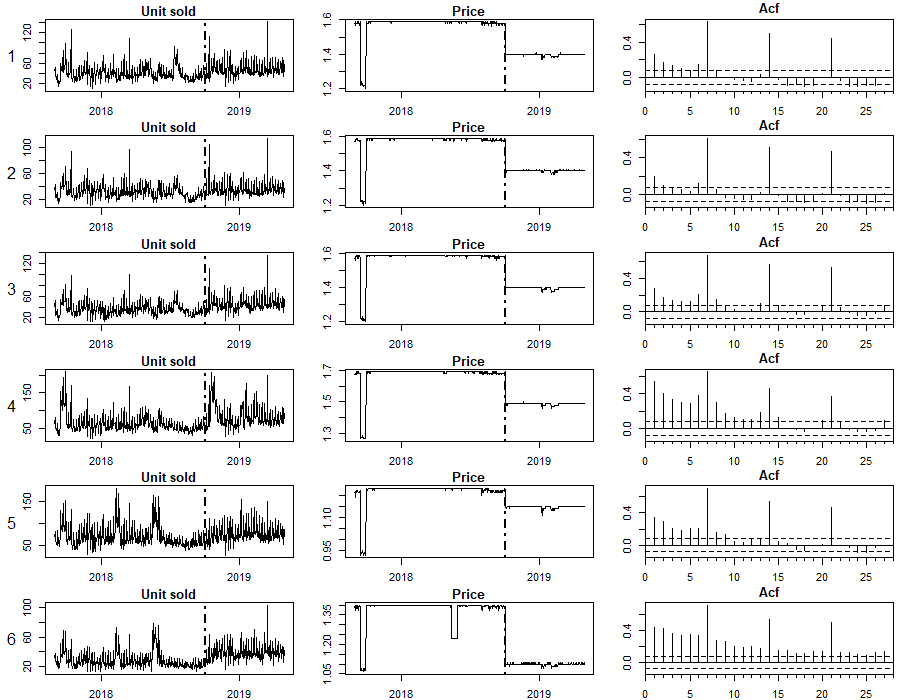}
 \includegraphics[scale=0.5]{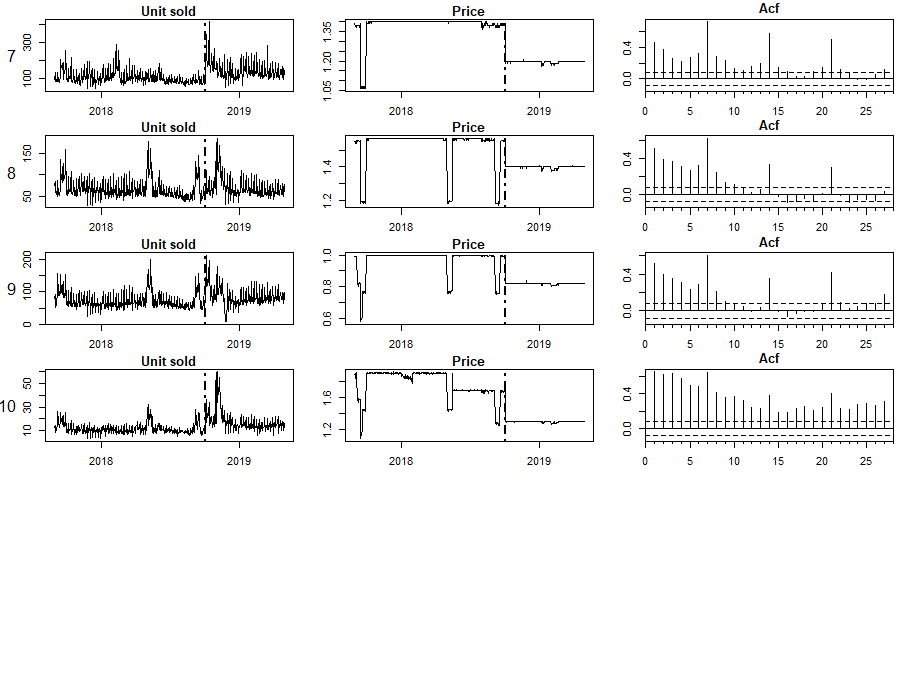}
 \end{figure}

\restoregeometry

 \clearpage

 \begin{figure}[ht]
 \centering
 \caption{\textbf{Competitor brands}. time series of the average unit sold per hour; evolution of price per unit; evolution of price relative to the store brand (the dashed horizontal line indicates a relative price equal to $1$); autocorrelation function. The price plot shows the temporary promotions these brands are subject to, i.e., both before and after the intervention date (indicated by the vertical dashed line) the price of competitor brands is reduced for a while and then bounces back to the original level.} 
 \label{fig:sp_des_all}
 \includegraphics[scale=0.6]{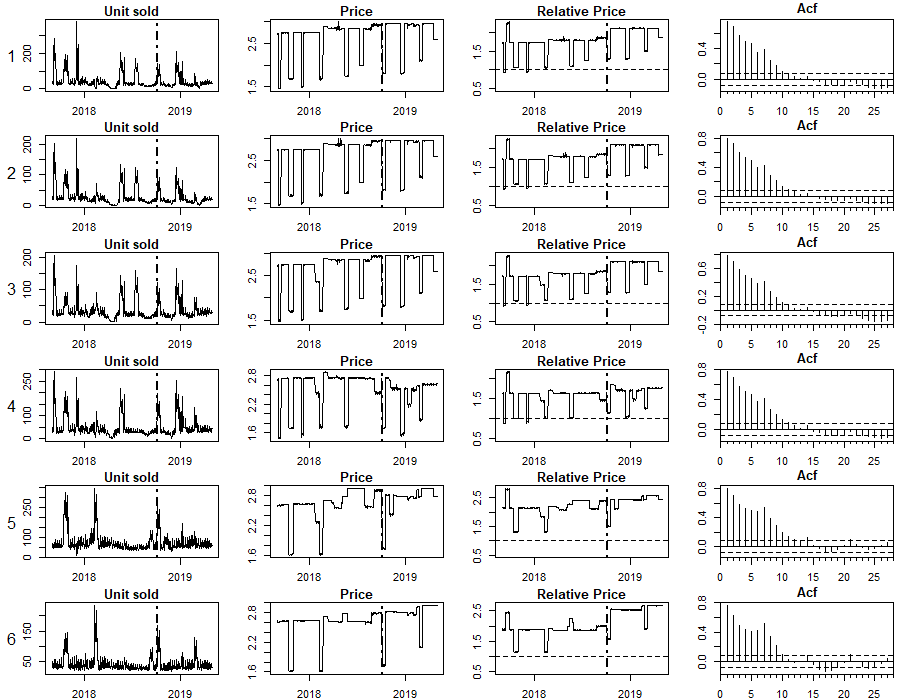}
 \includegraphics[scale=0.6]{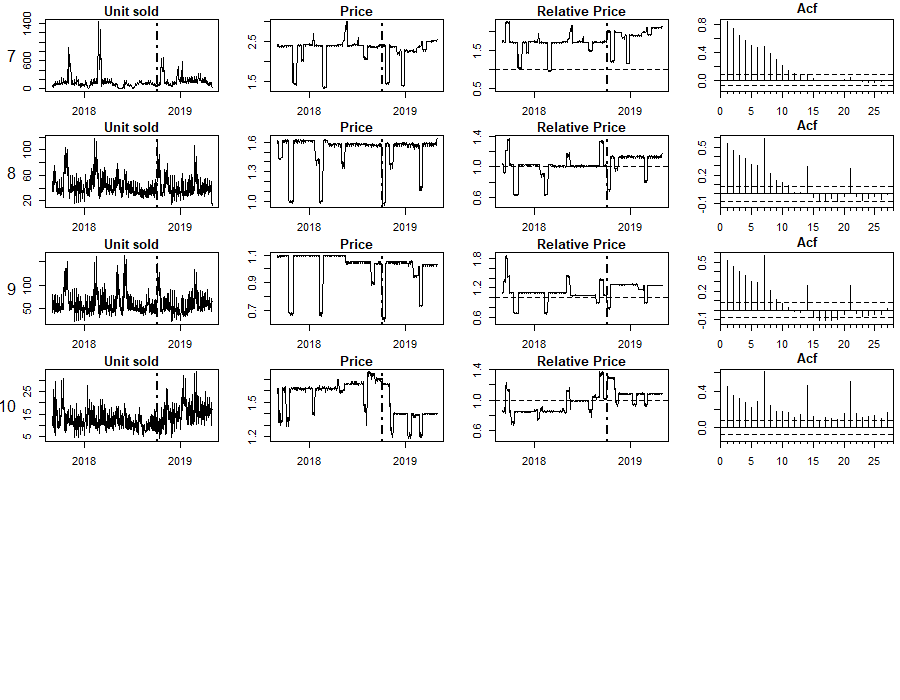}
 \end{figure}

\begin{figure}[t]
\centering
\caption{For one of the simulated data sets at $3$-month horizon, the figure plots: (a) simulated time series assuming an effect size of $+50\%$ vs true counterfactual series generated under model (\ref{eqn:appl_model}); (b) true counterfactual vs predicted counterfactual series under M1; (c) true effect vs the inferred effect under M1.}
\label{fig:sim_M1_y1}
\includegraphics[width = 13cm, height = 4cm]{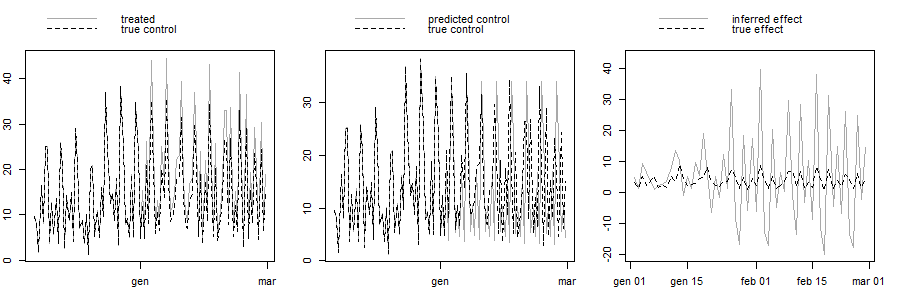} \\
\end{figure}

\begin{figure}[t]
\centering
\caption{For one of the simulated data sets at $3$-month horizon, the figure plots: (a) simulated time series assuming an effect size of $-50\%$ vs true counterfactual series generated under model (\ref{eqn:appl_model}); (b) true counterfactual vs predicted counterfactual series under M1; (c) true effect vs the inferred effect under M1.}
\label{fig:sim_M1_y2}
\includegraphics[width = 13cm, height = 4cm]{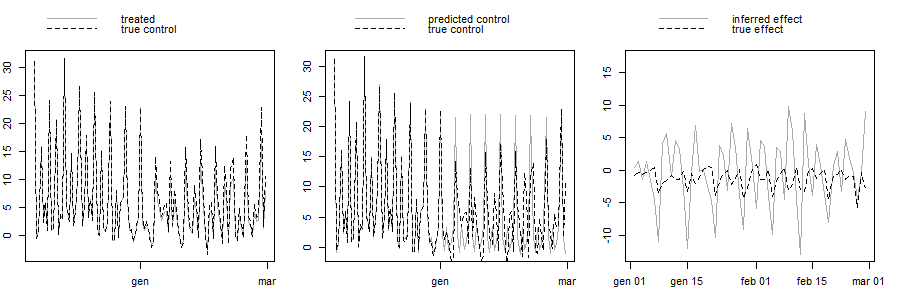} \\
\end{figure}

\clearpage

\begin{figure}[h!]
\centering
\caption{Posterior predictive checks under M1 for $\Y_1$ (first row) and $\Y_2$ (second row) for one of the simulated data sets. Starting from the left: i) density of observed data plotted against the posterior predictive mean; ii) observed maximum compared to the distribution of the maximum from the posterior draws; iii) Normal QQ-Plot of standardized residuals; iv) autocorrelation function of standardized residuals.}
\label{fig:sim_M1_ppc}
\includegraphics[scale=0.5]{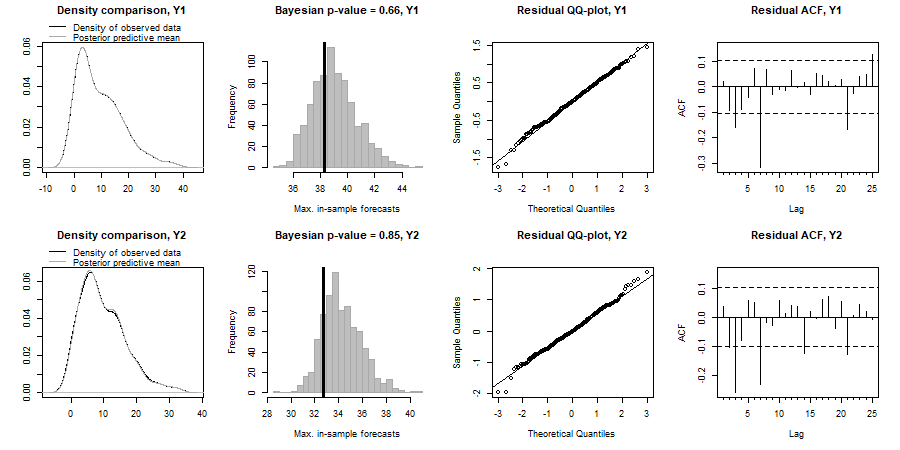}
\end{figure}

\begin{figure}[h!]
\centering
\caption{Posterior predictive checks under M2 for $\Y_1$ (first row) and $\Y_2$ (second row) for one of the simulated data sets. Starting from the left: i) density of observed data plotted against the posterior predictive mean; ii) observed maximum compared to the distribution of the maximum from the posterior draws; iii) Normal QQ-Plot of standardized residuals; iv) autocorrelation function of standardized residuals.}
\label{fig:sim_M2_ppc}
\includegraphics[scale=0.5]{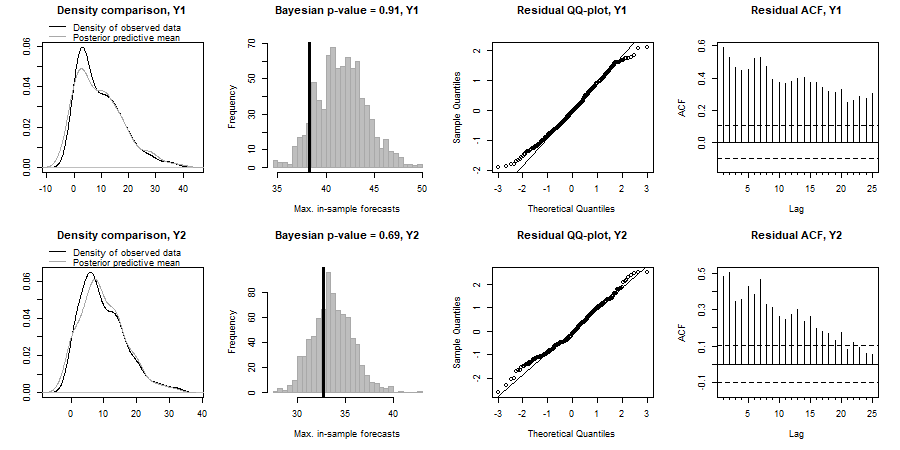}
\end{figure}

 \begin{figure}[h!]
 \centering
 \caption{Pointwise causal effect of the permanent price reduction on each store-competitor pair at $1$ month, $3$ months and $6$ months after the intervention.}
 \begin{tabular}{m{1cm}m{12cm}}
 (1) & \includegraphics[width=12cm, height=4.5cm]{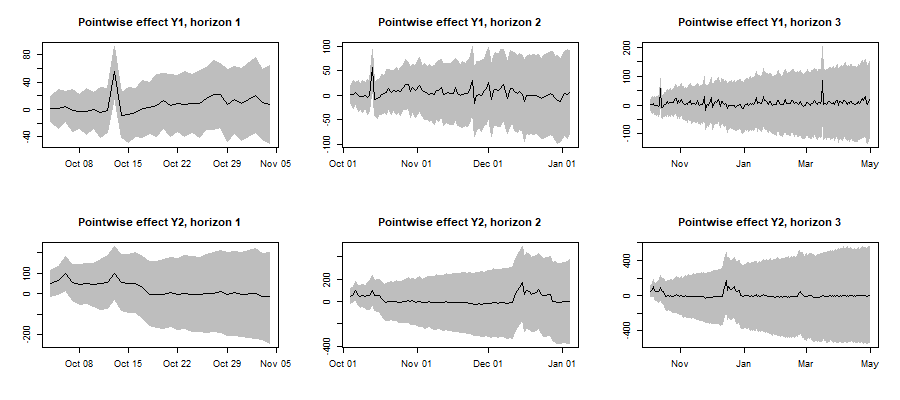} \\
 (2) & \includegraphics[width=12cm, height=4.5cm]{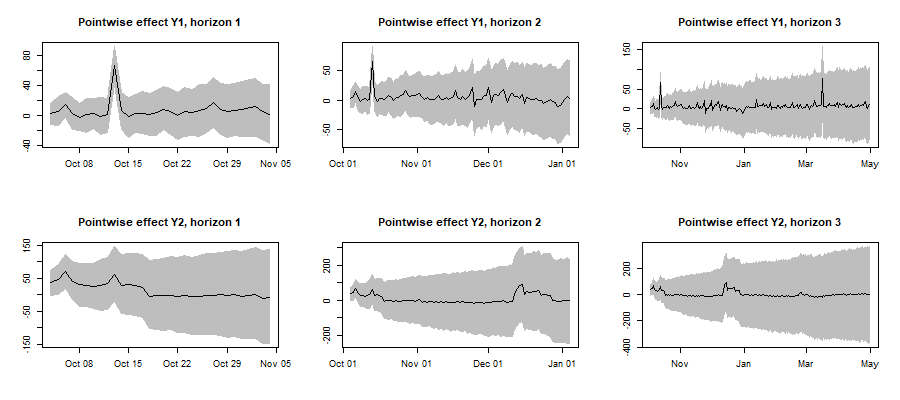} \\
 (3) & \includegraphics[width=12cm, height=4.5cm]{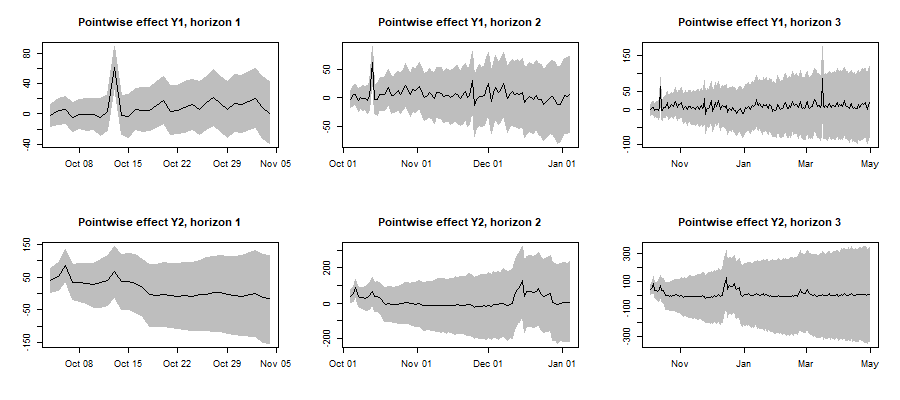} \\
 (4) & \includegraphics[width=12cm, height=4.5cm]{trendseas_joint_pair_4} \\
 (5) & \includegraphics[width=12cm, height=4.5cm]{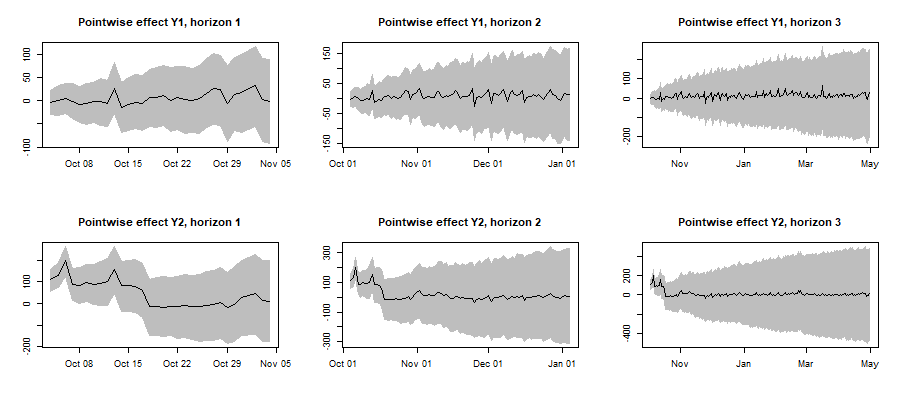} \\
 \end{tabular}
 \end{figure}

 \begin{figure}
 \centering
 \begin{tabular}{m{1cm}m{12cm}}
 (6) & \includegraphics[width=12cm, height=4.5cm]{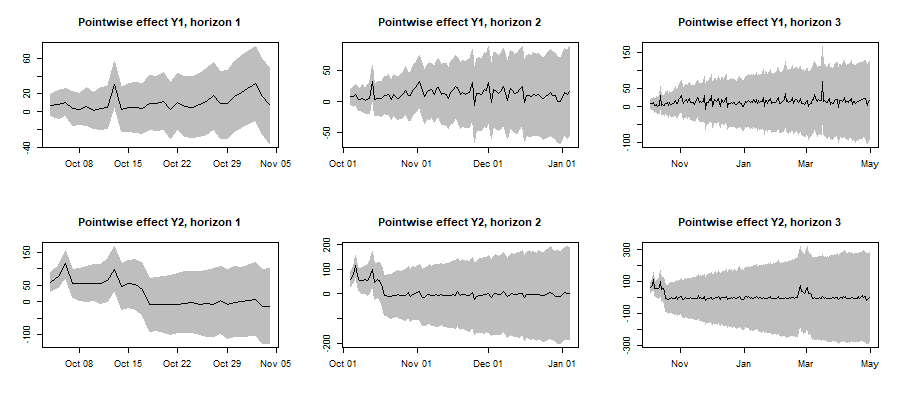} \\
 (7) & \includegraphics[width=12cm, height=4.5cm]{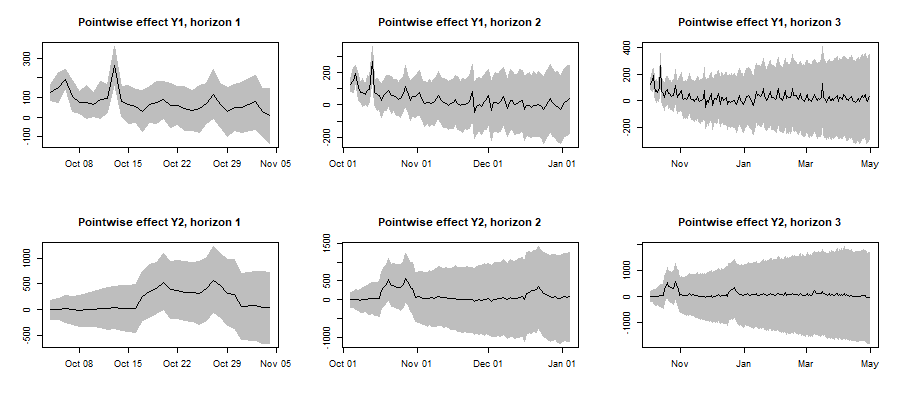} \\
 (8) & \includegraphics[width=12cm, height=4.5cm]{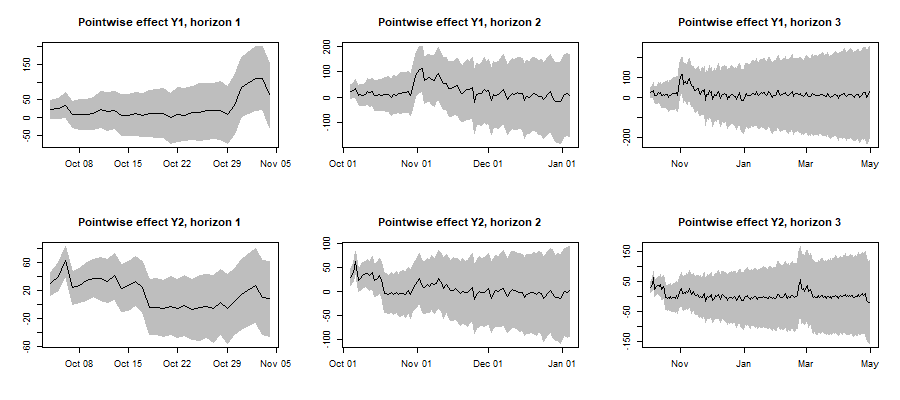} \\
 (9) & \includegraphics[width=12cm, height=4.5cm]{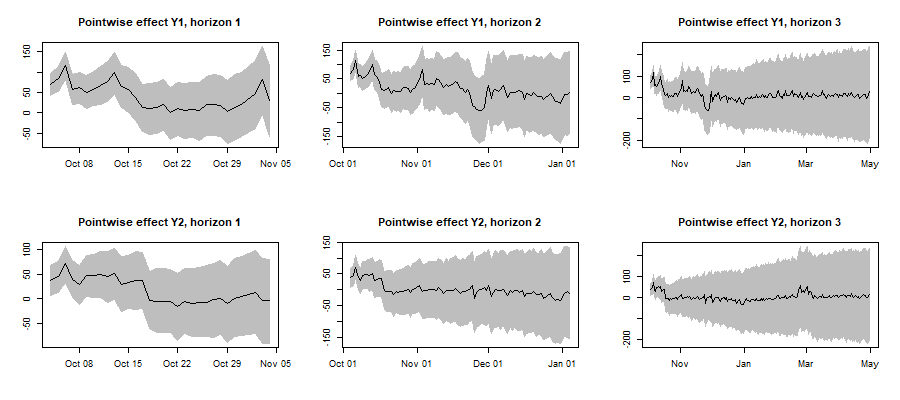} \\
 (10) & \includegraphics[width=12cm, height=4.5cm]{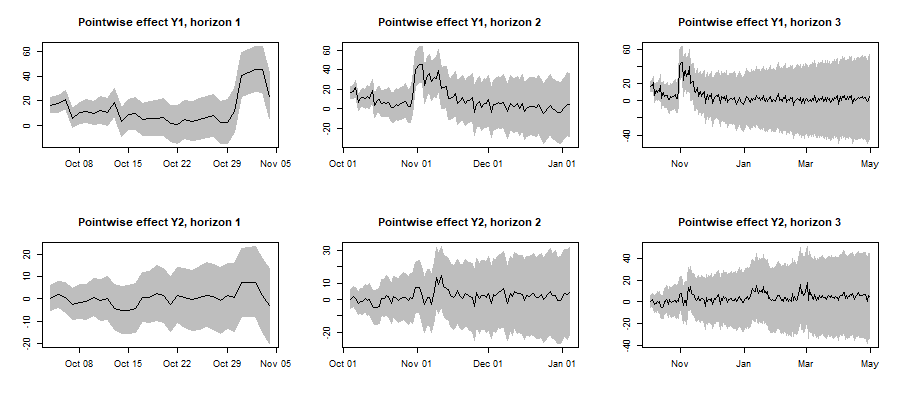} \\
 \end{tabular}
 \end{figure}
 \begin{figure}[h!]
 \centering
 \caption{For each store-competitor pair, observed outcome plotted against the counterfactual outcome in the absence of intervention after $1$ month, $3$ months and $6$ months from the intervention, indicated by the vertical line.}
 \begin{tabular}{m{1cm}m{12cm}}
 (1) & \includegraphics[width=12cm, height=4.5cm]{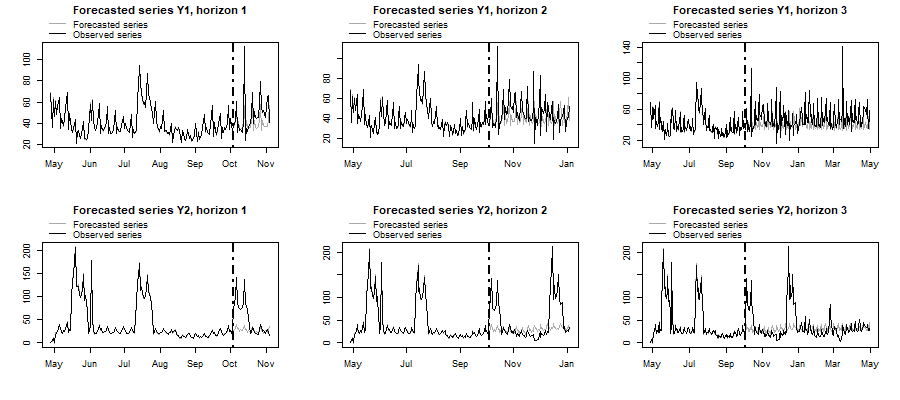} \\
 (2) & \includegraphics[width=12cm, height=4.5cm]{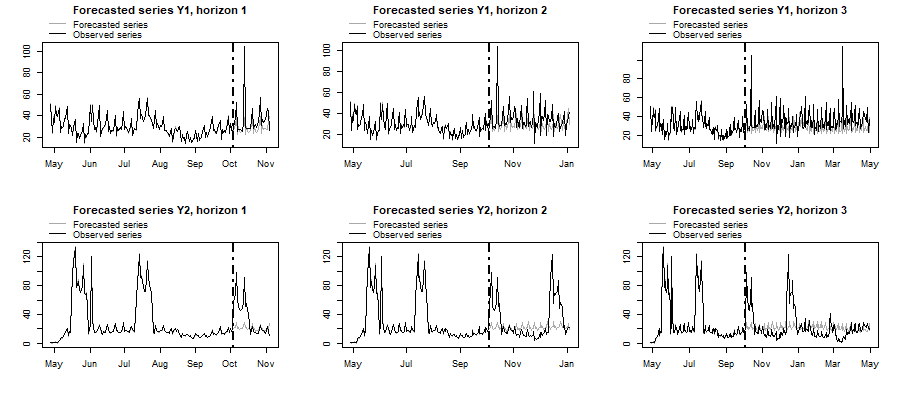} \\
 (3) & \includegraphics[width=12cm, height=4.5cm]{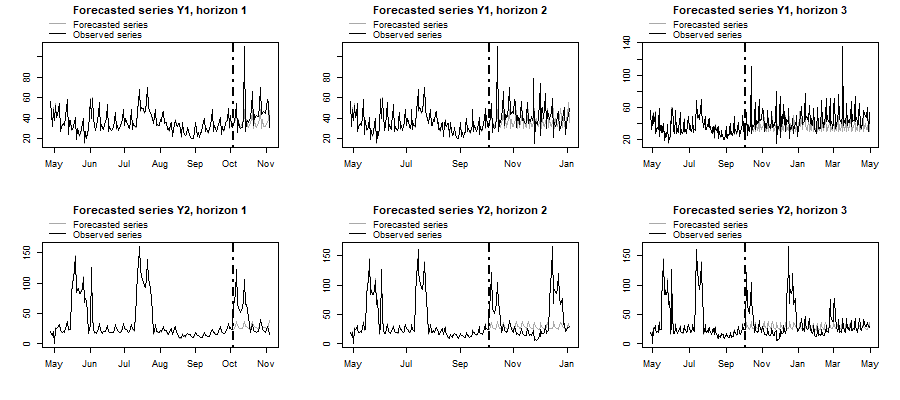} \\
 (4) & \includegraphics[width=12cm, height=4.5cm]{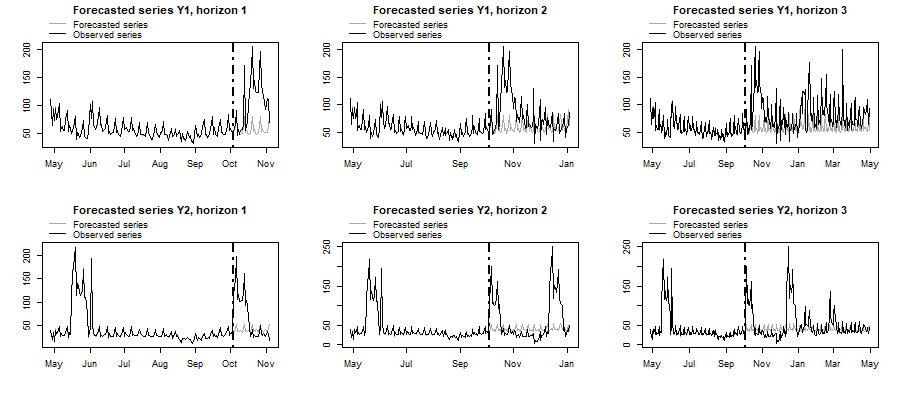} \\
 (5) & \includegraphics[width=12cm, height=4.5cm]{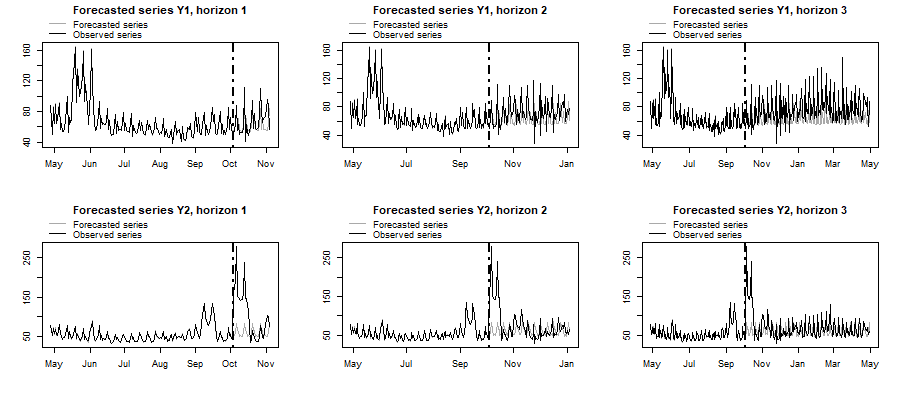} \\
 \end{tabular}
 \end{figure}
 \begin{figure}
 \centering
 \begin{tabular}{m{1cm}m{12cm}}
 (6) & \includegraphics[width=12cm, height=4.5cm]{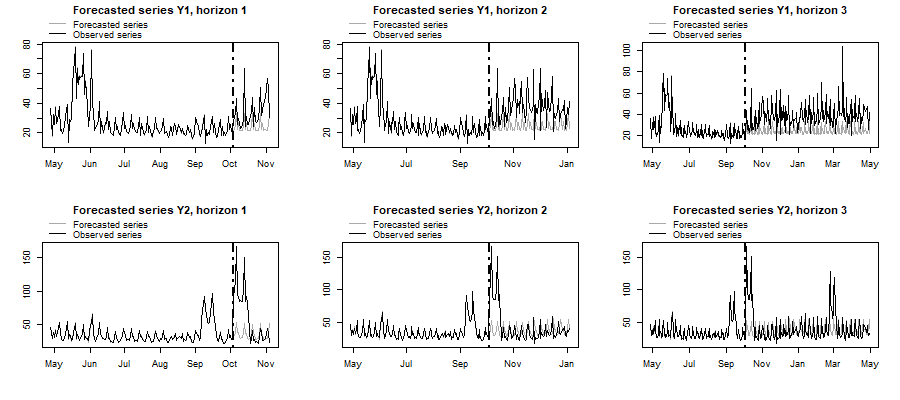} \\
 (7) & \includegraphics[width=12cm, height=4.5cm]{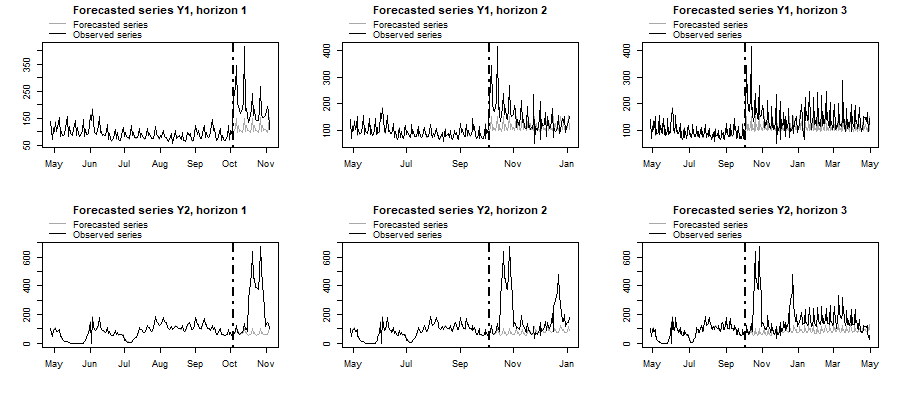} \\
 (8) & \includegraphics[width=12cm, height=4.5cm]{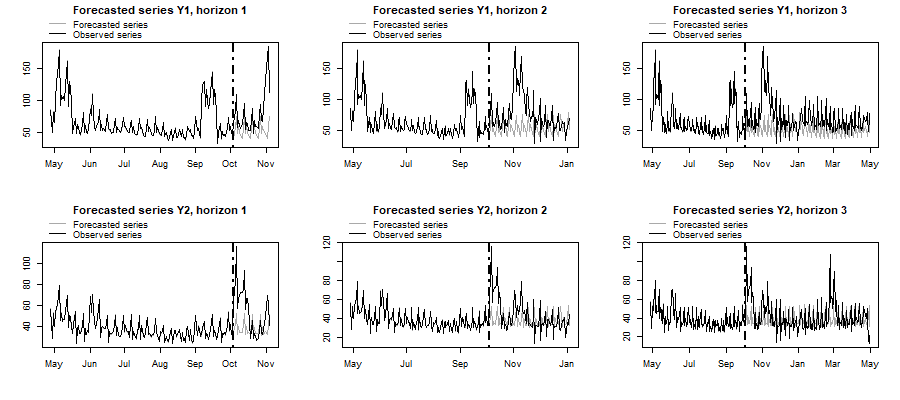} \\
 (9) & \includegraphics[width=12cm, height=4.5cm]{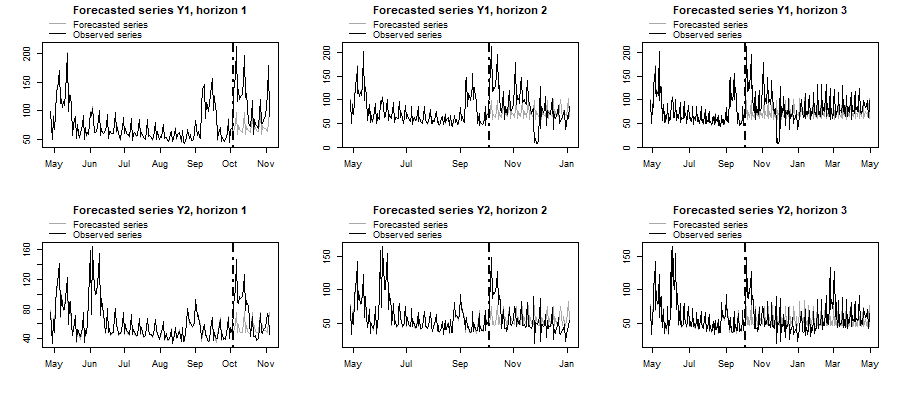} \\
 (10) & \includegraphics[width=12cm, height=4.5cm]{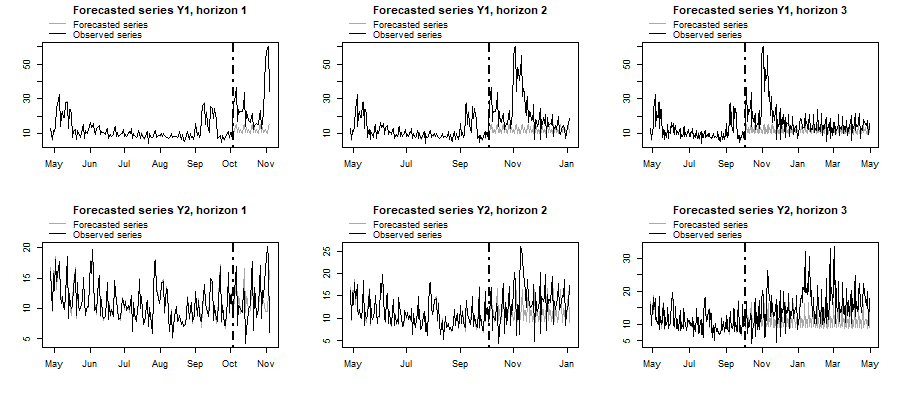} \\
 \end{tabular}
 \end{figure}

 \begin{figure}[h!]
 \centering
 \caption{Posterior predictive checks for each pair. Starting from the left: i) density of observed data plotted against the posterior predictive mean; ii) observed maximum compared to the distribution of the maximum from the posterior draws; iii) Normal QQ-Plot of standardized residuals; iv) autocorrelation function of standardized residuals.}
 \begin{tabular}{m{1cm}m{13cm}}
 (1) & \includegraphics[width=13cm, height=4.5cm]{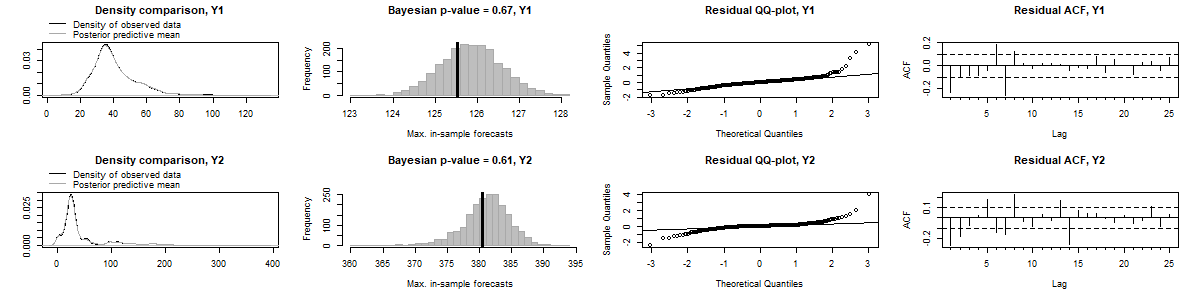} \\
 (2) & \includegraphics[width=13cm, height=4.5cm]{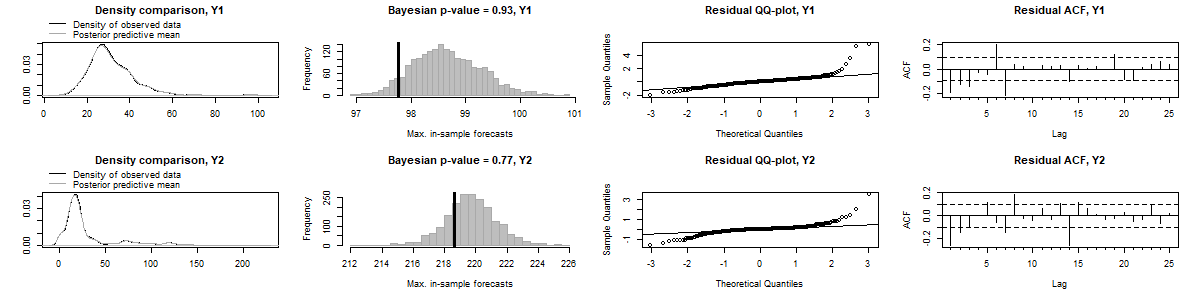} \\
 (3) & \includegraphics[width=13cm, height=4.5cm]{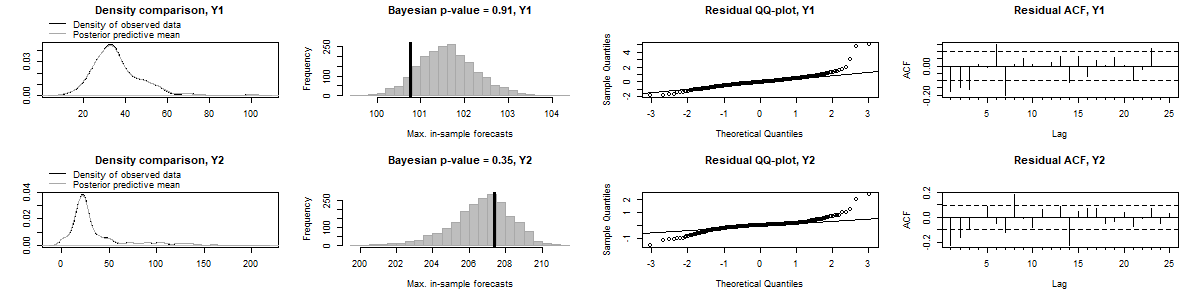} \\
 (4) & \includegraphics[width=13cm, height=4.5cm]{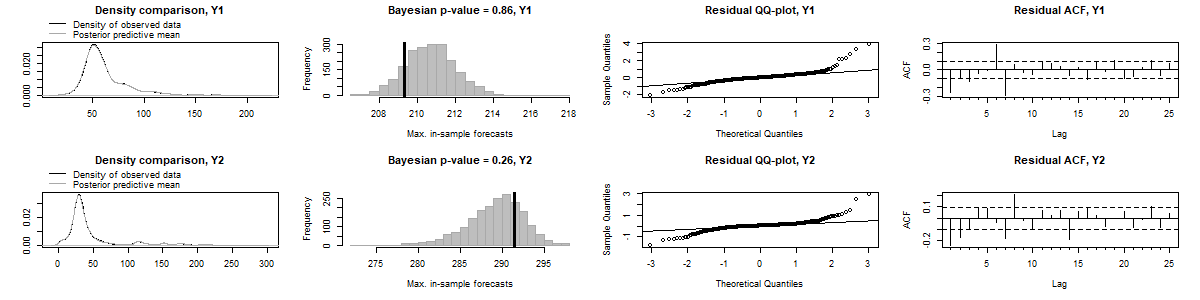} \\
 (5) & \includegraphics[width=13cm, height=4.5cm]{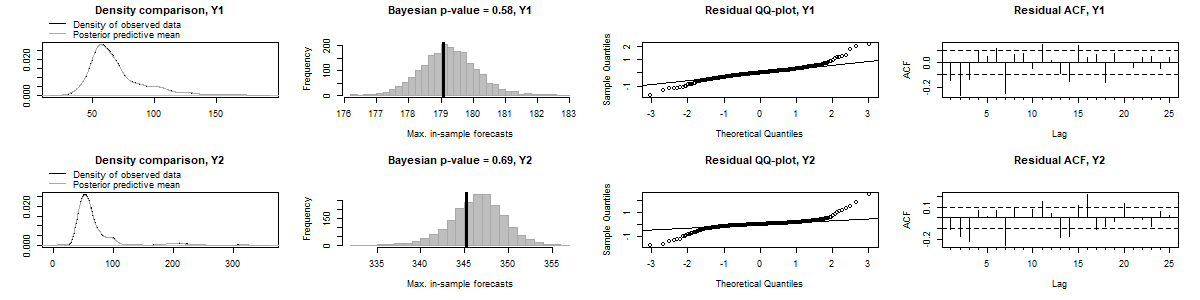} \\
 \end{tabular}
 \end{figure}
 \begin{figure}
 \begin{tabular}{m{1cm}m{13cm}}
 (6) & \includegraphics[width=13cm, height=4.5cm]{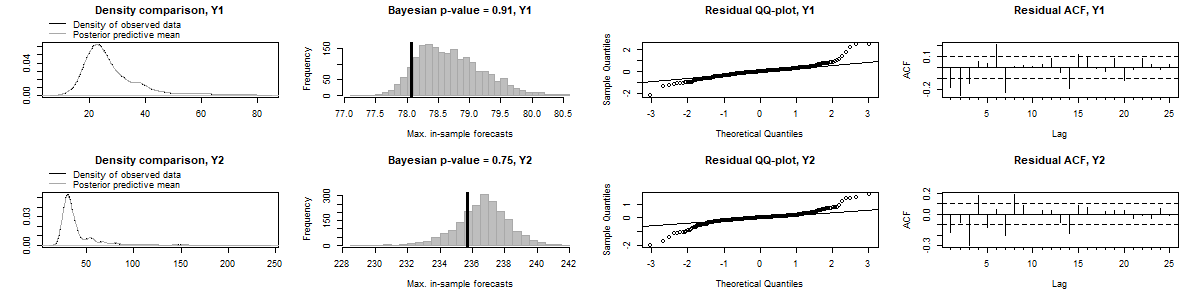} \\
 (7) & \includegraphics[width=13cm, height=4.5cm]{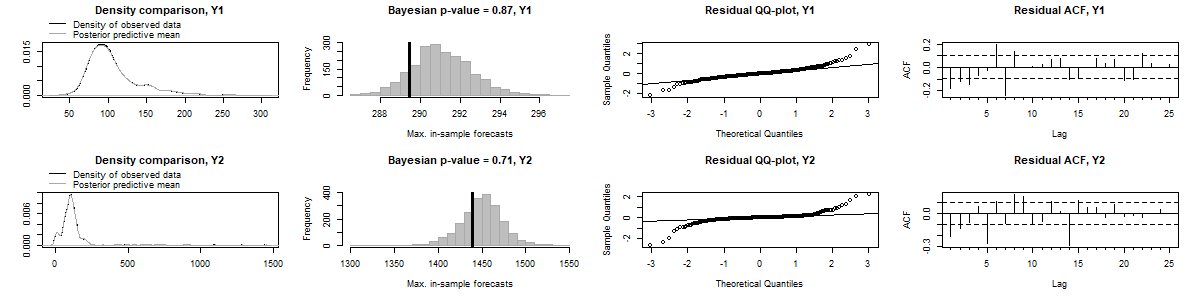} \\
 (8) & \includegraphics[width=13cm, height=4.5cm]{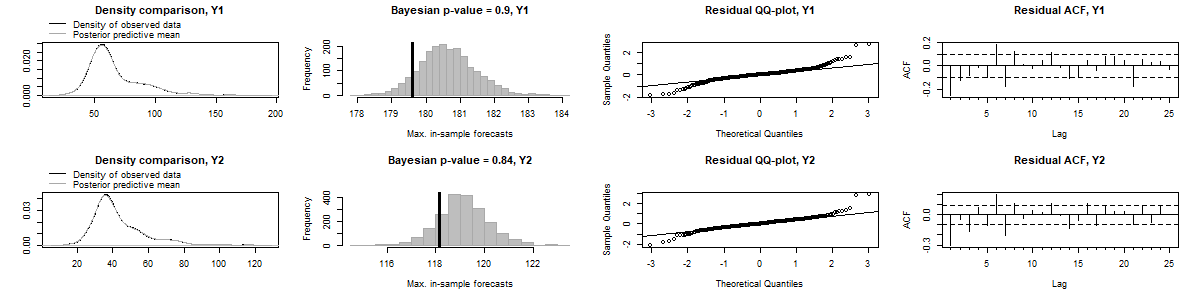} \\
 (9) & \includegraphics[width=13cm, height=4.5cm]{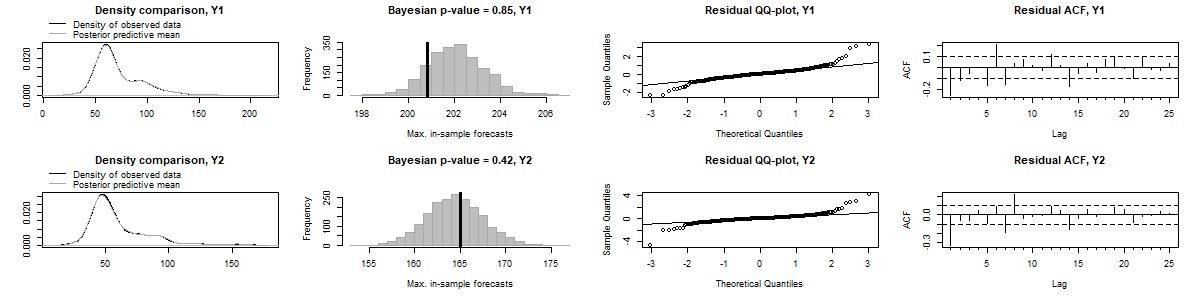} \\
 (10) & \includegraphics[width=13cm, height=4.5cm]{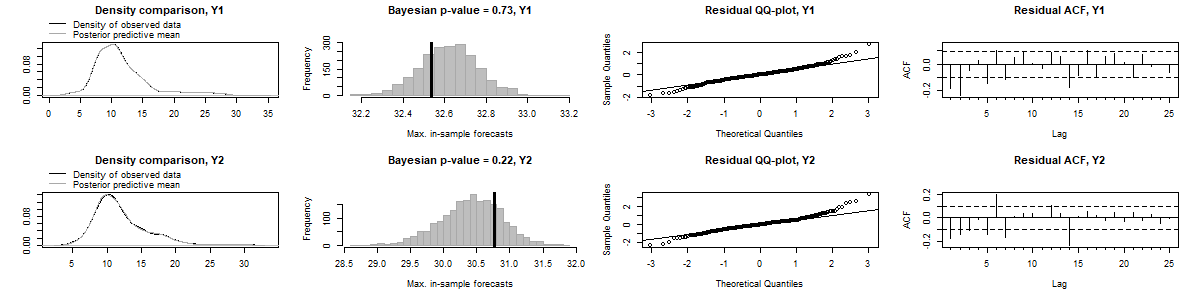} \\
 \end{tabular}
 \end{figure}

 \begin{figure}
 \centering
 \caption{Inclusion probabilities above the $0.5$ threshold of the regressors included in the MBSTS models estimated on each store-competitor pair.}
 \begin{tabular}{cc}
 \includegraphics[width=5cm,height=4.5cm]{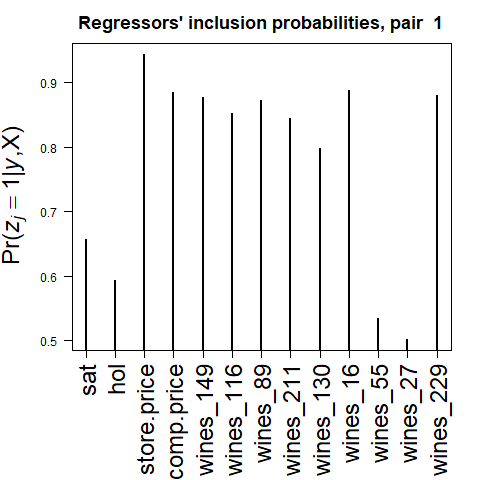} & \includegraphics[width=5cm,height=4.5cm]{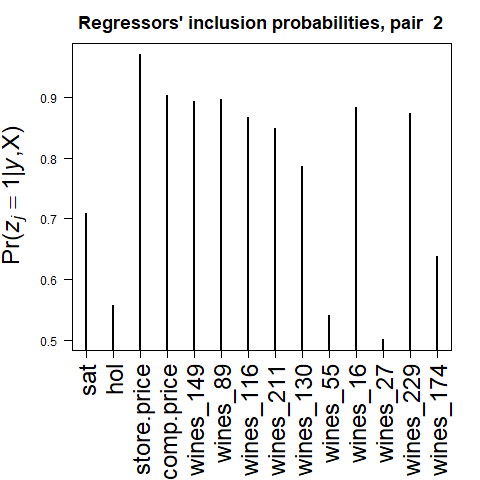} \\ \includegraphics[width=5cm,height=4.5cm]{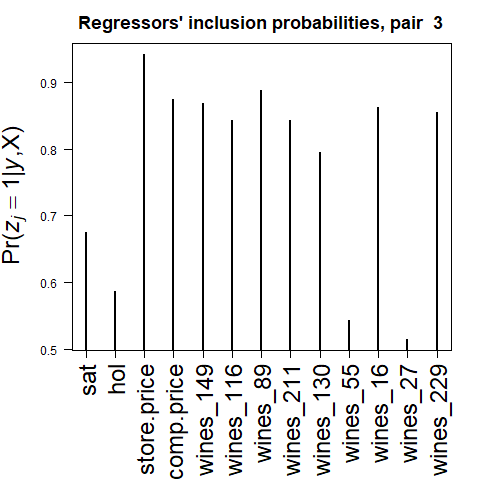} & \includegraphics[width=5cm,height=4.5cm]{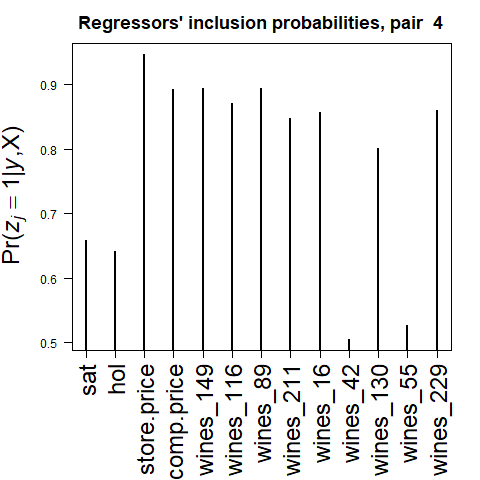} \\ \includegraphics[width=5cm,height=4.5cm]{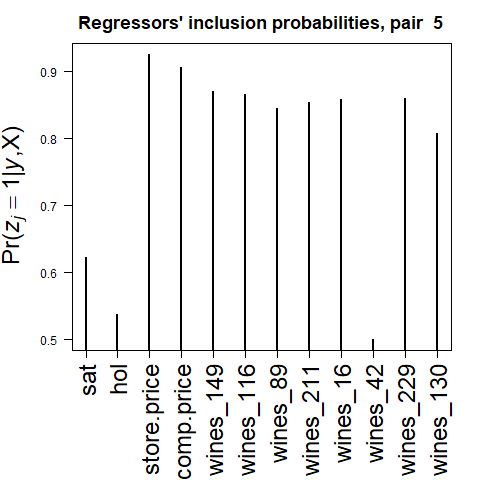} &
 \includegraphics[width=5cm,height=4.5cm]{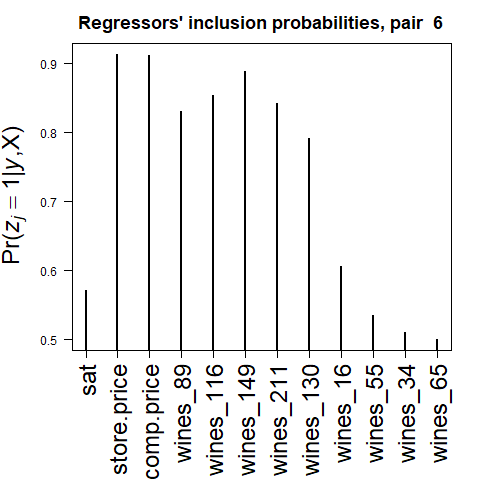} \\ \includegraphics[width=5cm,height=4.5cm]{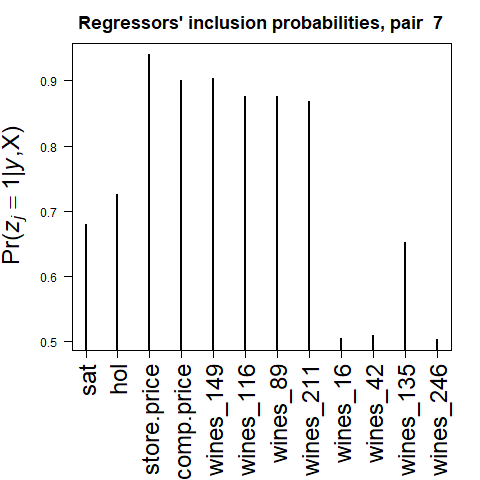} & \includegraphics[width=5cm,height=4.5cm]{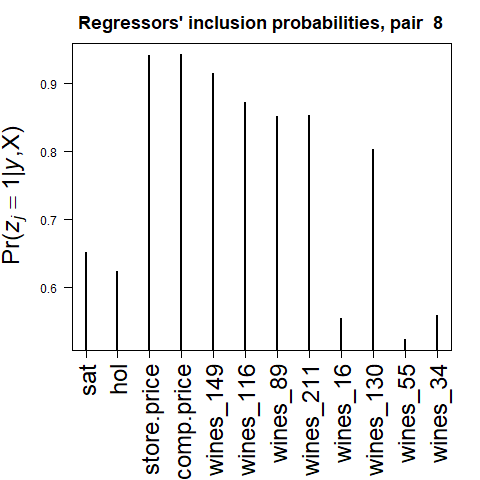} \\ \includegraphics[width=5cm,height=4.5cm]{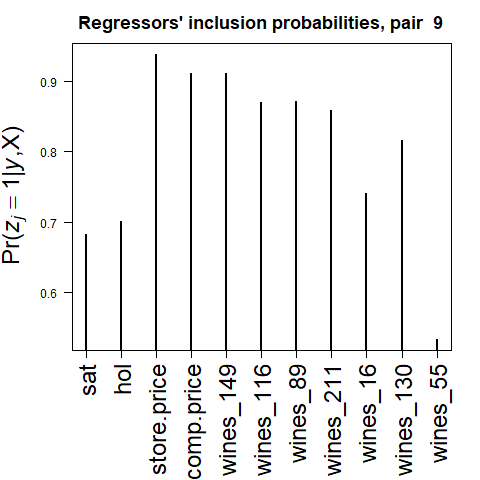} & \includegraphics[width=5cm,height=4.5cm]{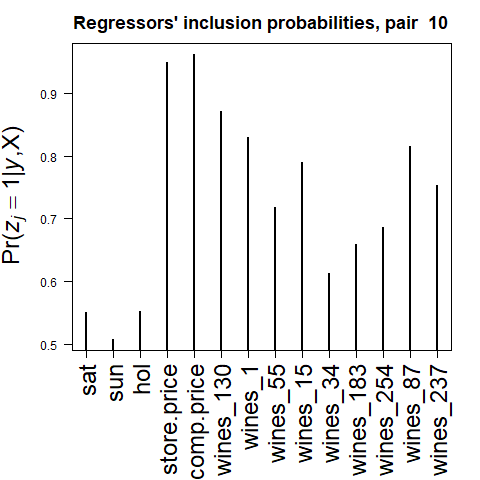} \\
 \end{tabular}
 \end{figure}

\clearpage

\newgeometry{bottom=2cm, top=2cm, left=2cm, right=2cm}
\subsection{Posterior predictive checks of alternative models}
\label{appA_postpred}

 \begin{figure}[h!]
 \centering
 \caption{Posterior predictive checks for a seasonal MBSTS model. Starting from the left: i) density of observed data plotted against the posterior predictive mean; ii) observed maximum vs distribution of the maximum from the posterior draws; iii) Normal QQ-Plot of standardized residuals; iv) autocorrelation function of standardized residuals.}
 \begin{tabular}{m{1cm}m{13cm}}
 (1) & \includegraphics[width=13cm, height=3.7cm]{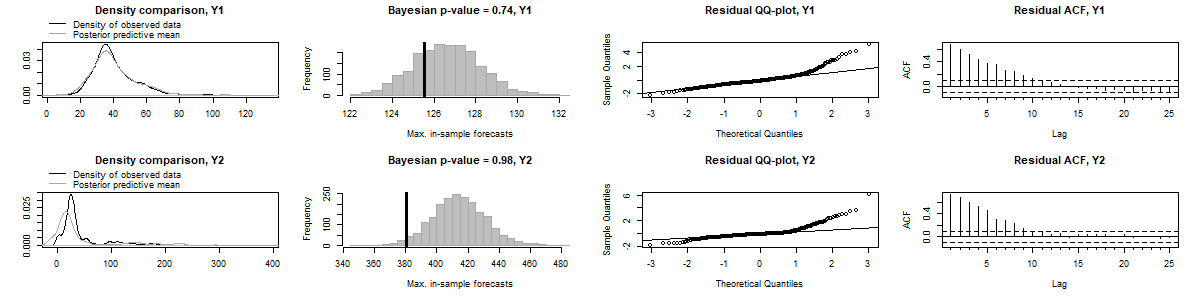} \\
 (2) & \includegraphics[width=13cm, height=3.7cm]{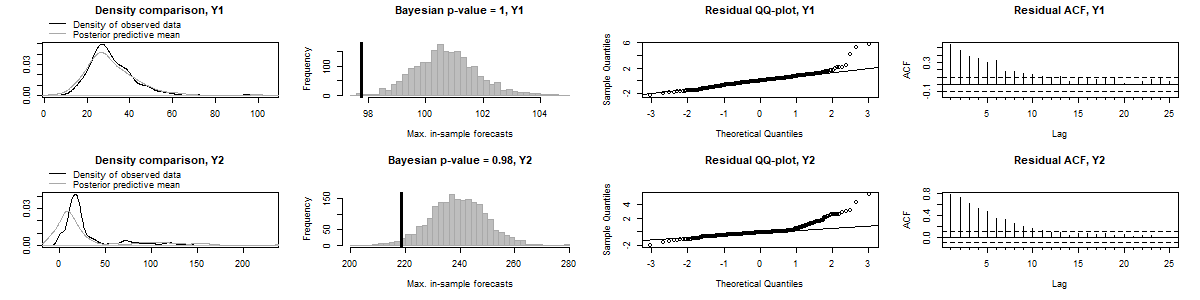} \\
 (3) & \includegraphics[width=13cm, height=3.7cm]{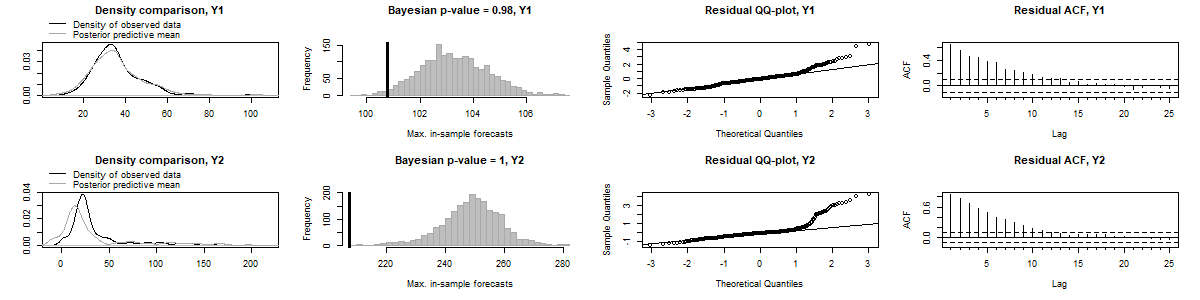} \\
 (4) & \includegraphics[width=13cm, height=3.7cm]{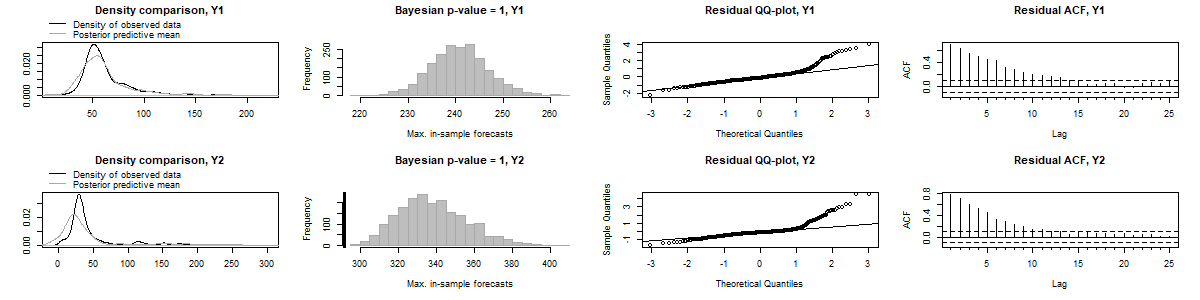} \\
 (5) & \includegraphics[width=13cm, height=3.7cm]{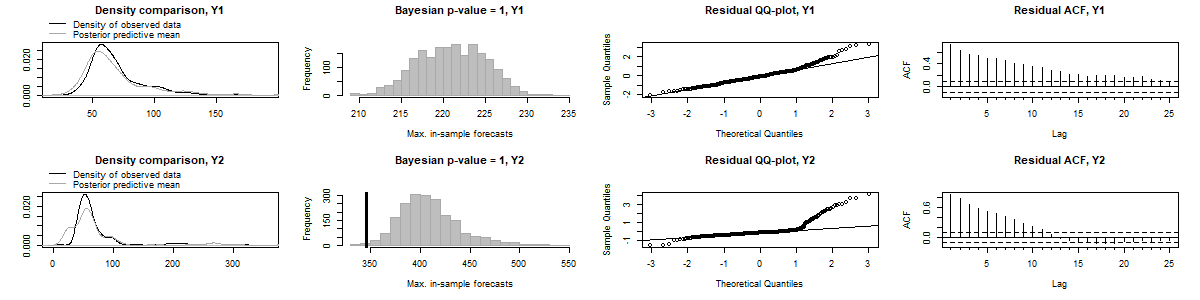} \\
 \end{tabular}
 \end{figure}

\restoregeometry

 \begin{figure}
 \centering
 \begin{tabular}{m{1cm}m{13cm}}
 (6) & \includegraphics[width=13cm, height=4.0cm]{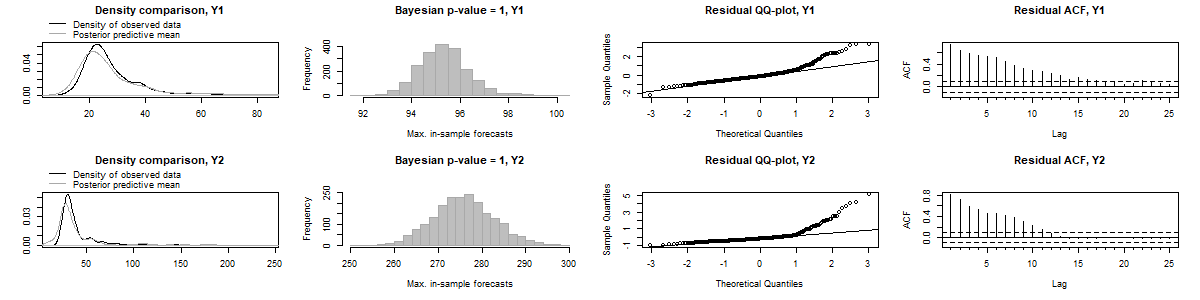} \\
 (7) & \includegraphics[width=13cm, height=4.0cm]{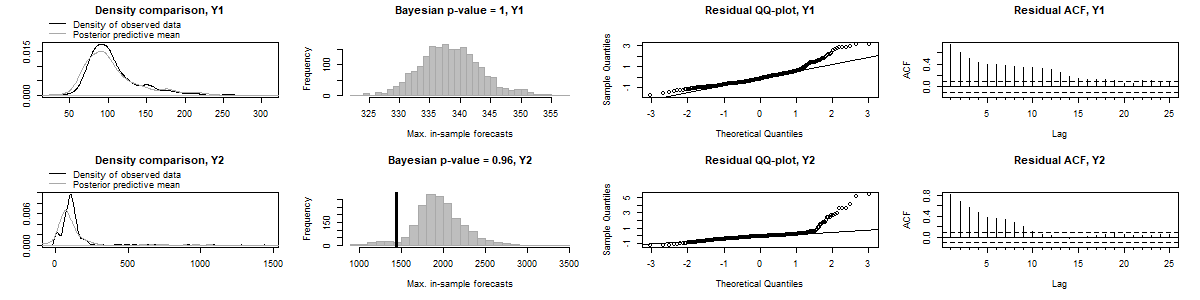} \\
 (8) & \includegraphics[width=13cm, height=4.0cm]{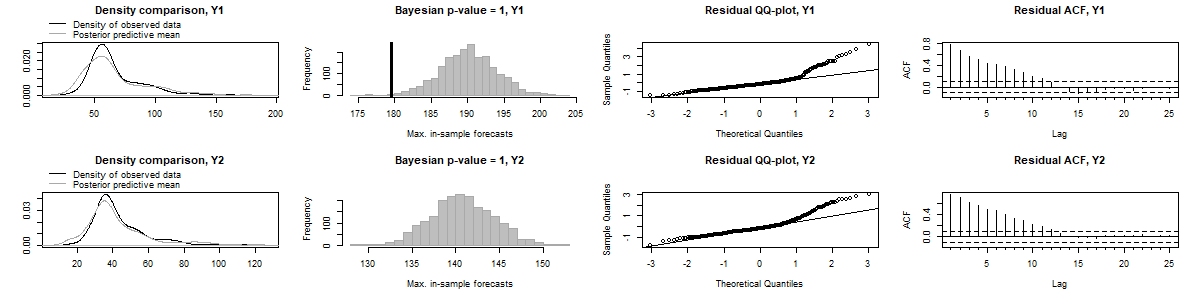} \\
 (9) & \includegraphics[width=13cm, height=4.0cm]{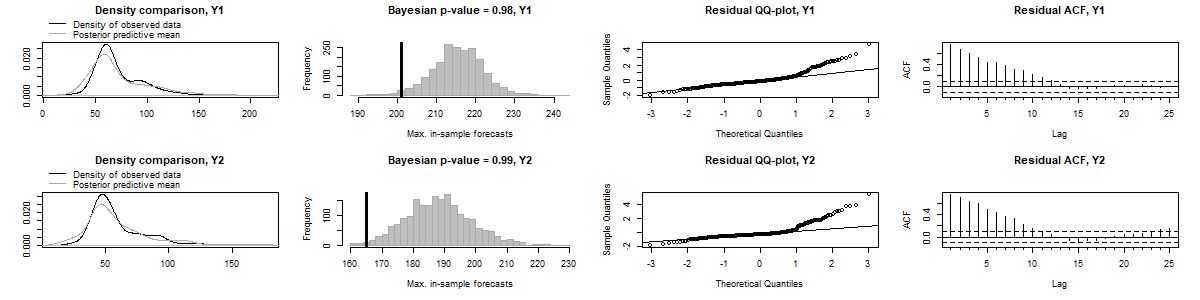} \\
 (10) & \includegraphics[width=13cm, height=4.0cm]{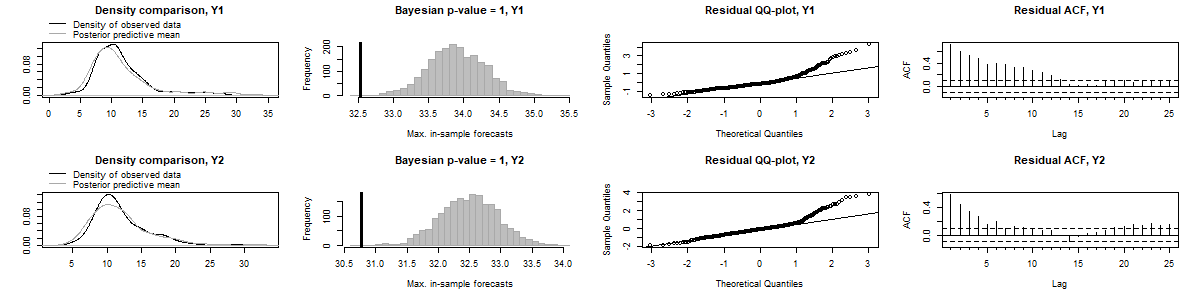} \\
 \end{tabular}
 \end{figure}

 \begin{figure}[h!]
 \centering
 \caption{Posterior predictive checks for a trend MBSTS model. Starting from the left: i) density of observed data plotted against the posterior predictive mean; ii) observed maximum compared to the distribution of the maximum from the posterior draws; iii) Normal QQ-Plot of standardized residuals; iv) autocorrelation function of standardized residuals.}
 \begin{tabular}{m{1cm}m{13cm}}
 (1) & \includegraphics[width=13cm, height=4.5cm]{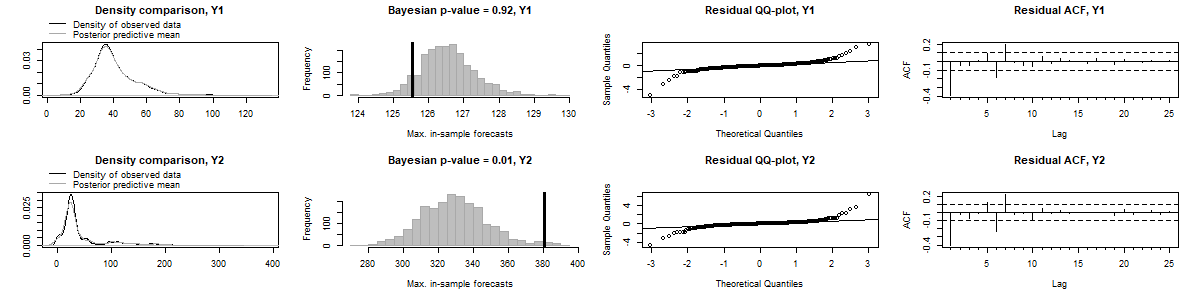} \\
 (2) & \includegraphics[width=13cm, height=4.5cm]{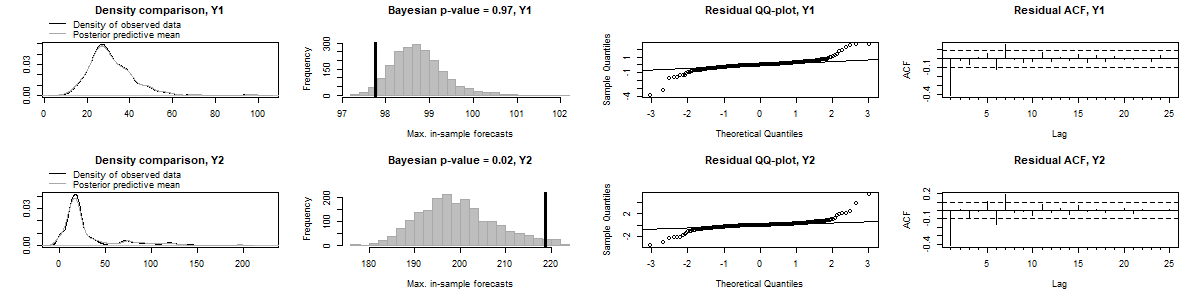} \\
 (3) & \includegraphics[width=13cm, height=4.5cm]{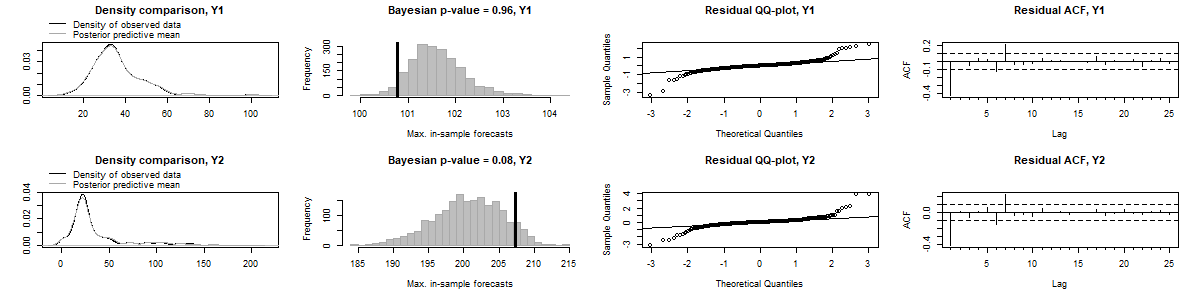} \\
 (4) & \includegraphics[width=13cm, height=4.5cm]{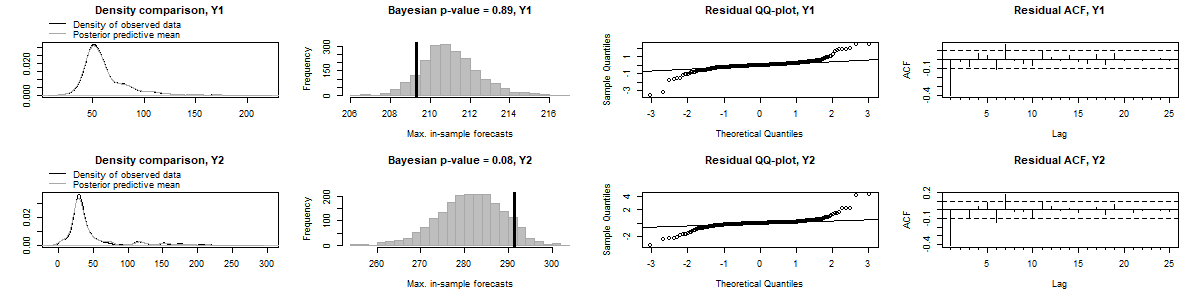} \\
 (5) & \includegraphics[width=13cm, height=4.5cm]{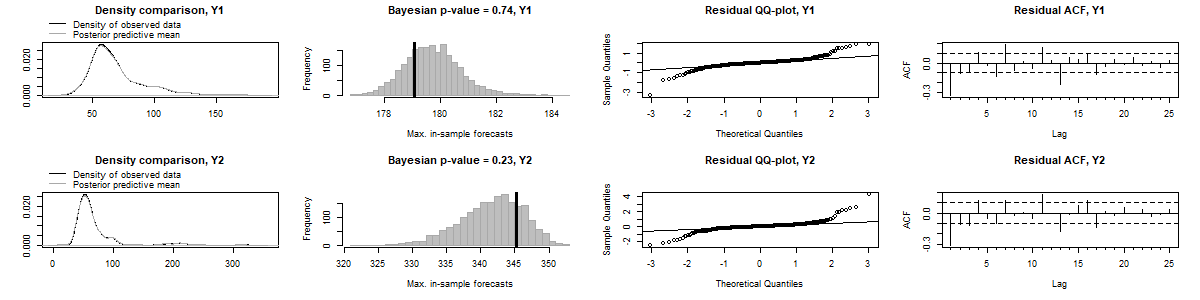} \\
 \end{tabular}
 \end{figure}
 \begin{figure}
 \centering
 \begin{tabular}{m{1cm}m{13cm}}
 (6) & \includegraphics[width=13cm, height=4.5cm]{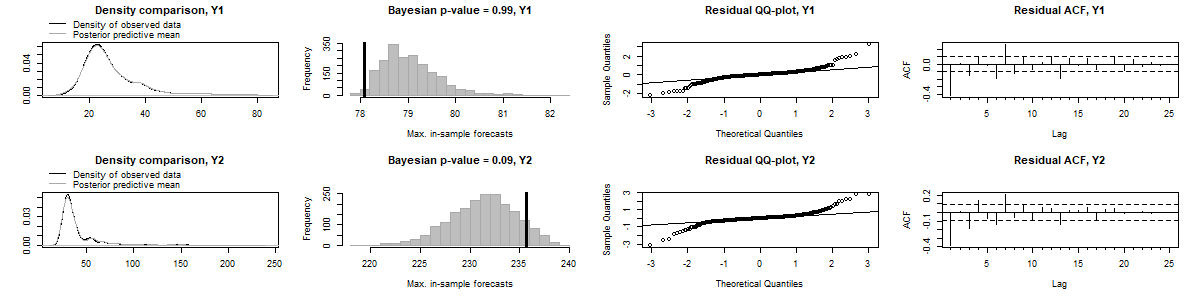} \\
 (7) & \includegraphics[width=13cm, height=4.5cm]{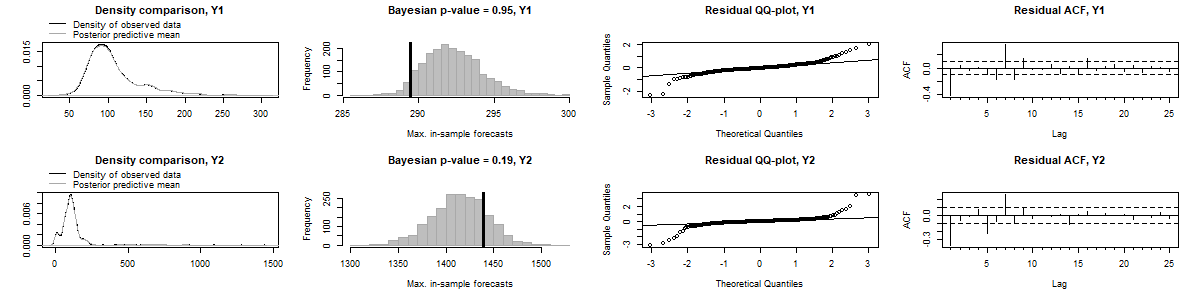} \\
 (8) & \includegraphics[width=13cm, height=4.5cm]{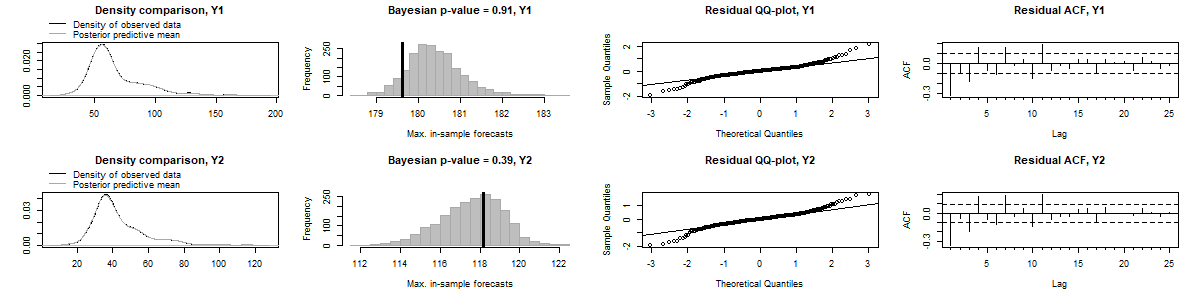} \\
 (9) & \includegraphics[width=13cm, height=4.5cm]{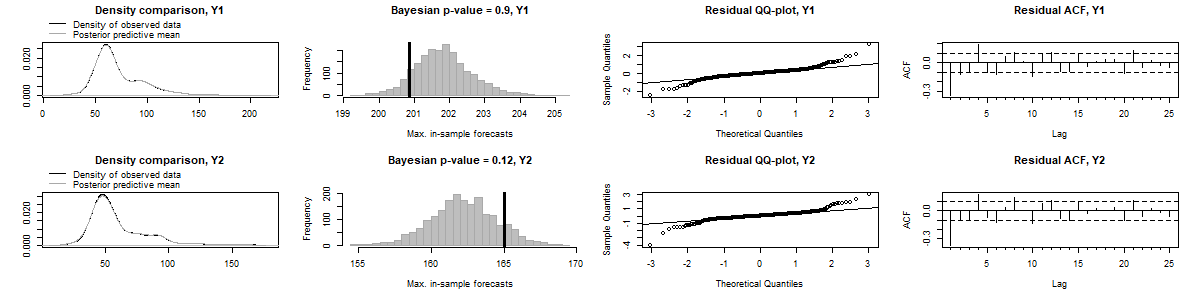} \\
 (10) & \includegraphics[width=13cm, height=4.5cm]{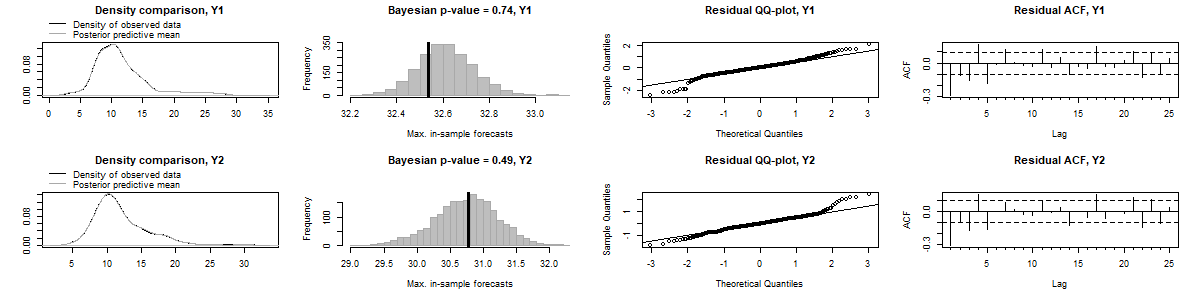} \\
 \end{tabular}
 \end{figure}

 \begin{figure}[h!]
 \centering
 \caption{Posterior predictive checks for a trend and seasonal MBSTS model estimated on the daily units sold. Starting from the left: i) density of observed data plotted against the posterior predictive mean; ii) observed maximum compared to the distribution of the maximum from the posterior draws; iii) Normal QQ-Plot of standardized residuals; iv) autocorrelation function of standardized residuals.}
 \begin{tabular}{m{1cm}m{13cm}}
 (1) & \includegraphics[width=13cm, height=4.5cm]{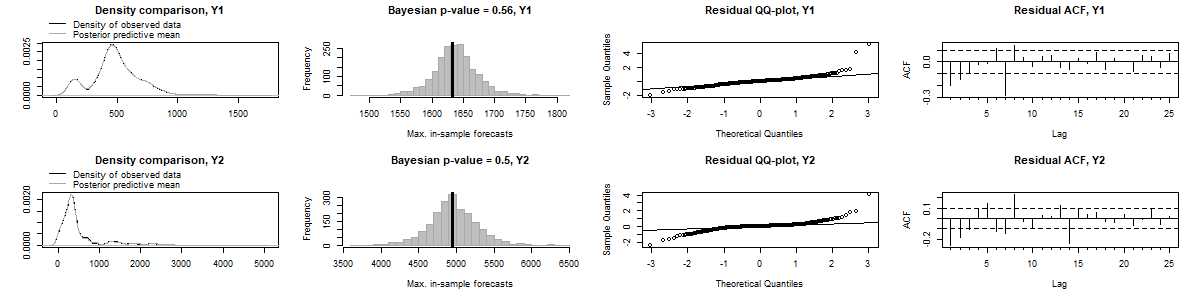} \\
 (2) & \includegraphics[width=13cm, height=4.5cm]{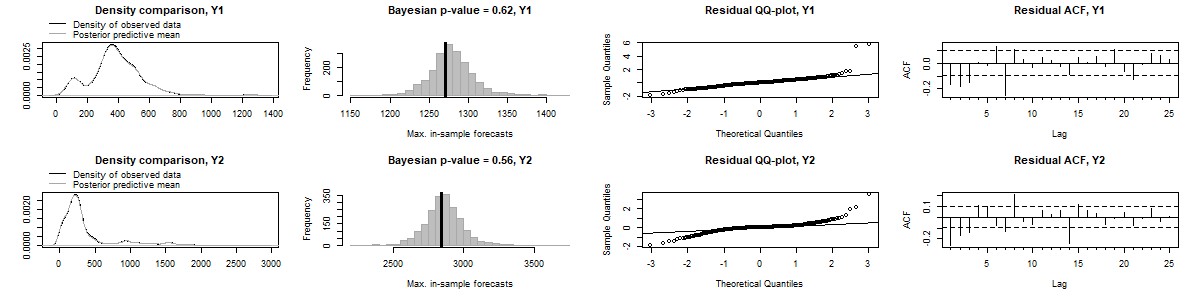} \\
 (3) & \includegraphics[width=13cm, height=4.5cm]{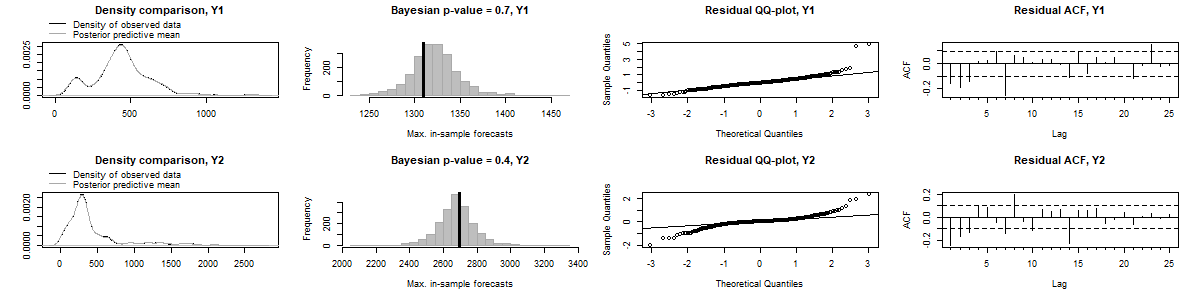} \\
 (4) & \includegraphics[width=13cm, height=4.5cm]{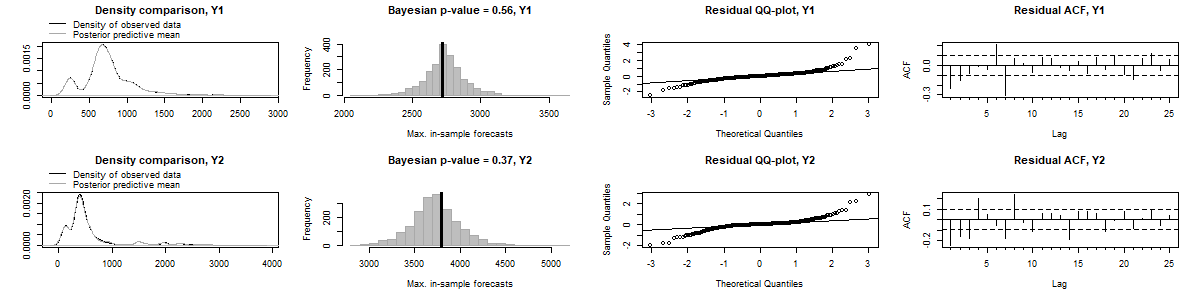} \\
 (5) & \includegraphics[width=13cm, height=4.5cm]{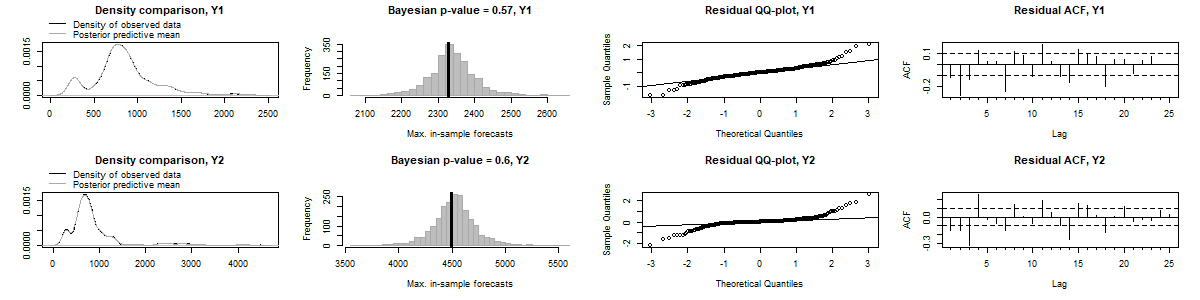} \\
 \end{tabular}
 \end{figure}
 \begin{figure}
 \centering
 \begin{tabular}{m{1cm}m{13cm}}
 (6) & \includegraphics[width=13cm, height=4.5cm]{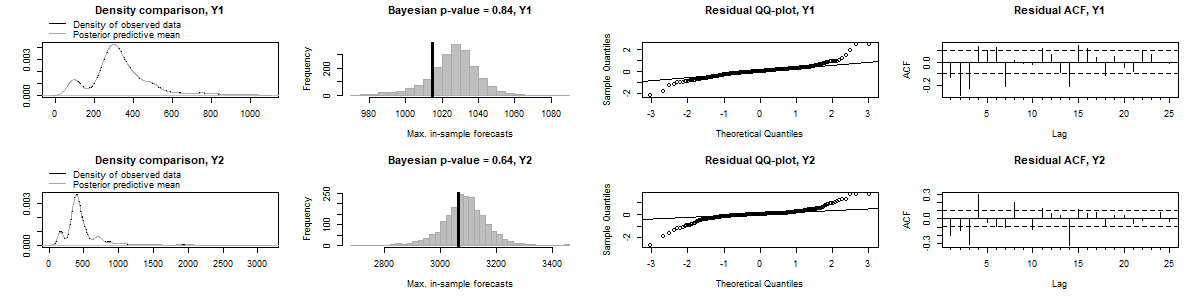} \\
 (7) & \includegraphics[width=13cm, height=4.5cm]{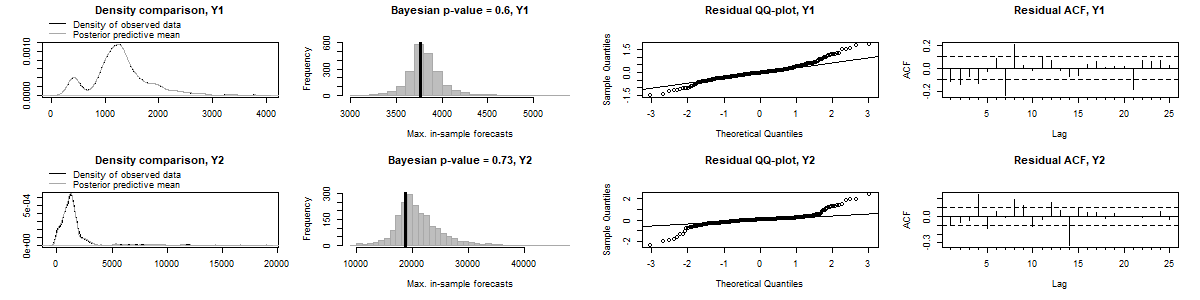} \\
 (8) & \includegraphics[width=13cm, height=4.5cm]{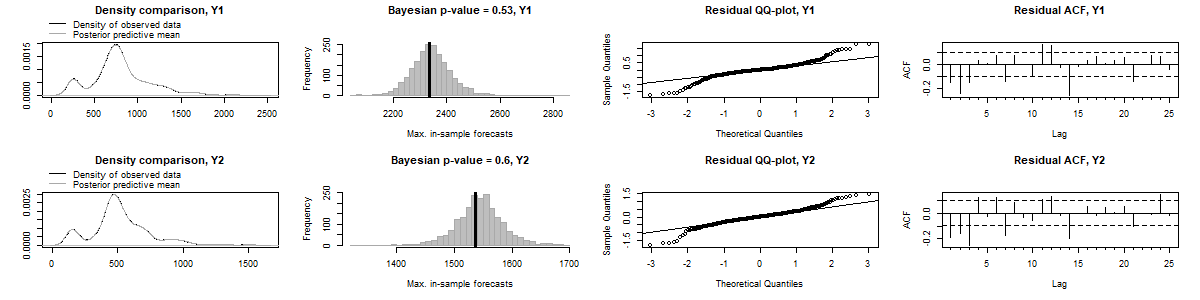} \\
 (9) & \includegraphics[width=13cm, height=4.5cm]{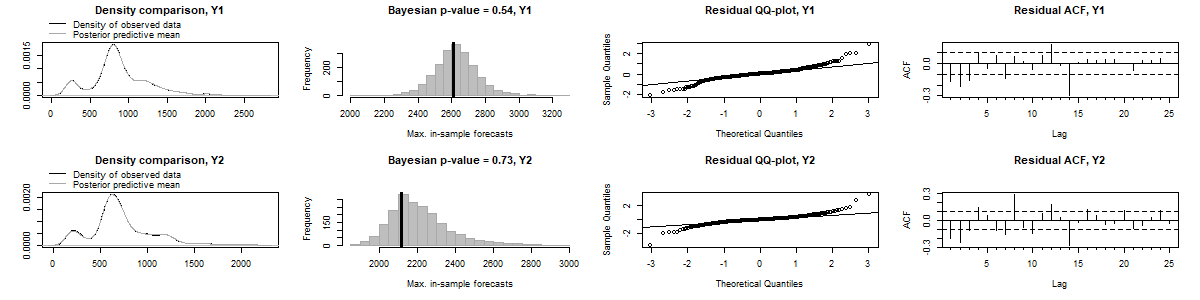} \\
 (10) & \includegraphics[width=13cm, height=4.5cm]{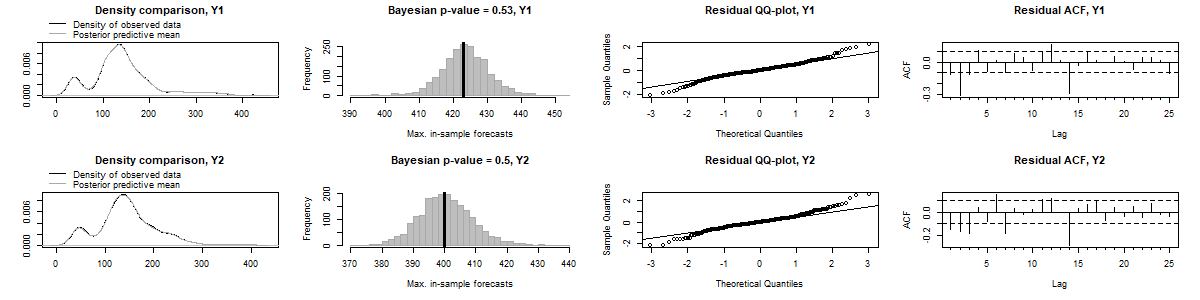} \\
 \end{tabular}
 \end{figure}

\clearpage

\section{}
\label{appB}

\subsection{Marginal and conditional causal effects}
\label{appB_other_effects}

We can combine the general causal effects to define the marginal causal effect that captures the impact of changing a single unit within a group across all possible treatment combinations the group could have received.

\begin{defn} 
Let $\mathcal{A}_i \subset \{0,1\}^d$ be the subset of all treatment paths $\bl{w}$ such that $\w^{(i)}=1$ and $\mathcal{B}_i \subset \{0,1\}^d$ be the subset of all  treatment paths $\tilde{\bl{w}}$ such that $\w^{(i)}=0$. The \textbf{marginal causal effect} on the $i^\text{th}$ series is the sum of the $i^\text{th}$ elements of $\bs{\tau}_t(\bl{w},\tilde{\bl{w}})$ computed across all the possible realizations in $\mathcal{A}_i \times \mathcal{B}_i$,

\begin{equation}
\label{eqn:marginal}
\tau_t(i) = \sum\limits_{(\bl{w},\tilde{\bl{w}}) \in \mathcal{A} \times \mathcal{B}} \tau_t^{(i)}(\bl{w},\tilde{\bl{w}})
\end{equation}

The cumulative marginal causal effect at time point $t' > t^*$ is 

\begin{equation}
\label{eqn:cum_marginal}
\Delta_{t'}(i) =  \sum\limits_{t=t^*+1}^{t'} \tau_t(i) 
\end{equation}

The temporal average marginal causal effect at time point $t'$ is 

\begin{equation}
\label{eqn:avg_marginal}
\bar{\tau}_{t'}(i) = \frac{1}{t'-t^*} \sum\limits_{t=t^*+1}^{t'} \tau_t(i) = \frac{1}{t'-t^*} \Delta_{t'}(i)
\end{equation}

Now, let $N_{\mathcal{A}_i \times \mathcal{B}_i}$ denote the total number of possible assignments in $\mathcal{A}_i \times \mathcal{B}_i$; the \textbf{mean marginal causal effect} can be defined as,

\begin{equation}
\label{eqn:mean_marginal}
\tau_t(i,N_{\mathcal{A}_i \times \mathcal{B}_i}) = \frac{1}{N_{\mathcal{A}_i \times \mathcal{B}_i}} \sum\limits_{(\bl{w},\tilde{\bl{w}}) \in \mathcal{A}_i \times \mathcal{B}_i} \tau_t^{(i)}(\bl{w},\tilde{\bl{w}})
\end{equation}

The cumulative and temporal average mean marginal effects can be then derived as in equations (\ref{eqn:cum_marginal}) and (\ref{eqn:avg_marginal}).
\end{defn}

\vspace{5pt}

The marginal causal effect captures the impact of assigning the $i^\text{th}$ unit to treatment, averaged over all possible interventions that could have been applied to the other units. %Thus, the marginal effect extends to the time series setting the average distributional shift effect in \citet{Savje:Aronow:Hudgens:2020}, with the difference that it is averaged across units whereas the marginal effect is individual-specific and, in its temporal average version, averaged across times. 
Thus, the marginal effect can be considered an extension to the time series setting of the average distributional shift effect in \citet{Savje:Aronow:Hudgens:2020}, with one main difference: the average distributional shift effect is averaged across units whereas the marginal effect is individual-specific and, in its temporal average version, it is averaged across times. 
We could make this effect slightly more general by introducing non-stochastic weights in the summation to up-weight or down-weight particular treatment combinations. However, this makes the notation somewhat more cumbersome without adding new insights. 

\begin{example}
Suppose that we are interested in estimating the marginal effect of the active treatment on the store brand, then $\mathcal{A} = \{ (1,0),(1,1) \}$, $\mathcal{B} = \{ (0,0),(0,1) \}$, and 
$\mathcal{A} \times \mathcal{B} = \{ (1,0)(0,0); (1,0)(0,1);
(1,1)(0,0); (1,1)(0,1) \}
$. In words, the set $\mathcal{A}$ indicates all the possible combinations where the store brand gets to be treated and the set $\mathcal{B}$ denotes all the combinations where it receives control; thus, the marginal causal effect compares all the scenarios where the store brand is permanently discounted with the situations where it is not discounted. Furthermore, $\bs{\tau}_t(\bl{w},\tilde{\bl{w}}) = (\tau_t^{(s)}(\bl{w},\tilde{\bl{w}}),\tau_t^{(c)}(\bl{w},\tilde{\bl{w}}))$ and hence,  \\
$
\tau_t(s) = \tau_{t}^{(s)}((1,0),(0,0))+ \tau_{t}^{(s)}((1,0),(0,1)) + \tau_{t}^{(s)}((1,1),(0,0)) + \tau_{t}^{(s)}((1,1),(0,1))
$. Finally, the mean marginal effect of the active treatment on the store brand is  
$
\tau_t(s,4) = 1/4 \cdot \tau_t(s).
$ 
\end{example}

\vspace{5pt}

A special case of the general causal effect is the conditional causal effect that fixes the treatments for all units within the group except for the $i^\text{th}$ unit. 
\begin{defn} 
For $\bl{w} \in \mathcal{W}^{d-1}$, the \textbf{conditional causal effect} is the effect of assigning the $i^\text{th}$ series to treatment as opposed to control, fixing the treatments of the other series to equal $\bl{w}$

\begin{equation}
\label{eqn:conditional}
% \bs{\tau}_t^\dagger(i) = \bl{Y}_t(\bl{w}^{\dagger}) - \bl{Y}_t(\tilde{\bl{w}}^{\dagger})
\bs{\tau}_t^\dagger(i,\bl{w}) = \bl{Y}_t((\w_1,\dots, \w_{i-1}, 1, \w_{i},\dots, \w_{d-1})) - \bl{Y}_t((\w_1,\dots, \w_{i-1}, 0, \w_{i},\dots, \w_{d-1}))
\end{equation}
Similar to the marginal and mean marginal causal effects, we can define the cumulative and temporal average conditional causal effect at time point $t'>t^\ast$.
\end{defn}

The conditional effect can also be seen as the generalization to the time-series setting of the assignment-conditional unit-level treatment effect in \citet{Savje:Aronow:Hudgens:2020}.

\begin{example}
The general effect defined in Example 1 is already a conditional effect, since it measures the impact of the permanent reduction on the store brand given that the competitor is always assigned to control. However, we may also be interested in the conditional effect of the permanent price reduction on the store brand when the competitor brand is permanently discounted as well, that is, $\bl{w}^{\dagger} = (1,1)$, $\tilde{\bl{w}}^{\dagger} = (0,1)$ and $\bs{\tau}_t^{\dagger}(s,(1,1)) = \bl{Y}_t(1,1) - \bl{Y}_t(0,1)$. 
\end{example}

We report below the results for the mean marginal effect as defined in (\ref{eqn:mean_marginal}) and the conditional effect $\hat{\bar{\bs{\tau}}}_t((1,1),(0,1))$.

\begin{table}[htbp]
\centering
\caption{Temporal average mean marginal causal effect of the new price policy on the ten store brands computed at three time horizons.}
\label{tab:mean_marginal_effect}
\begin{tabular}{rrrrrrrrrrrr} \toprule
  & \multicolumn{3}{c}{$1$ month} & & \multicolumn{3}{c}{$3$ months} & & \multicolumn{3}{c}{$6$ months} \\ 
  & $\hat{\bs{\tau}}_t(s,4)$ & $2.5\%$ & $97.5\%$ & & $\hat{\bs{\tau}}_t(s,4)$ & $2.5\%$ & $97.5\%$ & & $\hat{\bs{\tau}}_t(s,4)$ & $2.5\%$ & $97.5\%$ \\ \cmidrule{2-4} \cmidrule{6-8} \cmidrule{10-12}
  1 & 3.53 & -12.27 & 19.33 & & 2.39 & -21.98 & 26.88 & & 3.55 & -32.92 & 39.98 \\ 
  2 & 3.55 & -7.39  & 14.51 & & 2.51 & -15.09 & 19.39 & & 3.34 & -22.05 & 29.27 \\ 
  3 & 4.02 & -7.00  & 16.14 & & 2.71 & -15.91 & 20.70 & & 3.97 & -24.13 & 31.25 \\ 
  4 &\bf{24.06} & \bf{2.07} & \bf{49.43} & & 11.58 & -26.51 & 50.12 & & 12.22 & -46.22 & 68.82 \\ 
  5 & 1.98 & -24.69 & 28.29 & & 3.73 & -40.92 & 48.08 & & 5.74 & -60.05 & 73.52 \\ 
  6 & 4.85 & -7.97  & 17.53 & & 5.94 & -14.73 & 26.51 & & 6.78 & -24.61 & 38.53 \\ 
  7 & \bf{39.19} & \bf{0.04} & \bf{77.11} & & 17.33 & -40.86 & 76.01 & & 14.84 & -74.46 & 102.94 \\ 
  8 & 12.67 & -14.32 & 39.30 & & 11.71 & -34.58 & 54.99 & & 8.63 & -57.56 & 73.01 \\ 
  9 & 20.46 & -9.44 & 50.67 & & 8.19 & -39.87 & 57.46 & & 6.44 & -65.70 & 82.99 \\ 
  10& \bf{6.26} & \bf{0.52} & \bf{11.98} & & 4.86 & -4.22 & 14.22 & & 2.72 & -11.39 & 16.84 \\
\bottomrule
\end{tabular}
\end{table}

\begin{table}[h!]
\centering
\caption{Temporal average conditional causal effect of the new price policy on the ten store (s) - competitor (c) pairs computed at three time horizons. In this table, $\hat{\bar{\tau}}_t$ stands for the conditional effect $\hat{\bar{\bs{\tau}}}_t((1,1),(0,1))$.}
\label{tab:cond_effect}
\begin{tabular}{rrrrrrrrrrrrr} \toprule
 & & \multicolumn{3}{c}{$1$ month} & & \multicolumn{3}{c}{$3$ months} & & \multicolumn{3}{c}{$6$ months} \\ 
 & & $\hat{\bar{\bs{\tau}}}_t$ & $2.5\%$ & $97.5\%$ & & $\hat{\bar{\bs{\tau}}}_t$ & $2.5\%$ & $97.5\%$ & & $\hat{\bar{\bs{\tau}}}_t$ & $2.5\%$ & $97.5\%$ \\ \cmidrule{3-5} \cmidrule{7-9} \cmidrule{11-13}
 \multirow{2}{*}{(1)}
& s & 0.09  & -0.14 & 0.39 & & 0.11  & -0.14 & 0.38 & & 0.12 & -0.12 & 0.37 \\ 
& c & -0.34 & -1.68 & 0.75 & & -0.63 & -1.64 & 0.78 & & -0.78 & -1.55 & 0.64 \\ 
\multirow{2}{*}{(2)}
& s & 0.08  & -0.12 & 0.30 & & 0.10  & -0.14 & 0.31 & & 0.11 & -0.10 & 0.29 \\ 
& c & -0.30 & -0.86 & 0.34 & & -0.54 & -0.93 & 0.34 & & -0.65 & -0.74 & 0.24 \\
\multirow{2}{*}{(3)}
& s & 0.11  & -0.15 & 0.36 & & 0.12  & -0.12 & 0.36 & & 0.11 & -0.09 & 0.35 \\ 
& c & -0.22 & -0.79 & 0.67 & & -0.37 & -0.94 & 0.58 & & -0.21 & -0.74 & 0.23 \\  
\multirow{2}{*}{(4)}
& s & 0.28  & -0.97 & 1.50 & & 0.50  & -0.92 & 1.62 & & 0.71 & -0.47 & 4.17 \\ 
& c & -1.04 & -3.92 & 2.64 & & -2.13 & -4.18 & 2.36 & & -3.22 & -19.10 & 1.11 \\ 
\multirow{2}{*}{(5)}
& s & -0.15 & -2.69 & 4.01 & & -0.12 & -7.83 & 1.30 & & -0.27 & -23.15 & 1.39 \\ 
& c & -0.08 & -2.48 & 2.53 & & -0.08 & -2.73 & 2.67 & & -0.07 & -2.23 & 3.66 \\  
\multirow{2}{*}{(6)}
& s & 0.17  & -0.28 & 0.57 & & 0.12  & -0.31 & 0.67 & & -0.02 & -0.27 & 0.55 \\ 
& c & -0.34 & -1.62 & 0.84 & & -0.31 & -1.90 & 0.73 & & -0.29 & -1.50 & 0.70 \\
\multirow{2}{*}{(7)}
& s & 0.20  & -1.16 & 1.60 & & 0.21  & -1.14  & 1.62  & & 0.20 & -1.15 & 1.63 \\ 
& c & -1.09 & -21.89& 18.58& & -1.46 & -21.83 & 18.40 & & -1.26 & -22.02 & 18.63 \\
\multirow{2}{*}{(8)}
& s & 0.12  & -2.75  & 2.79  & & 0.09 & -4.46  & 4.22  & & 0.18 & -6.69 & 7.86 \\ 
& c & -0.02 & -12.99 & 14.38 & & 0.15 & -19.31 & 23.83 & & -0.31 & -39.60 & 32.96 \\ 
\multirow{2}{*}{(9)}
& s & 0.64  & -42.52 & 43.54 & & 1.00  & -70.38 & 78.18 & & 0.81 & -106.58 & 119.18 \\ 
& c & -0.29 & -45.17 & 44.19 & & -0.25 & -74.86 & 72.15 & & 0.28 & -112.62 & 115.76 \\ 
\multirow{2}{*}{(10)}
& s & 0.09 & -2.76 & 3.08 & & 0.08 & -5.16 & 4.39  & & 0.13 & -7.60 & 7.00 \\ 
& c & 0.04 & -5.83 & 6.93 & & 0.07 & -8.62 & 10.90 & & -0.02 & -17.00 & 16.37 \\   
\bottomrule
\end{tabular}
\end{table}
\clearpage

\newgeometry{top = 3cm, bottom = 2cm, left = 2cm, right = 2cm}
\subsection{Additional results}
\label{appB_addres}

\begin{table}[h!]
\centering
\caption{Temporal average general causal effects of the new price policy on the ten store (s) - competitor (c) pairs computed at three time horizons. In this table, $\hat{\bar{\bs{\tau}}}_t$ stands for the general effect $\hat{\bar{\bs{\tau}}}_t((1,0),(0,0))$ and the results are obtained including in the set of covariates the difference in price between the store and competitor brand prior to the intervention (in the post-intervention period the difference in price is computed from the prior price).}
\label{tab:general_effect_diff}
\scalebox{0.8}{
\begin{tabular}{rrrrrrrrrrrrr} \toprule
 & & \multicolumn{3}{c}{$1$ month} & & \multicolumn{3}{c}{$3$ months} & & \multicolumn{3}{c}{$6$ months} \\ 
 & & $\hat{\bar{\bs{\tau}}}_t$ & $2.5\%$ & $97.5\%$ & & $\hat{\bar{\bs{\tau}}}_t$ & $2.5\%$ & $97.5\%$ & & $\hat{\bar{\bs{\tau}}}_t$ & $2.5\%$ & $97.5\%$ \\ \cmidrule{3-5} \cmidrule{7-9} \cmidrule{11-13}
\multirow{2}{*}{(1)}   
& s & 7.86 & -22.72 & 39.39 & & 6.01 & -44.36 & 54.53 & & 8.69 & -62.66 & 81.65 \\ 
& c & 24.76 & -101.23 & 154.16 & & 18.14 & -189.20 & 223.43 & & 7.94 & -299.89 & 322.01 \\ 
  \multirow{2}{*}{(2)}
& s & 6.32 & -15.06 & 27.87 & & 4.64 & -27.51 & 36.56 & & 5.78 & -43.30 & 55.55 \\ 
& c & 14.36 & -65.53 & 97.56 & & 8.08 & -129.50 & 142.40 & & -1.55 & -206.59 & 198.41 \\ 
  \multirow{2}{*}{(3)}
& s & 7.74 & -15.37 & 31.07 & & 5.76 & -32.76 & 40.91 & & 8.98 & -45.53 & 64.71 \\ 
& c & 17.60 & -60.32 & 98.08 & & 12.58 & -116.06 & 142.92 & & 6.48 & -182.11 & 198.27 \\
  \multirow{2}{*}{(4)}
& s & \bf{47.39} & \bf{0.94} & \bf{96.95} & & 23.29 & -49.15 & 104.14 & & 24.21 & -88.64 & 136.26 \\ 
& c & 31.44 & -74.80 & 140.15 & & 23.04 & -156.67 & 205.96 & & 14.52 & -259.18 & 280.48 \\  
  \multirow{2}{*}{(5)}
& s & 4.51 & -46.29 & 57.07 & & 8.11 & -75.41 & 91.55 & & 13.40 & -108.70 & 136.45 \\ 
& c & 48.56 & -55.74 & 160.97 & & 18.78 & -155.55 & 199.51 & & 11.59 & -255.06 & 276.53 \\ 
  \multirow{2}{*}{(6)}
& s & 10.05 & -14.63 & 35.36 & & 12.24 & -28.79 & 54.40 & & 14.69 & -45.35 & 76.51 \\ 
& c & 25.66 & -39.05 & 92.53 & & 7.03 & -101.58 & 117.02 & & 5.53 & -159.96 & 167.62 \\  
  \multirow{2}{*}{(7)}
& s & \bf{80.83} & \bf{6.45} & \bf{158.56} & & 38.12 & -82.24 & 154.90 & & 34.47 & -137.44 & 209.06 \\ 
& c & 184.75 & -216.88 & 596.71 & & 106.78 & -553.29 & 757.07 & & 92.10 & -904.77 & 1086.75 \\
  \multirow{2}{*}{(8)}
& s & 25.29 & -25.76 & 77.12 & & 23.02 & -62.62 & 103.02 & & 14.70 & -111.95 & 135.90 \\ 
& c & 15.27 & -14.96 & 45.95 & & 5.17 & -44.71 & 53.87 & & 3.01 & -68.34 & 73.61 \\   
  \multirow{2}{*}{(9)}
& s & 41.09 & -8.93 & 89.23 & & 16.95 & -61.21 & 99.53 & & 13.91 & -102.74 & 132.98 \\ 
& c & 18.71 & -30.61 & 71.21 & & 2.68 & -77.27 & 80.47 & & 3.93 & -114.88 & 122.98 \\  
  \multirow{2}{*}{(10)}
& s & \bf{12.16} & \bf{1.06} & \bf{23.02} & & 9.42 & -8.54 & 26.50 & & 5.12 & -21.80 & 32.30 \\ 
& c & -0.21 & -8.89 & 8.87 & & 1.64 & -13.12 & 17.01 & & 3.64 & -17.52 & 24.97 \\  
\bottomrule
\end{tabular}
}
\end{table}

\begin{table}[h!]
\centering
\caption{Temporal average general causal effects of the new price policy on the ten store (s) - competitor (c) pairs computed at three time horizons. In this table, $\hat{\bar{\bs{\tau}}}_t$ stands for the general effect $\hat{\bar{\bs{\tau}}}_t((1,0),(0,0))$ and the results are obtained including in the set of covariates the price ratio between the store and competitor brand prior to the intervention (in the post-intervention period the ratio is computed from the prior price).}
\label{tab:general_effect_ratio}
\scalebox{0.8}{
\begin{tabular}{rrrrrrrrrrrrr} \toprule
 & & \multicolumn{3}{c}{$1$ month} & & \multicolumn{3}{c}{$3$ months} & & \multicolumn{3}{c}{$6$ months} \\ 
 & & $\hat{\bar{\bs{\tau}}}_t$ & $2.5\%$ & $97.5\%$ & & $\hat{\bar{\bs{\tau}}}_t$ & $2.5\%$ & $97.5\%$ & & $\hat{\bar{\bs{\tau}}}_t$ & $2.5\%$ & $97.5\%$ \\ \cmidrule{3-5} \cmidrule{7-9} \cmidrule{11-13}
\multirow{2}{*}{(1)}   
& s & 7.86 & -23.99 & 40.25 & & 5.57 & -43.61 & 56.18 & & 7.60 & -65.59 & 81.24 \\ 
& c & 24.24 & -103.31 & 149.08 & & 18.24 & -190.07 & 236.58 & & 9.94 & -302.70 & 321.87 \\ 
  \multirow{2}{*}{(2)}
& s & 6.29 & -15.08 & 27.85 & & 4.58 & -28.01 & 36.62 & & 5.78 & -43.58 & 55.33 \\ 
& c & 14.43 & -65.19 & 97.88 & & 8.04 & -129.58 & 142.52 & & -1.94 & -206.72 & 198.81 \\ 
  \multirow{2}{*}{(3)}
& s & 7.69 & -15.61 & 31.11 & & 5.69 & -33.02 & 41.00 & & 8.94 & -45.58 & 65.00 \\ 
& c & 17.67 & -60.31 & 98.22 & & 12.55 & -116.11 & 142.85 & & 6.40 & -182.21 & 198.30 \\
  \multirow{2}{*}{(4)}
& s & 47.59 & -1.43 & 95.37 & & 23.49 & -52.91 & 99.97 & & 26.11 & -85.00 & 143.55 \\ 
& c & 30.86 & -76.21 & 142.37 & & 21.79 & -156.22 & 203.90 & & 12.56 & -247.89 & 285.21 \\   
  \multirow{2}{*}{(5)}
& s & 4.93 & -45.95 & 56.46 & & 8.44 & -74.91 & 93.47 & & 13.63 & -107.63 & 138.26 \\ 
& c & 48.63 & -58.86 & 160.54 & & 18.78 & -161.04 & 203.72 & & 11.66 & -267.79 & 280.47 \\ 
  \multirow{2}{*}{(6)}
& s & 9.89 & -14.74 & 34.85 & & 12.05 & -29.01 & 54.04 & & 14.37 & -46.42 & 75.06 \\ 
& c & 25.76 & -38.76 & 92.99 & & 7.05 & -100.67 & 117.62 & & 5.59 & -155.74 & 167.47 \\ 
  \multirow{2}{*}{(7)}
& s & \bf{80.67} & \bf{1.53} & \bf{161.11} & & 36.73 & -84.22 & 156.80 & & 31.45 & -150.41 & 207.70 \\ 
& c & 183.01 & -222.65 & 583.47 & & 108.84 & -559.14 & 799.66 & & 102.14 & -892.35 & 1113.15 \\
  \multirow{2}{*}{(8)}
& s & 23.54 & -28.05 & 73.80 & & 22.06 & -59.32 & 103.49 & & 14.64 & -113.07 & 140.54 \\ 
& c & 14.98 & -15.50 & 44.80 & & 4.46 & -44.03 & 53.53 & & 2.35 & -69.75 & 75.51 \\   
  \multirow{2}{*}{(9)}
& s & 41.00 & -7.02 & 87.54 & & 16.93 & -64.31 & 97.09 & & 14.35 & -106.63 & 136.62 \\ 
& c & 18.68 & -32.60 & 69.15 & & 2.66 & -82.03 & 83.46 & & 4.81 & -113.13 & 120.65 \\  
  \multirow{2}{*}{(10)}
& s & \bf{12.50} & \bf{1.45} & \bf{23.71} & & 9.62 & -9.64 & 27.65 & & 5.07 & -23.35 & 31.58 \\ 
& c & -0.11 & -9.77 & 9.72 & & 1.72 & -13.10 & 16.31 & & 3.77 & -18.52 & 25.31 \\  
\bottomrule
\end{tabular}
}
\end{table}
\restoregeometry

\newpage
\subsection{Posterior inference}
\label{appB_inference}

We provide below the proof of relations (\ref{eqn:full_beta}), (\ref{eqn:full_sigma}) and (\ref{eqn:full_sigma_eta}). \\

$\bs{\beta}$ has prior density function given by ,

\begin{small}
\begin{align*}
%\label{eqn:matrix_beta}
\Pr(\bs{\beta}_{\varrho}|\bs{\Sigma}_{\varepsilon}, \bs{\varrho}, \bs{\theta}) & =  (2\pi)^{-p_{\varrho}d/2} \det{(\bl{H}_{\varrho})}^{-d/2} \det{(\bs{\Sigma}_{\varepsilon})}^{-p_{\varrho}/2} \exp{\left\lbrace -\frac{1}{2} \tr{ \left[\bl{H}_{\varrho}^{-1}\bs{\beta}_{\varrho} \bs{\Sigma}_{\varepsilon}^{-1}\bs{\beta}_{\varrho} ' \right]}  \right\rbrace } \\
& = (2\pi)^{-p_{\varrho}d/2} \det{(\bl{H}_{\varrho})}^{-d/2} \det{(\bs{\Sigma}_{\varepsilon})}^{-p_{\varrho}/2} \exp{\left\lbrace -\frac{1}{2} \tr{ \left[\bs{\beta}_{\varrho}'\bl{H}_{\varrho}^{-1}\bs{\beta}_{\varrho} \bs{\Sigma}_{\varepsilon}^{-1} \right]} \right\rbrace }
\end{align*}
\end{small}

Where $p_{\varrho}$ is the number of selected regressors. Similarly, the density function $\Pr(\tilde{\bl{Y}}_{1:t^*})$ can be written as,

\begin{small}
\begin{align*}
\Pr(\tilde{\bl{Y}}_{1:t^*} | \bs{\beta}_{\varrho}, \bs{\Sigma}_{\varepsilon}, \bs{\varrho}, \bs{\theta}) & = (2\pi)^{-d t^*/2} \det{(\bs{\Sigma}_{\varepsilon})^{-t^*/2}} \exp{ \left\lbrace -\frac{1}{2} \sum\limits_{t=1}^{t^*}(\tilde{\bl{Y}}_{1:t^*} - \bl{X}_{\varrho}\bs{\beta})\bs{\Sigma}_{\varepsilon}^{-1}(\tilde{\bl{Y}}_{1:t^*} - \bl{X}_{\varrho}\bs{\beta})' \right\rbrace } \\
& = (2\pi)^{-dt^*/2} \det{(\bs{\Sigma}_{\varepsilon})^{-t^*/2}} \exp{ \left\lbrace -\frac{1}{2} \tr{\left[(\tilde{\bl{Y}}_{1:t^*} - \bl{X}_{\varrho}\bs{\beta})'(\tilde{\bl{Y}}_{1:t^*} - \bl{X}_{\varrho}\bs{\beta})\bs{\Sigma}_{\varepsilon}^{-1} \right]} \right\rbrace }
\end{align*}
\end{small}

Now we can derive the posterior distribution for the regression coefficients as follows,

\begin{small}
\begin{align*}
\Pr(\bs{\beta}_{\varrho}| \tilde{\bl{Y}}_{1:t^*}, \bs{\Sigma}_{\varepsilon}, \bs{\varrho},\bs{\theta}) & \propto \Pr(\tilde{\bl{Y}}_{1:t^*} | \bs{\beta}_{\varrho}, \bs{\Sigma}_{\varepsilon}, \bs{\varrho}, \bs{\theta} ) \Pr(\bs{\beta}_{\varrho}| \bs{\Sigma}_{\varepsilon}, \bs{\varrho},\bs{\theta}) \\
& \propto \exp{\left\lbrace -\frac{1}{2} \tr{\left[(\tilde{\bl{Y}}_{1:t^*} - \bl{X}_{\varrho} \bs{\beta}_{\varrho})'(\tilde{\bl{Y}}_{1:t^*} - \bl{X}_{\varrho} \bs{\beta}_{\varrho})\bs{\Sigma}_{\varepsilon}^{-1}\right]} \right\rbrace} \exp{\left\lbrace -\frac{1}{2} \tr{\left[\bs{\beta}_{\varrho}' \bl{H}_{\varrho}^{-1}\bs{\beta}_{\varrho}\bs{\Sigma}_{\varepsilon}^{-1} \right]} \right\rbrace} \\
& \propto \exp{\left\lbrace -\frac{1}{2} \tr{\left[ \bs{\beta}_{\varrho}'\bl{X}_{\varrho}' \bl{X}_{\varrho} \bs{\beta}_{\varrho}\bs{\Sigma}_{\varepsilon}^{-1} -2 \bs{\beta}_{\varrho}'\bl{X}_{\varrho}' \tilde{\bl{Y}}_{1:t^*} \bs{\Sigma}_{\varepsilon}^{-1} + \bs{\beta}_{\varrho}' \bl{H}_{\varrho}^{-1}\bs{\beta}_{\varrho}\bs{\Sigma}_{\varepsilon}^{-1} \right]} \right\rbrace} \\
& \propto \exp{\left\lbrace -\frac{1}{2} \tr{ \left[ \bs{\beta}_{\varrho}'(\bl{X}_{\varrho}' \bl{X}_{\varrho} + \bl{H}_{\varrho}^{-1})\bs{\beta}_{\varrho}\bs{\Sigma}_{\varepsilon}^{-1} -2\bs{\beta}_{\varrho}'\bl{X}_{\varrho}' \tilde{\bl{Y}}_{1:t^*} \bs{\Sigma}_{\varepsilon}^{-1}  \right]} \right\rbrace}
\end{align*}
\end{small}

Which is the kernel of a matrix-normal distribution $\mathcal{N}(\bl{M}, \bl{W}, \bs{\Sigma}_{\varepsilon})$, with $\bl{W} = (\bl{X}_{\varrho}' \bl{X}_{\varrho}+\bl{H}_{\varrho}^{-1})^{-1}$ and $\bl{M} = (\bl{X}_{\varrho}' \bl{X}_{\varrho}+\bl{H}_{\varrho}^{-1})^{-1}\bl{X}_{\varrho}'\tilde{\bl{Y}}_{1:t^*}$.

Integration of the above quantity is necessary to derive the posterior distribution of $\bs{\Sigma}_{\varepsilon}$ and yields the inverse of the normalization constant, which is $\kappa = (2\pi)^{p_{\varrho}d/2}\det{(\bl{W})}^{d/2} \det{(\bs{\Sigma}_{\varepsilon})}^{p_{\varrho}/2}$. However, $\kappa$  simplifies with the constants singled out from the integral, which are \\ $(2 \pi)^{-p_{\varrho}d/2}\det{(\bs{\Sigma}_{\varepsilon})}^{-p_{\varrho}/2} \det{(\bl{H}_{\varrho})}^{-d/2}$ and $(2 \pi)^{-dt^*/2}\det{(\bs{\Sigma}_{\varepsilon})}^{-t^*/2}$, leaving \\ $\det{(\bl{H}_{\varrho})}^{-d/2}\det{(\bl{W})}^{d/2} (2 \pi)^{-dt^*/2}\det{(\bs{\Sigma}_{\varepsilon})}^{-t^*/2}$.

\begin{small}
\begin{align*}
\Pr(\bs{\Sigma}_{\varepsilon}|\tilde{\bl{Y}}_{1:t^*}, \bs{\varrho}, \bs{\theta}) & \propto \Pr(\tilde{\bl{Y}}_{1:t^*} | \bs{\Sigma}_{\varepsilon}, \bs{\varrho}, \bs{\theta})\Pr(\bs{\Sigma}_{\varepsilon}|\bs{\varrho}, \bs{\theta}) \\
& \propto \Pr(\bs{\Sigma}_{\varepsilon}|\bs{\varrho}, \bs{\theta}) \int \Pr(\tilde{\bl{Y}}_{1:t^*} | \bs{\beta}_{\varrho}, \bs{\Sigma}_{\varepsilon}, \bs{\varrho}, \bs{\theta} ) \Pr(\bs{\beta}| \bs{\Sigma}_{\varepsilon}, \bs{\varrho},\bs{\theta}) d\bs{\beta}  \\
& \propto \det{(\bs{\Sigma}_{\varepsilon})}^{-(d+\nu_{\varepsilon}+t^*+1)/2} \exp{\left\lbrace -\frac{1}{2} \tr{(\bl{S}_{\varepsilon} \bs{\Sigma}_{\varepsilon}^{-1})} \right\rbrace} \exp{\left\lbrace -\frac{1}{2} \tr{\left[(\tilde{\bl{Y}}_{1:t^*}' \tilde{\bl{Y}}_{1:t^*} + \bl{M}'\bl{W}^{-1}\bl{M}) \bs{\Sigma}_{\varepsilon}^{-1} \right]} \right\rbrace} \\
& \propto \det{(\bs{\Sigma}_{\varepsilon})}^{-(d+\nu_{\varepsilon}+t^*+1)/2} \exp{ \left\lbrace -\frac{1}{2} \tr{\left[(\bl{S}_{\varepsilon} + \tilde{\bl{Y}}_{1:t^*}' \tilde{\bl{Y}}_{1:t^*} - \bl{M}'\bl{W}^{-1}\bl{M})\bs{\Sigma}_{\varepsilon}^{-1} \right]} \right\rbrace }
\end{align*}
\end{small}

This is the kernel of an Inverse-Wishart distribution with $\nu = \nu_{\epsilon} +t^* $ degrees of freedom and scale matrix $\bl{SS}_{\varepsilon} = (\bl{S}_{\varepsilon} + \tilde{\bl{Y}}_{1:t^*}' \tilde{\bl{Y}}_{1:t^*} - \bl{M}'\bl{W}^{-1} \bl{M})$. We can also derive the posterior of the latent vector $\bs{\varrho}$,

$$
\Pr(\bs{\varrho} | \tilde{\bl{Y}}_{1:t^*}, \bs{\theta}) = \frac{\Pr(\tilde{\bl{Y}}_{1:t^*} |\bs{\varrho}, \bs{\theta})\Pr(\bs{\varrho}|\bs{\theta})}{\sum\limits_{\varrho} \Pr(\tilde{\bl{Y}}_{1:t^*} |\bs{\varrho}, \bs{\theta})\Pr(\bs{\varrho}|\bs{\theta}) }
$$

where,

\begin{small}
\begin{align*}
\Pr(\tilde{\bl{Y}}_{1:t^*} |\bs{\varrho}, \bs{\theta}) & = \int \int \Pr(\tilde{\bl{Y}}_{1:t^*} | \bs{\beta}_{\varrho}, \bs{\Sigma}_{\varepsilon}, \bs{\varrho}, \bs{\theta})\Pr(\bs{\beta}_{\varrho}|\bs{\Sigma}_{\varepsilon}, \bs{\varrho}, \bs{\theta})\Pr(\bs{\Sigma}_{\varepsilon}|\bs{\varrho},\bs{\theta}) d\bs{\beta} d\bs{\Sigma}_{\varepsilon} \\
& = \int \left( \int \Pr(\tilde{\bl{Y}}_{1:t^*} | \bs{\beta}_{\varrho}, \bs{\Sigma}_{\varepsilon}, \bs{\varrho}, \bs{\theta})\Pr(\bs{\beta}_{\varrho}|\bs{\Sigma}_{\varepsilon}, \bs{\varrho}, \bs{\theta}) d\bs{\beta} \right) \Pr(\bs{\Sigma}_{\varepsilon}|\bs{\varrho},\bs{\theta}) d \bs{\Sigma}_{\varepsilon} \\
& = \int \det{(\bl{H}_{\varrho})}^{-d/2} \det{(\bl{W})}^{d/2} (2 \pi)^{-dt^*/2}\det{(\bs{\Sigma}_{\varepsilon})}^{-t^*/2} \exp{ \left\lbrace -\frac{1}{2}\tr{\left[(\tilde{\bl{Y}}_{1:t^*}'\tilde{\bl{Y}}_{1:t^*} - \bl{M}'\bl{W}^{-1}\bl{M})\bs{\Sigma}_{\varepsilon}^{-1} \right]}\right\rbrace } \\
& \hspace{30pt} \frac{\det{(\bl{S}_{\varepsilon})}^{\nu_{\varepsilon}/2}}{2^{\nu_{\varepsilon}d/2} \Gamma_d(\nu_{\varepsilon}/2)} \det{(\bs{\Sigma}_{\varepsilon})}^{-(\nu_{\varepsilon}+d+1)/2} \exp{\left\lbrace -\frac{1}{2} \tr{(\bl{S}_{\varepsilon} \bs{\Sigma}_{\varepsilon}^{-1})} \right\rbrace} d\bs{\Sigma}_{\varepsilon} \\
& = \frac{\det{(\bl{H}_{\varrho})}^{-d/2} \det{(\bl{W})}^{d/2} (2 \pi)^{-dt^*/2}\det{(\bl{S}_{\varepsilon})}^{\nu_{\varepsilon}/2}}{2^{\nu_{\varepsilon}d/2} \Gamma_d(\nu_{\varepsilon}/2)} \cdot   \frac{2^{(\nu_{\varepsilon}+t^*)d/2}\Gamma_d(\nu_{\varepsilon}+t^*/2)}{\det{(\bl{SS}_{\varepsilon})}^{\nu_{\varepsilon}+t^*/2}} \\
& = \frac{\det{(\bl{H}_{\varrho})}^{-d/2} \det{(\bl{W})}^{d/2}(\pi)^{-dt^*/2}\det{(\bl{S}_{\varepsilon})}^{\nu_{\varepsilon}/2}\Gamma_d(\nu_{\varepsilon}+t^*/2)}{\Gamma_d(\nu_{\varepsilon}/2)\det{(\bl{SS}_{\varepsilon})}^{\nu_{\varepsilon}+t^*/2}}
\end{align*}
\end{small}

Notice that if we set $\bl{H}_{\varrho} = (\bl{X}_{\varrho}' \bl{X}_{\varrho})^{-1}$, the above expressions simplify to $\bl{W} = \frac{1}{2}(\bl{X}_{\varrho}' \bl{X}_{\varrho})^{-1}$, $\bl{M} = \frac{1}{2}(\bl{X}_{\varrho}' \bl{X}_{\varrho})^{-1} \bl{X}_{\varrho}'\tilde{\bl{Y}}_{1:t^*}$ and $\bl{SS}_{\varepsilon} = \bl{S}_{\varepsilon} + \tilde{\bl{Y}}_{1:t^*}' \tilde{\bl{Y}}_{1:t^*} - \frac{1}{2} \tilde{\bl{Y}}_{1:t^*}'(\bl{X}_{\varrho}'\bl{X}_{\varrho})^{-1} \bl{X}_{\varrho}'\tilde{\bl{Y}}_{1:t^*}$.

\vspace{25pt}

In order to evaluate the posterior distribution $\Pr(\bs{\varrho} | \tilde{\bl{Y}}_{1:t^*}, \bs{\theta})$ we can resort to the odds and update the elements of the selection vector one component at a time, while the others are held fixed. This ensures that at each step only the most likely model is retained, either the one with $\X_p$ in it or the one without. More formally, let $\varrho_p = 1$ and indicate with $\bs{\varrho}_{-p}$ the vector of all the elements in $\bs{\varrho}$ except $\varrho_p$. The full conditional of $\varrho_p$ is given by, 

\begin{small}
\begin{align}
\label{eqn:selection_postprob}
\Pr(\varrho_p = 1 | \tilde{\bl{Y}}_{1:t^*}, \bs{\varrho}_{-p}, \bs{\theta}) & = \frac{\Pr(\varrho_p = 1 | \bs{\theta})\Pr(\tilde{\bl{Y}}_{1:t^*} | \varrho_p = 1, \bs{\varrho}_{-p}, \bs{\theta})}{\Pr(\varrho_p = 1 | \bs{\theta})\Pr(\tilde{\bl{Y}}_{1:t^*} | \varrho_p = 1, \bs{\varrho}_{-p}, \bs{\theta}) + \Pr(\varrho_p = 0 | \bs{\theta})\Pr(\tilde{\bl{Y}}_{1:t^*} | \varrho_p = 0, \bs{\varrho}_{-p}, \bs{\theta}) }  \\ \nonumber
& = \frac{1}{1+o_p^{-1}} 
\end{align}
\end{small}

Where, assuming equal prior probabilities $\Pr(\varrho_p = 1 | \bs{\theta}) = \Pr(\varrho_p = 0 | \bs{\theta})$ we have,

$$o_p = \frac{\Pr(\varrho_p = 1 | \bs{\theta})}{\Pr(\varrho_p = 0 | \bs{\theta})} \frac{\Pr(\tilde{\bl{Y}}_{1:t^*} | \varrho_p = 1, \bs{\varrho}_{-p}, \bs{\theta})}{\Pr(\tilde{\bl{Y}}_{1:t^*} | \varrho_p = 0, \bs{\varrho}_{-p}, \bs{\theta})} = \frac{\Pr(\tilde{\bl{Y}}_{1:t^*} | \varrho_p = 1, \bs{\varrho}_{-p}, \bs{\theta})}{\Pr(\tilde{\bl{Y}}_{1:t^*} | \varrho_p = 0, \bs{\varrho}_{-p}, \bs{\theta})}$$

\vspace{5pt}

Finally, let $\bs{\eta}_{1:t^*}^{(r)}$ indicate the disturbances up to time $t^*$ of the $r$-th state. Then, $\bs{\eta}_{1:t^*}^{(r)}$ is a $(t^* \times d)$ matrix independently drawn from a $\mathcal{N}(0, I_{t^*}, \bs{\Sigma}_r)$.  Thus we have,

\begin{small}
\begin{align*}
\Pr(\bs{\Sigma}_r | \bs{\eta}_{1:t^*}^{(r)}, \bs{\theta}) & \propto \Pr(\bs{\eta}_{1:t^*}^{(r)}| \bs{\Sigma}_r, \bs{\theta})\Pr(\bs{\Sigma}_r |\bs{\theta}) \\
 & \propto \det{(\bs{\Sigma}_r)}^{-t^*/2} \exp{ \left\lbrace -\frac{1}{2} \tr{(\bs{\eta}_{1:t^*}^{(r)} \bs{\Sigma}_r^{-1} \bs{\eta}_{1:t^*}^{'(r)})} \right\rbrace} \det{(\bs{\Sigma}_r)}^{-\frac{\nu_{r} +d+1}{2}} \exp{ \left\lbrace -\frac{1}{2} \tr{(\bl{S}_{r} \bs{\Sigma}_r^{-1})} \right\rbrace } \\
 & \propto \det{(\bs{\Sigma}_r)}^{-\frac{\nu_{r} +d+t^*+1}{2}} \exp{ \left\lbrace -\frac{1}{2} \tr{\left[ (\bl{S}_{r} + \bs{\eta}_{1:t^*}^{(r)} \bs{\Sigma}_r^{-1} \bs{\eta}_{1:t^*}^{(r)} ) \right] } \right\rbrace }
\end{align*}
\end{small}

Which is the kernel of an Inverse-Wishart distribution with $\nu_{r}+t^*$ degrees of freedom and scale matrix $\bl{S}_{r} + \bs{\eta}_{1:t^*}^{'(r)} \bs{\eta}_{1:t^*}^{(r)}$.

To sample from the joint posterior distribution of the states and model parameters we can employ the following MCMC algorithm. 

\begin{algorithm}[h!]
\begin{algorithmic}[1]
\caption{Gibbs sampler to draw from the joint posterior distribution of the states and model parameters}
\label{algorithm}
\REQUIRE $\bs{\Sigma}_{\varepsilon}^{(0)}$, $\bs{\Sigma}_{r}^{(0)}$, $\bs{\theta}$, $\bl{H}_{\varrho}$, \texttt{niter}
\FOR{s in $1:\texttt{niter}$} 
\STATE draw $\bs{\alpha}_{t}^{(s)}$ from $\Pr(\bs{\alpha}_t | \bl{Y}_{1:t^*}, \bs{\Sigma}_{\varepsilon}^{(s-1)}, \bs{\Sigma}_r^{(s-1)}, \bs{\theta}) $ using the simulation smoothing by \citet{Durbin:Koopman:2002}\footnote{An implementation of the simulation smoothing algorithm can be found in the R package \texttt{KFAS} \citep{KFAS}.}
\STATE draw $\bs{\Sigma}_r^{(s)}$ from $\Pr(\bs{\Sigma}_r |  \bs{\eta}_{1:t^*}^{(r, s)}, \bs{\theta})$ according to equation (\ref{eqn:full_sigma_eta})
\STATE compute $\tilde{\bl{Y}}_{1:t^*}^{(s)}$ and draw $\bs{\varrho}^{(s)}$ from $\Pr(\varrho_p | \tilde{\bl{Y}}_{1:t^*}^{(s)}, \bs{\varrho}_{-p}^{(s)}, \bs{\theta})$ by changing $\bs{\varrho}$ one component at a time and computing its posterior probability (this ensures that every time a component $\varrho_p$ is changed, the most likely model is retained, i.e. either the one with $\X_p$ in or the one without $\X_p$)
\STATE draw $\bs{\Sigma}_{\varepsilon}^{(s)}$ from $\Pr(\bs{\Sigma}_{\varepsilon} | \tilde{\bl{Y}}_{1:t^*}^{(s)} , \bs{\varrho}^{(s)}, \bs{\theta})$ according to equation (\ref{eqn:full_sigma})
\STATE draw $\bs{\beta}_{\varrho}^{(s)}$ from $\Pr(\bs{\beta}_{\varrho} | \tilde{\bl{Y}}_{1:t^*}^{(s)}, \bs{\Sigma}_{\varepsilon}^{(s)}, \bs{\varrho}^{(s)}, \bs{\theta})$ according to equation (\ref{eqn:full_beta})
\ENDFOR
\end{algorithmic}
\end{algorithm}

\newpage

\subsection{Unbiased causal effects}
\label{appB_unbiased}

\begin{theorem} \label{thrm} For a positive integer $k$,
define $\widehat{\bl{Y}}_{t^*+k}(\bl{w}) = \E[\Pr (\bl{Y}_{t^*+k}(\bl{w}) | \bl{Y}_{1:t^*}(0,0))]$ and \\ $\widehat{\bl{Y}}_{t^*+k}(\tilde{\bl{w}}) = \E[\Pr (\bl{Y}_{t^*+k}(\tilde{\bl{w}}) | \bl{Y}_{1:t^*}(0,0))]$; under model (\ref{eqn:model}), $\widehat{\bl{Y}}_{t^*+k}(\bl{w})$ and $\widehat{\bl{Y}}_{t^*+k}(\tilde{\bl{w}})$  are the $k$-step ahead forecasts of $\bl{Y}_{t^*+k}(\bl{w})$ and $\bl{Y}_{t^*+k}(\tilde{\bl{w}})$ given the information set up to time $t^*$, $\info{I}{t^*} = \{ \bl{Y}_{1:t^*}, \bl{X}_{1:t^*} \}$. Then, $\hat{\bs{\tau}}_{t^*+k}(\bl{w},\tilde{\bl{w}}) = \widehat{\bl{Y}}_{t^*+k}(\bl{w}) - \widehat{\bl{Y}}_{t^*+k}(\tilde{\bl{w}})$ is the point estimator of the general causal effect and, conditionally on $\info{I}{t^*}$ we have,

\begin{align}
\bs{\tau}_{t^*+k}(\bl{w},\tilde{\bl{w}}) - \hat{\bs{\tau}}_{t^*+k}(\bl{w},\tilde{\bl{w}}) & \sim  N(\bs{0}, \bs{\Sigma}_{\bl{w}} + \bs{\Sigma}_{\tilde{\bl{w}}}) \label{eqn:distrib_poitwise}\\ 
\Delta _{t^*+k} - \widehat{\Delta }_{t^*+k} & \sim  \mathcal{N}(\bs{0}, \bs{\Sigma}_{D(\bl{w})} + \bs{\Sigma}_{D(\tilde{\bl{w}})}, \bs{\Sigma}) \label{eqn:distrib_cum} \\
\bar{\bs{\tau}}_{t^*+k}(\bl{w},\tilde{\bl{w}}) - \hat{\bar{\bs{\tau}}}_{t^*+k}(\bl{w},\tilde{\bl{w}}) & \sim  \mathcal{N} \left(\bs{0}, \frac{1}{k^2} (\bs{\Sigma}_{D(\bl{w})} + \bs{\Sigma}_{D(\tilde{\bl{w}})}), \bs{\Sigma} \right) \label{eqn:distrib_avg} 
\end{align}  

where, $\bs{\Sigma}_{\underline{\bl{w}}} = Var \left[ \bl{Y}_{t^*+k}(\underline{\bl{w}})- \widehat{\bl{Y}}_{t^*+k}(\underline{\bl{w}}) \;\middle|\; \info{I}{t^*} \right]$, $\bs{\Sigma}_{D(\underline{\bl{w}})} = Var \left[\sum\limits_k (\bl{Y}_{t^*+k}(\underline{\bl{w}})- \widehat{\bl{Y}}_{t^*+k}(\underline{\bl{w}}) ) \;\middle|\; \info{I}{t^*} \right]$ with $\underline{\bl{w}} \in \{ \bl{w}, \tilde{\bl{w}} \}$ are defined as follows

\begin{align}
\bs{\Sigma}_{\underline{\bl{w}}} & =  \bs{Z}_t \bl{P}_t \bs{Z}_t' + \bs{\Sigma}_{\varepsilon} \label{eqn:var_pointwise} \\
\bs{\Sigma}_{D(\underline{\bl{w}})} & =  \left( \bl{D}_{t^*+1} \bl{P}_{t^*+1} \bl{D}_{t^*+1}' + \sum\limits_{k} \left(\bl{D}_{t^*+k} \bl{R}_{t^*+K-1} \bl{C}_{t^*+K-1} \bl{R}_{t^*+K-1}' \bl{D}_{t^*+k}' \right) \label{eqn:var_cum} \right) + KH_t 
\end{align}

and 

\begin{align*}
\bl{D}_{t^*+k} & = \bl{Z}_{t^*+k} +  \bl{D}_{t^*+k+1} \bl{T}_{t^*+k} \hspace{5pt}, \hspace{5pt} k = 1,\dots,K-1 \\
\bl{D}_{t^* + K} & = \bl{Z}_{t^*+K}
\end{align*}

\textbf{Proof.}

The difference between the general causal effect and its point estimator can be written as,

\begin{align*}
\bs{\tau}_{t^*+k}(\bl{w},\bl{\tilde{w}}) - \hat{\bs{\tau}}_{t^*+k}(\bl{w},\bl{\tilde{w}}) & = \bl{Y}_{t^*+k}(\bl{w}) - \bl{Y}_{t^*+k}(\tilde{\bl{w}}) - \left[ \widehat{\bl{Y}}_{t^*+k}(\bl{w}) - \widehat{\bl{Y}}_{t^*+k}(\tilde{\bl{w}}) \right] \\
& = \underbrace{\bl{Y}_{t^*+k}(\bl{w}) - \widehat{\bl{Y}}_{t^*+k}(\bl{w})}_{\text{A}} -  \underbrace{ \left[ \bl{Y}_{t^*+k}(\tilde{\bl{w}}) - \widehat{\bl{Y}}_{t^*+k}(\tilde{\bl{w}}) \right]}_{\text{B}} 
\end{align*}

Let's focus our attention on A and define $\bl{a}_{t^*+k} = E[\bs{\alpha}_{t^*+k}|\info{I}{t^*}]$ and $\bl{P}_{t^*+k} = Var[\bs{\alpha}_{t^*+k}|\info{I}{t^*}]$. Under model (\ref{eqn:model}) we have,

\begin{align*}
\bl{Y}_{t^*+k}(\bl{w}) - \widehat{\bl{Y}}_{t^*+k}(\bl{w}) & = \bl{Z}_{t^*+k} \bs{\alpha}_{t^*+k} + \bl{X}_{t^*+k} \bs{\beta} + \bs{\varepsilon}_{t^*+k} - E[\bl{Y}_{t^*+k}(\bl{w})|\info{I}{t^*}] \\
& = \bl{Z}_{t^*+k} \bs{\alpha}_{t^*+k} + \bl{X}_{t^*+k} \bs{\beta} + \bs{\varepsilon}_{t^*+k} - \bl{Z}_{t^*+k} \bl{a}_{t^*+k} - \bl{X}_{t^*+k} \bs{\beta} \\
& = \bl{Z}_{t^*+k} \bs{\alpha}_{t^*+k} - \bl{Z}_{t^*+k} \bl{a}_{t^*+k} + \bs{\varepsilon}_{t^*+k}
\end{align*} 

Then,

$$E[\bl{Y}_{t^*+k}(\bl{w}) - \widehat{\bl{Y}}_{t^*+k}(\bl{w}) | \info{I}{t^*}] = 0$$
$$Var[\bl{Y}_{t^*+k}(\bl{w}) - \widehat{\bl{Y}}_{t^*+k}(\bl{w}) | \info{I}{t^*}] = \bl{Z}_{t^*+k} \bl{P}_{t^*+k} \bl{Z}'_{t^*+k} + \bs{\Sigma}_{\varepsilon} = \bs{\Sigma}_{\bl{w}}$$

Following the exact same steps for B we can show that $\bl{Y}_{t^*+k}(\tilde{\bl{w}}) - \widehat{\bl{Y}}_{t^*+k}(\tilde{\bl{w}}) \sim N(\bs{0}, \bs{\Sigma}_{\tilde{\bl{w}}})$. Since the potential paths are independent of each other, relation (\ref{eqn:distrib_poitwise}) follows from the properties of the difference of two independent multivariate Normal random variables. 

Based on the above result, we can easily show that the expectation of the difference between the cumulative effect and its estimator is zero. In what follows we derive the proof for $t'=t^*+K$ but it could be shown for every $k=1,\dots,K$.

%By Definition \ref{eqn:cum_joint}, $\Delta C_{K} = \sum\limits_{k=1}^{K} \bs{\tau}_{t^*+k}(\bl{w},\bs{\tilde{w}}) $ and  by Definition \ref{eqn:hat_cum_joint} $\widehat{\Delta C}_{K} = \sum\limits_{k=1}^{K} \hat{\bs{\tau}}_{t^*+k}(\bl{w},\bs{\tilde{w}})$ 
 
\begin{small}
\begin{align*}
E \left[ \Delta_{t^*+K} - \widehat{\Delta}_{t^*+K} \right | \info{I}{t^*}] & = 
E\left[\sum\limits_{k=1}^{K} \left( \bs{\tau}_{t^*+k}(\bl{w},\bl{\tilde{w}})-\hat{\bs{\tau}}_{t^*+k}(\bl{w},\bl{\tilde{w}}) \right)\;\middle|\; \info{I}{t^*}\right]  \\
& = \sum_{k=1}^K E[\bs{\tau}_{t^*+k}(\bl{w},\bl{\tilde{w}})-\hat{\bs{\tau}}_{t^*+k}(\bl{w},\bl{\tilde{w}}) |\info{I}{t^*}] = 0
\end{align*}
\end{small}

The derivation of the variance may be somewhat more cumbersome, because the time dependency also come into play. So we have three dependence structures to take into account: the one between the $d$ series, the one between times and the one between the states. To address this issue it is useful to re-define $\bs{\varepsilon}_t \sim \mathcal{N}(0,H_t,\bs{\Sigma})$; in this way, $\bs{\varepsilon}_t$ can be seen as a single-row matrix following a matrix Normal distribution, which is in line with the definition provided in Section \ref{sect:MBSTS}. Thus, we have 
 
\begin{small}
\begin{align*}
Var \left[ \Delta_{t^*+K} - \widehat{\Delta}_{t^*+K}  \;\middle|\; \info{I}{t^*} \right] & = 
Var \left[ \sum\limits_{k=1}^{K} \bs{\tau}_{t^*+k}(\bl{w},\bl{\tilde{w}})  - \sum\limits_{k=1}^{K} \hat{\bs{\tau}}_{t^*+k}(\bl{w},\bl{\tilde{w}}) \;\middle|\; \info{I}{t^*}\right] \\
& = Var \left[\sum\limits_{k=1}^{K} \left( \bl{Y}_{t^*+k}(\bl{w}) - \widehat{\bl{Y}}_{t^*+k}(\bl{w}) \right) - \sum\limits_{k=1}^{K} \left( \bl{Y}_{t^*+k}(\tilde{\bl{w}}) - \widehat{\bl{Y}}_{t^*+k}(\tilde{\bl{w}}) \right) \;\middle|\; \info{I}{t^*} \right]
\end{align*}
\end{small}

Focusing on the first term, 

\begin{small}
\begin{align*}
Var \left[\sum\limits_{k=1}^{K} \left( \bl{Y}_{t^*+k}(\bl{w}) - \widehat{\bl{Y}}_{t^*+k}(\bl{w}) \right) \;\middle|\; \info{I}{t^*} \right] & = Var \left[ \sum\limits_{k=1}^{K} \bl{Z}_{t^*+k}\bs{\alpha}_{t^*+k} - \sum\limits_{k=1}^{K} \bl{Z}_{t^*+k}\bl{a}_{t^*+k} + \sum\limits_{k=1}^{K} \bs{\varepsilon}_{t^*+k} \;\middle|\; \info{I}{t^*}\right] \\
& = Var \left[ \sum\limits_{k=1}^{K} \bl{Z}_{t^*+k}\bs{\alpha}_{t^*+k} \;\middle|\; \info{I}{t^*}\right] + K H_t \\
\end{align*}
\end{small}

where,

\begin{small}
\begin{align*}
Var \left[ \sum\limits_{k=1}^{K} \bl{Z}_{t^*+k}\bs{\alpha}_{t^*+k} \;\middle|\; \info{I}{t^*}\right] & = Var \left[ \bl{Z}_{t^*+1} \bs{\alpha}_{t^*+1} +  \bl{Z}_{t^*+2} \bs{\alpha}_{t^*+2}+ \dots + \bl{Z}_{t^*+K} \bs{\alpha}_{t^*+K} | \info{I}{t^*}\right] \\
& = Var \left[ \bl{Z}_{t^*+1} \bs{\alpha}_{t^*+1} +  \bl{Z}_{t^*+2} (\bl{T}_{t^*+1} \bs{\alpha}_{t^*+1} + \bl{R}_{t^*+1} \bs{\eta}_{t^*+1})+ \dots + \bl{Z}_{t^*+K} \bs{\alpha}_{t^*+K} | \info{I}{t^*}\right] \\
& = Var [(\bl{Z}_{t^*+1}+ \bl{Z}_{t^*+2} \bl{T}_{t^*+1} + \dots + \bl{Z}_{t^*+K} \bl{T}_{t^*+K-1}\cdots \bl{T}_{t^*+1}) \bs{\alpha}_{t^* +1} + \\
& + (\bl{Z}_{t^*+2}+ \bl{Z}_{t^*+3} \bl{T}_{t^*+2} + \dots + \bl{Z}_{t^*+K} \bl{T}_{t^*+K-1}\cdots \bl{T}_{t^*+2}) \bl{R}_{t^*+1} \bs{\eta}_{t^*+1} + \\
& + \dots + \bl{Z}_{t^*+K} \bl{R}_{t^*+K-1}\bs{\eta}_{t^*+K-1}  | \info{I}{t^*}]
\end{align*}
\end{small}

Then, defining $\bl{D}_{t^*+1} = \bl{Z}_{t^*+1}+\bl{Z}_{t^*+2} \bl{T}_{t^*+1} + \dots + \bl{Z}_{t^*+K} \bl{T}_{t^*+K-1}\cdots \bl{T}_{t^*+1}$ we can notice that $\bl{D}_{t^*+1} = \bl{Z}_{t^*+1}+(\bl{Z}_{t^*+2} + \bl{Z}_{t^*+3} \bl{T}_{t^*+2} \dots \bl{Z}_{t^*+K} \bl{T}_{t^*+K-1}\cdots \bl{T}_{t^*+2} ) \bl{T}_{t^*+1} = \bl{Z}_{t^*+1}+ \bl{D}_{t^*+2} \bl{T}_{t^*+1}$. Thus, in general we have

\begin{align*}
\bl{D}_{t^*+k} & = \bl{Z}_{t^*+k} +  \bl{D}_{t^*+k+1} \bl{T}_{t^*+k} \hspace{5pt}, \hspace{5pt} k = 1,\dots,K-1 \\
\bl{D}_{t^* + K} & = \bl{Z}_{t^*+K}
\end{align*}

and 

\begin{align*}
Var \left[ \sum\limits_{k=1}^{K} \bl{Z}_{t^*+k}\bs{\alpha}_{t^*+k} \;\middle|\; \info{I}{t^*} \right] & = \left( \bl{D}_{t^*+1} \bl{P}_{t^*+1} \bl{D}_{t^*+1}' + \sum\limits_{k=2}^K \left(\bl{D}_{t^*+k} \bl{R}_{t^*+K-1} \bl{C}_{t^*+K-1} \bl{R}_{t^*+K-1}' \bl{D}_{t^*+k}' \right) \right) 
\end{align*}

This yields to the final result in equation (\ref{eqn:var_cum}). Repeating these steps for the second term we obtain equation (\ref{eqn:distrib_cum}). Finally, applying the usual properties of variance we obtain relation (\ref{eqn:distrib_avg}) for the temporal average causal effect,

\begin{align*}
Var \left[ \bar{\bs{\tau}}_{t^*+K}(\bl{w},\tilde{\bl{w}}) - \hat{\bar{\bs{\tau}}}_{t^*+K}(\bl{w},\tilde{\bl{w}})  \;\middle|\; \info{I}{t^*} \right] & = Var \left[ \frac{1}{K} \sum\limits_{k=1}^K \bs{\tau}_{t^*+k}(\bl{w},\tilde{\bl{w}}) - \frac{1}{K} \sum\limits_{k=1}^K \hat{\bs{\tau}}_{t^*+k}(\bl{w},\tilde{\bl{w}}) \;\middle|\; \info{I}{t^*} \right] \\
& = \frac{1}{K^2} Var \left[ \Delta_{t^*+K} - \widehat{\Delta}_{t^*+K} \;\middle|\; \info{I}{t^*} \right]
\end{align*}

\end{theorem}

Theorem \ref{thrm}, states that the point estimator of the general causal effect and, by extension, the marginal and the conditional causal effect estimators are unbiased. From equation (\ref{eqn:var_cum}) we can infer that the variance of the difference between the cumulative effect and its estimator increases with the variance of both $\bs{\varepsilon}_t$ and $\bs{\eta}_t$. Furthermore, the variance is an increasing function of $\bl{D}_t$, therefore, our uncertainty increases with time, reflecting our intuition that we have less information about potential outcomes that are further from the time of the intervention.

\newpage

\subsection{Posterior predictive checks}
\label{appB_postpred}

To produce reliable causal effect estimates from the model-based predictions, the assumed model has to adequately describe the data. One way to check the quality of the model fit within a Bayesian framework is to use posterior predictive checks \citep{Rubin:1981, Rubin:1984, Gelman:Carlin:Stern:Dunson:Vehtari:Rubin:2013}. Intuitively, this entails generating synthetic data sets from the fitted model and comparing them to the observed data. 

Typically, we generate replicated data by drawing multiple times from the posterior predictive distribution; then, we compare these draws with the observed data using both numerical and graphical checks \citep{Gelman:Carlin:Stern:Dunson:Vehtari:Rubin:2013}. More specifically, let $T(\bl{Y}_{1:t^*}, \bs{\vartheta})$ be a test quantity that depends on the data and the unknown model parameters and denote with $\bl{Y}_{1:t^*}^{new}$ a new vector of observations sampled from the posterior predictive distribution, as outlined in equation (\ref{eqn:post_pred_distrib}). To describe the degree of the discrepancy, we use the Bayesian $p$-value, which is the probability of observing a test quantity at least as extreme as the observed data, $T(\bl{Y}_{1:t^*}^{new}, \bs{\vartheta})$, we denote this by

\begin{equation}
\label{eqn:Bayesian_pvalue}
p_B = \Pr(T(\bl{Y}_{1:t^*}^{new}, \bs{\vartheta}) \geq T(\bl{Y}_{1:t^*}, \bs{\vartheta}) | \bl{Y}_{1:t^*}). 
\end{equation}

Unlike in frequentist statistics where a $p$-value near $0$ indicates that the corresponding null hypothesis can be rejected, an extreme Bayesian $p$-value denotes that the specific feature of the data captured by the test quantity is inconsistent with the assumed model. For example, if we suspect that our model may not be able to reproduce the large values observed in the data, a suitable test quantity could be the observations' maximum. In this case, a $p$-value near $0$ indicates that, under the assumed model, it is unlikely to encounter a value larger than the observed maximum; so, if the replicated data were generated under a Normal model, a heavy tail distribution may be more appropriate. A Bayesian $p$-value can be estimated by computing the proportion of replicated data sets satisfying (\ref{eqn:Bayesian_pvalue}). 

We can also provide a graphical representation by plotting the distribution of the test quantity against the observed test quantity; as in a classical setting, the Bayesian $p$-value is the right tail-area probability. Another graphical check consists of computing the posterior predictive mean (i.e., the mean of the posterior predictive distribution) and then plotting it against the distribution of the observed data. Generally, graphical model checks are useful for highlighting the systematic discrepancies between the observed and the simulated data.

Finally, for both linear and non-linear regression models, we can also assess the goodness of fit using residual plots. We can think of Bayesian model residuals as a generalization of classical residuals that accounts for the uncertainty in the model parameters.
%, whereas the latter are typically based on a point estimate of the parameters. 

In Section \ref{sect:empirical_analysis}, we extensively used posterior predictive checks to select and validate the model used for our empirical analysis.

\newpage

\subsection{Sensitivity analysis}
\label{appB_sensitivity}
Model validation performed through posterior predictive checks shows that the structural time series model with a trend and seasonal component adequately describe the data (see \ref{appA_plots} and \ref{appA_postpred} for the details). Nonetheless, posterior inference might still be affected by prior assumptions. Thus, to strengthen our confidence in the assumed model, we performed a sensitivity analysis in order to evaluate to what extent our inferred causal effect changes to different values of the prior hyperparameters. 

As described in Section \ref{subsect:priors}, for the unknown scale matrices of the Inverse-Wishart distributions we chose the following variance-covariance matrix,

$$\bl{S}_{\varepsilon} = \bl{S}_r  = \begin{bmatrix}
hs^2_1 & \sqrt{hk} s_1 s_2 \rho \\
\sqrt{hk} s_1 s_2 \rho & k s^2_2 
\end{bmatrix},$$
where, $s^2_1$,$s^2_2$ are the sample variances, which can be scaled by some positive values $h$ and $k$, and $\rho$ is the correlation coefficient. Linking the scale matrix to the sample variances is in line with an objective Bayesian approach and can ensure a reasonable scale for the prior \citep{Brodersen:Gallusser:Koehler:Remy:Scott:2015}. In our empirical analysis, we set $h = k = 1$ but we could have used different values. For example, since the sample variance of the competitor brands is, on average, ten times higher than the sample variance of the store brands, another reasonable scaling can be obtained by setting $h = 0.1$, $k = 1$. Table \ref{tab:k_sensitivity} presents the estimated causal effects under different assumptions for the scaling factors. 

Another parameter that can influence our posterior inference is the linear correlation coefficient. We set $\rho = -0.8$ based on our prior belief that the two products in the pair are perfect substitutes, but the correlation might be smaller than what assumed or even positive. Table \ref{tab:rho_sensitivity} shows  the estimated causal effects under different combinations of the correlation and the scaling factors.  

Finally, we assumed $\bl{S}_{\varepsilon} = \bl{S}_r$ but we can also allow the state disturbances to vary more (less) freely than the observation disturbances. The estimated effects under different assumptions for $\bl{S}_r$ are reported in Table \ref{tab:sr_sensitivity}.

Overall, our estimates seem to be robust to different prior assumptions: even if in some instances we find only one or two significant effects, this still supports our general conclusion that the new price policy had only a minor impact on the sales of store brands.

\begin{table}[ht]
\centering
\caption{Temporal average general effect estimates at one month horizon under different prior assumptions for the scaling factors $h$ and $k$.}
\label{tab:k_sensitivity}
\begin{tabular}{lrrrrrrrrrrrr}
  \toprule
& & \multicolumn{3}{c}{$h = 1$, $k = 0.01$} & & \multicolumn{3}{c}{$h = 1$, $k = 0.1$} & & \multicolumn{3}{c}{$h = 1$, $k = 1$} \\ \cmidrule{3-5} \cmidrule{7-9} \cmidrule{11-13}
& & $\hat{\bar{\bs{\tau}}}_t$ & $2.5\%$ & $97.5\%$ & & $\hat{\bar{\bs{\tau}}}_t$ & $2.5\%$ & $97.5\%$ & & $\hat{\bar{\bs{\tau}}}_t$ & $2.5\%$ & $97.5\%$ \\ \midrule
 \multirow{2}{*}{(1)} 
 & s & 7.24  & -23.37  & 36.89  & & 7.25  & -23.42  & 37.53  & & 6.97  & -24.25  & 38.47 \\ 
 & c & 23.55 & -126.97 & 178.70 & & 24.34 & -118.60 & 168.55 & & 24.89 & -101.30 & 153.64 \\ 
 \multirow{2}{*}{(2)}
 & s & 7.41  & -13.63  & 30.46  & & 7.15  & -13.80  & 29.48  & & 7.02  & -14.79  & 28.90 \\ 
 & c & 12.33 & -87.37  & 114.44 & & 13.33 & -77.80  & 106.46 & & 14.71 & -62.26  & 99.44 \\ 
 \multirow{2}{*}{(3)}
 & s & 7.46  & -15.24  & 29.66  & & 7.68  & -15.17  & 30.33  & & 7.94  & -14.08  & 32.26 \\ 
 & c & 13.57 & -76.17  & 100.46 & & 14.39 & -70.34  & 95.71  & & 15.42 & -62.17  & 90.81 \\
 \multirow{2}{*}{(4)} 
 & s & \bf{47.19} & \bf{0.25}    & \bf{94.40}  & & \bf{47.25} & \bf{1.05}    & \bf{94.28}  & & \bf{47.84} & \bf{4.71}    & \bf{96.82} \\ 
 & c & 26.28 & -101.06 & 150.07 & & 26.93 & -93.87  & 142.41 & & 28.86 & -77.93  & 135.93 \\ 
 \multirow{2}{*}{(5)}
 & s & 3.60  & -44.33  & 52.00  & & 3.46  & -45.09  & 53.48  & & 4.11  & -46.65  & 54.64 \\ 
 & c & 41.69 & -82.11  & 159.68 & & 43.67 & -72.81  & 157.08 & & 45.47 & -63.13  & 154.24 \\ 
 \multirow{2}{*}{(6)}
 & s & 9.43  & -13.15  & 33.27  & & 9.48  & -13.45  & 33.57  & & 9.53  & -14.45  & 33.68 \\ 
 & c & 22.83 & -52.50  & 95.71  & & 23.33 & -47.52  & 92.92  & & 25.64 & -37.88  & 93.36 \\ 
 \multirow{2}{*}{(7)}
 & s & \bf{79.87} & \bf{12.19}   & \bf{151.16} & & \bf{78.25} & \bf{5.65}    & \bf{148.78} & & \bf{78.19} & \bf{0.15}    & \bf{154.08} \\ 
 & c & 165.50& -313.51 & 621.07 & & 180.33& -262.04 & 644.09 & & 182.70& -221.16 & 600.08 \\ 
 \multirow{2}{*}{(8)}
 & s & 24.79 & -25.48  & 78.87  & & 25.20 & -28.56  & 75.59  & & 25.23 & -28.60  & 78.16 \\ 
 & c & 14.90 & -16.30  & 47.43  & & 15.83 & -15.80  & 47.50  & & 15.91 & -15.15  & 47.53 \\ 
 \multirow{2}{*}{(9)}
 & s & 40.54 & -9.93   & 91.72  & & 40.34 & -10.24  & 89.36  & & 40.29 & -9.84   & 90.38 \\ 
 & c & 15.91 & -31.49  & 64.54  & & 16.63 & -31.47  & 66.75  & & 17.17 & -30.76  & 68.56 \\ 
 \multirow{2}{*}{(10)}
 & s & \bf{12.39} & \bf{0.81}    & \bf{23.67}  & & \bf{12.43} & \bf{1.00}    & \bf{23.82}  & & \bf{12.43} & \bf{1.35}    & \bf{23.64} \\ 
 & c & 0.06  & -9.02   & 9.56   & & 0.16  & -8.78   & 9.37   & & 0.04  & -9.36   & 9.79 \\ \midrule
& & \multicolumn{3}{c}{$h = 0.1$, $k = 0.01$} & & \multicolumn{3}{c}{$h = 0.1$, $k = 0.1$} & & \multicolumn{3}{c}{$h = 0.1$, $k = 1$} \\ \cmidrule{3-5} \cmidrule{7-9} \cmidrule{11-13}
& & $\hat{\bar{\bs{\tau}}}_t$ & $2.5\%$ & $97.5\%$ & & $\hat{\bar{\bs{\tau}}}_t$ & $2.5\%$ & $97.5\%$ & & $\hat{\bar{\bs{\tau}}}_t$ & $2.5\%$ & $97.5\%$ \\ \midrule
 \multirow{2}{*}{(1)}
 & s & 7.96  & -19.49 & 37.36  & & 7.82  & -20.49 & 38.49  & & 7.56  & -21.54  & 38.87 \\ 
 & c & 19.89 & -122.41& 158.93 & & 19.07 & -119.11& 151.84 & & 20.61 & -104.27 & 142.84 \\ 
 \multirow{2}{*}{(2)}
 & s & 7.13  & -11.29 & 27.24  & & 6.75  & -12.31 & 26.95  & & 6.71  & -12.91  & 26.21 \\ 
 & c & 13.14 & -82.89 & 111.07 & & 13.44 & -80.64 & 102.26 & & 14.72 & -62.25  & 99.20 \\ 
 \multirow{2}{*}{(3)}
 & s & 7.50  & -13.08 & 27.75  & & 7.60  & -13.25 & 28.14  & & 7.83  & -11.88  & 29.91 \\ 
 & c & 13.80 & -74.26 & 98.68  & & 14.18 & -70.76 & 92.72  & & 15.30 & -61.50  & 91.00 \\ 
 \multirow{2}{*}{(4)}
 & s & \bf{47.72} & \bf{2.69}   & \bf{93.21}  & & \bf{47.66} & \bf{2.99}   & \bf{93.63}  & & \bf{47.99} & \bf{4.60}    & \bf{94.78} \\ 
 & c & 25.65 & -98.05 & 146.90 & & 26.33 & -92.66 & 139.63 & & 29.72 & -78.09  & 135.79 \\ 
 \multirow{2}{*}{(5)}
 & s & 5.15  & -49.86 & 60.98  & & 4.46  & -51.98 & 62.41  & & 5.46  & -53.97  & 66.87 \\ 
 & c & 46.15 & -74.29 & 174.59 & & 43.95 & -69.46 & 154.96 & & 48.31 & -55.84  & 157.87 \\ 
 \multirow{2}{*}{(6)}
 & s & 8.80  & -15.48 & 34.40  & & 9.04  & -17.71 & 36.31  & & 8.90  & -15.72  & 34.29 \\ 
 & c & 22.76 & -51.75 & 95.12  & & 23.10 & -47.28 & 91.12  & & 25.19 & -37.30  & 91.33 \\ 
 \multirow{2}{*}{(7)}
 & s & 75.17 & -5.10  & 156.68 & & 77.54 & -3.56  & 160.93 & & 74.41 & -9.53   & 158.00 \\ 
 & c & 186.83& -289.26& 671.61 & & 177.16& -250.44& 586.05 & & 190.17& -201.93 & 593.67 \\ 
 \multirow{2}{*}{(8)}
 & s & 24.07 & -32.97 & 77.91  & & 24.17 & -31.64 & 77.64  & & 24.35 & -31.78  & 78.01 \\ 
 & c & 15.42 & -16.28 & 46.81  & & 15.61 & -15.21 & 46.83  & & 15.80 & -15.42  & 46.50 \\ 
 \multirow{2}{*}{(9)}
 & s & 38.07 & -15.31 & 92.90  & & 38.30 & -14.32 & 91.26  & & 37.89 & -13.39  & 88.85 \\ 
 & c & 16.59 & -32.28 & 67.16  & & 16.44 & -31.49 & 64.43  & & 17.27 & -30.87  & 66.77 \\ 
 \multirow{2}{*}{(10)}
 & s & 11.56 & -1.61  & 25.12  & & 11.73 & -1.12  & 24.18  & & \bf{12.00} & \bf{0.02}    & \bf{23.85} \\ 
 & c & 0.25  & -8.65  & 9.83   & & 0.30  & -8.71  & 9.68   & & 0.00 & -9.46   & 9.77 \\ 
   \bottomrule
\end{tabular}
\end{table}

\newgeometry{bottom=2cm, top=1.5cm, left=2cm, right=2cm}
\begin{table}[h!]
\centering
\caption{Temporal average general effect estimates at one month horizon under different prior assumptions for the scaling factors $h$, $k$ and the linear correlation coefficient $\rho$.}
\label{tab:rho_sensitivity}
\scalebox{1}{
\begin{tabular}{lrrrrrrrrrrrr}
  \toprule
  & & \multicolumn{3}{c}{$h = 1$, $k = 0.1$, $\rho = -0.3$} & & \multicolumn{3}{c}{$h = 1$, $k = 1$, $\rho = -0.3$} & & \multicolumn{3}{c}{$h = 0.1$, $k = 1$, $\rho = +0.3$} \\ \cmidrule{3-5} \cmidrule{7-9} \cmidrule{11-13}
& & $\hat{\bar{\bs{\tau}}}_t$ & $2.5\%$ & $97.5\%$ & & $\hat{\bar{\bs{\tau}}}_t$ & $2.5\%$ & $97.5\%$ & & $\hat{\bar{\bs{\tau}}}_t$ & $2.5\%$ & $97.5\%$ \\ \midrule
  \multirow{2}{*}{(1)}
& s & 7.20 & -23.85 & 38.44 & & 7.13 & -24.72 & 39.13 & & 7.88 & -21.90 & 39.20 \\ 
& c & 23.73 & -116.98 & 167.21 & & 24.14 & -99.19 & 152.88 & & 20.89 & -101.96 & 144.23 \\ 
  \multirow{2}{*}{(2)}
& s & 7.24 & -14.58 & 30.35 & & 7.12 & -15.19 & 29.16 & & 6.85 & -13.39 & 28.50 \\ 
& c & 13.65 & -71.93 & 110.44 & & 14.63 & -61.80 & 99.32 & & 14.61 & -66.66 & 93.83 \\ 
  \multirow{2}{*}{(3)}
& s & 7.70 & -15.19 & 30.47 & & 7.94 & -13.96 & 32.43 & & 8.03 & -12.18 & 30.54 \\ 
& c & 14.52 & -69.13 & 93.31 & & 15.36 & -60.10 & 91.26 & & 15.07 & -60.56 & 90.46 \\ 
  \multirow{2}{*}{(4)}
& s & \bf{47.31} & \bf{0.14} & \bf{95.55} & & \bf{48.05} & \bf{4.46} & \bf{97.81} & & \bf{48.20} & \bf{3.77} & \bf{96.39} \\ 
& c & 26.54 & -92.09 & 141.12 & & 28.17 & -76.13 & 134.14 & & 27.36 & -80.92 & 133.40 \\ 
  \multirow{2}{*}{(5)}
& s & 3.92 & -42.58 & 51.27 & & 4.28 & -42.17 & 53.99 & & 5.26 & -48.09 & 59.97 \\ 
& c & 44.02 & -69.30 & 152.63 & & 48.36 & -54.29 & 155.36 & & 47.98 & -52.36 & 154.19 \\ 
  \multirow{2}{*}{(6)}
& s & 9.60 & -11.69 & 32.73 & & 9.55 & -12.56 & 31.58 & & 9.38 & -13.30 & 32.70 \\ 
& c & 23.86 & -44.24 & 89.96 & & 25.89 & -35.07 & 89.86 & & 25.68 & -35.01 & 89.41 \\ 
  \multirow{2}{*}{(7)}
& s & \bf{79.00} & \bf{6.63} & \bf{148.42} & & \bf{78.96} & \bf{1.67} & \bf{154.79} & & 76.86 & -9.14 & 165.34 \\ 
& c & 187.38 & -244.11 & 635.17 & & 190.74 & -198.58 & 596.36 & & 187.56 & -202.75 & 572.28 \\ 
  \multirow{2}{*}{(8)}
& s & 25.66 & -28.03 & 76.80 & & 25.65 & -26.71 & 79.36 & & 25.04 & -35.08 & 83.11 \\ 
& c & 16.24 & -15.76 & 48.84 & & 16.09 & -15.54 & 46.68 & & 16.17 & -14.73 & 48.06 \\ 
  \multirow{2}{*}{(9)}
& s & 40.33 & -10.75 & 90.49 & & 40.49 & -8.66 & 89.99 & & 38.33 & -13.22 & 90.34 \\ 
& c & 17.34 & -30.12 & 65.67 & & 17.64 & -31.95 & 67.90 & & 17.64 & -32.55 & 68.40 \\ 
  \multirow{2}{*}{(10)}
& s & \bf{12.37} & \bf{0.88} & \bf{23.58} & & \bf{12.39} & \bf{0.71} & \bf{23.66} & & 11.53 & -1.71 & 24.82 \\ 
& c & 0.28 & -8.44 & 9.53 & & -0.09 & -9.77 & 9.99 & & 0.05 & -9.42 & 9.85 \\ 
\bottomrule
\end{tabular}
}
\end{table}

\begin{table}[h!]
\centering
\caption{Temporal average general effect estimates at one month horizon for $h = k = 1$, $\rho = 1$ and under different prior assumptions for $\bl{S}_r$.}
\label{tab:sr_sensitivity}
\scalebox{1}{
\begin{tabular}{lrrrrrrrrrrrr}
  \toprule
  & & \multicolumn{3}{c}{$\bl{S}_r = 0.5 \bl{S}_{\varepsilon}$} & & \multicolumn{3}{c}{$\bl{S}_r = 2 \bl{S}_{\varepsilon}$} & & \multicolumn{3}{c}{$\bl{S}_r = 10 \bl{S}_{\varepsilon}$} \\ \cmidrule{3-5} \cmidrule{7-9} \cmidrule{11-13}
& & $\hat{\bar{\bs{\tau}}}_t$ & $2.5\%$ & $97.5\%$ & & $\hat{\bar{\bs{\tau}}}_t$ & $2.5\%$ & $97.5\%$ & & $\hat{\bar{\bs{\tau}}}_t$ & $2.5\%$ & $97.5\%$ \\ \midrule
  \multirow{2}{*}{(1)}
& s & 6.90 & -23.91 & 37.77 & & 7.10 & -26.11 & 39.83 & & 7.09 & -34.28 & 46.37 \\ 
& c & 23.86 & -102.42 & 152.62 & & 24.79 & -101.46 & 154.68 & & 25.29 & -115.70 & 171.43 \\ 
\multirow{2}{*}{(2)}
& s & 6.97 & -13.72 & 27.83 & & 7.06 & -16.70 & 31.65 & & 7.26 & -24.02 & 37.26 \\ 
& c & 14.22 & -63.03 & 100.72 & & 14.81 & -65.48 & 97.76 & & 15.80 & -72.62 & 110.79 \\ 
\multirow{2}{*}{(3)}
& s & 7.92 & -12.94 & 31.16 & & 7.94 & -15.49 & 34.35 & & 8.05 & -22.00 & 41.74 \\ 
& c & 15.09 & -63.38 & 91.06 & & 15.65 & -62.26 & 91.39 & & 15.68 & -73.12 & 103.23 \\ 
\multirow{2}{*}{(4)}
& s & \bf{48.08} & \bf{4.49} & \bf{95.57} & & \bf{47.62} & \bf{2.24} & \bf{98.14} & & 47.59 & -8.74 & 110.51 \\ 
& c & 27.52 & -78.23 & 131.69 & & 29.77 & -78.38 & 136.75 & & 29.36 & -90.11 & 152.06 \\ 
\multirow{2}{*}{(5)}
& s & 4.08 & -45.71 & 55.58 & & 3.32 & -47.07 & 55.74 & & 5.33 & -52.97 & 67.21 \\ 
& c & 45.10 & -63.01 & 152.22 & & 49.28 & -58.20 & 163.80 & & 46.12 & -81.66 & 170.30 \\ 
\multirow{2}{*}{(6)}
& s & 9.42 & -13.67 & 32.68 & & 9.60 & -14.88 & 34.67 & & 9.29 & -19.14 & 37.63 \\ 
& c & 23.53 & -42.49 & 87.81 & & 25.84 & -38.52 & 94.63 & & 26.15 & -47.89 & 102.64 \\ 
\multirow{2}{*}{(7)}
& s & \bf{77.94} & \bf{0.62} & \bf{153.30} & & 78.39 & -0.28 & 156.90 & & 82.02 & -9.49 & 180.16 \\ 
& c & 184.04 & -215.19 & 594.20 & & 181.14 & -233.10 & 602.97 & & 169.11 & -307.08 & 619.20 \\ 
\multirow{2}{*}{(8)}
& s & 24.70 & -27.11 & 75.79 & & 25.79 & -30.32 & 81.06 & & 26.76 & -37.56 & 90.71 \\ 
& c & 15.90 & -14.53 & 47.33 & & 16.08 & -15.65 & 49.14 & & 16.38 & -21.24 & 56.59 \\ 
\multirow{2}{*}{(9)}
& s & 39.83 & -9.11 & 89.22 & & 41.04 & -9.98 & 92.03 & & 41.00 & -20.73 & 103.17 \\ 
& c & 16.81 & -31.86 & 66.30 & & 16.84 & -36.00 & 69.79 & & 17.26 & -45.31 & 79.88 \\ 
\multirow{2}{*}{(10)}
& s & \bf{12.42} & \bf{1.28} & \bf{23.61} & & \bf{12.61} & \bf{1.34} & \bf{24.15} & & 12.71 & -0.72 & 25.90 \\ 
& c & 0.09 & -9.13 & 9.35 & & -0.12 & -10.10 & 10.10 & & -0.50 & -12.44 & 12.13 \\ \bottomrule
\end{tabular}
}
\end{table}

\restoregeometry

\clearpage

\subsection{Convergence diagnostics}
\label{appB_convergence}
To make inference with Markov Chain Monte Carlo (MCMC) methods we need to verify that our Markov chain has converged to the stationary distribution. Geweke's diagnostic test \citep{Geweke:1992} compares the sample means of two non-overlapping quantiles of the chain (for example, the first $10\%$ and the last $50\%$ of the draws). If the draws are sampled from the same stationary distribution, the sample means are equal and the test statistic is asympotically Normal. 

Table \ref{tab:geweke} shows the resulting p-value for the two-sided test for every parameter of the bivariate models estimated on the $10$ store-competitor pairs. The Geweke diagnostic fails to detect non-convergence of the chains to the stationary distribution (at the $5\%$ level, the test fails to reject the null hypothesis of the equality of means in $81$ cases out of $90$). 

\begin{figure}
\centering
\caption{Trace plots of the variance-covariance matrices of the model estimated on the first store-competitor pair.}
\label{fig:trace1}
\includegraphics[scale=0.5]{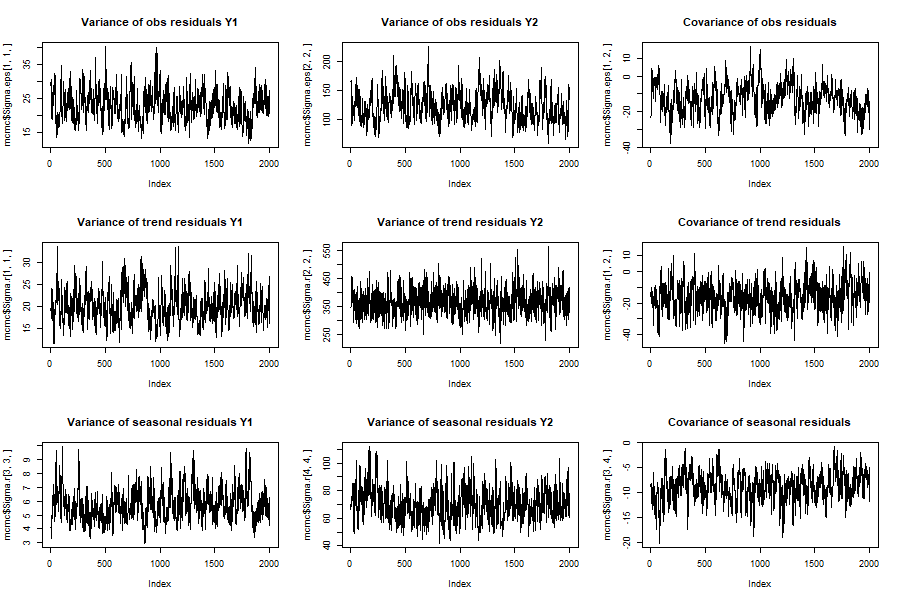}
\caption{Trace plots of the variance-covariance matrices of the model estimated on the second store-competitor pair.}
\label{fig:trace2}
\includegraphics[scale=0.5]{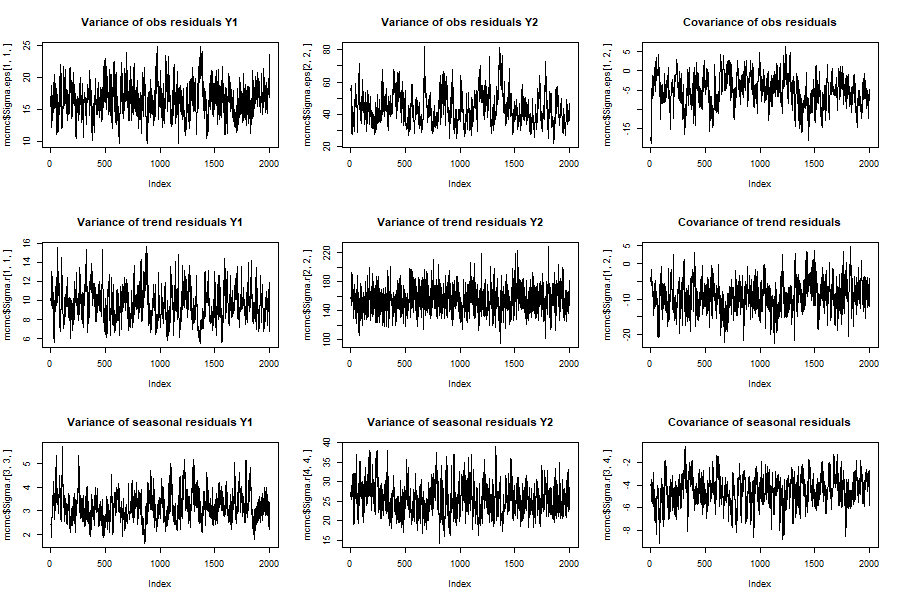}
\end{figure}

Finally, for a visual inspection of the chain convergence, we also include the trace plots for the parameters of the first two models (Figures \ref{fig:trace1} and \ref{fig:trace2}).
   
\begin{table}[ht]
\centering
\caption{Geweke's diagnostics at the lower $10\%$ and upper $50\%$ quantiles. In this table, $\sigma^2_i$, $i \in \{1,2 \}$ and $\sigma_{1,2}$ indicate, respectively, the variances and the covariance of the observation disturbances;  $\sigma^2_{\mu_i}$ and $\sigma_{\mu_{1,2}}$ the variances and the covariance of the trend disturbances; $\sigma^2_{\gamma_i}$ and $\sigma_{\gamma_{1,2}}$ the variances and the covariance of the disturbances of the seasonal component.}
\label{tab:geweke}
\begin{tabular}{rrrrrrrrrr}
  \toprule
 & $\sigma^2_1$ & $\sigma^2_2$ & $\sigma_{1,2}$ & $\sigma^2_{\mu_1}$ & $\sigma^2_{\mu_2}$ & $\sigma_{\mu_{1,2}}$ & $\sigma^2_{\gamma_1}$ & $\sigma^2_{\gamma_2}$ & $\sigma_{\gamma_{1,2}}$ \\ \midrule
  1 & 0.41 & 0.92 & 0.75 & 0.59 & 0.32 & 0.71 & 0.95 & 0.03 & 0.03 \\ 
  2 & 0.94 & 0.75 & 0.25 & 0.71 & 0.19 & 0.55 & 0.88 & 0.00 & 0.01 \\ 
  3 & 0.62 & 0.99 & 0.64 & 0.88 & 0.52 & 0.80 & 0.98 & 0.01 & 0.06 \\ 
  4 & 0.55 & 0.83 & 0.96 & 0.71 & 0.16 & 0.67 & 0.65 & 0.02 & 0.07 \\ 
  5 & 0.59 & 0.98 & 0.65 & 0.97 & 0.27 & 0.73 & 0.88 & 0.20 & 0.30 \\ 
  6 & 0.76 & 0.83 & 0.81 & 0.39 & 0.17 & 0.98 & 0.90 & 0.17 & 0.28 \\ 
  7 & 0.81 & 0.54 & 0.16 & 0.53 & 0.47 & 0.04 & 0.89 & 0.18 & 0.10 \\ 
  8 & 0.29 & 0.78 & 0.47 & 0.40 & 0.45 & 0.83 & 0.30 & 0.64 & 0.02 \\ 
  9 & 0.86 & 0.65 & 0.46 & 0.39 & 0.24 & 0.91 & 0.95 & 0.87 & 0.11 \\ 
  10 & 0.81 & 0.05 & 0.91 & 0.53 & 0.34 & 0.79 & 0.72 & 0.35 & 0.04 \\  
   \bottomrule
\end{tabular}
\end{table}

\clearpage

\end{appendices}

\end{document}